\documentclass[lettersize,journal]{IEEEtran}
\usepackage{graphicx}
\usepackage{dcolumn}
\usepackage{bm}
\usepackage{epstopdf}
 \usepackage[usenames,dvipsnames]{pstricks}
 \usepackage{epsfig}
 \usepackage{pst-grad} 
 \usepackage{pst-plot} 
 \usepackage[space]{grffile} 
 \usepackage{etoolbox} 
 \makeatletter 
 \patchcmd\Gread@eps{\@inputcheck#1 }{\@inputcheck"#1"\relax}{}{}
 \makeatother
\usepackage{booktabs}
\usepackage{boxhandler}
\usepackage{eurosym}
\usepackage[whole]{bxcjkjatype}

\usepackage{floatrow}
\usepackage[utf8]{inputenc}
\usepackage{braket}
\usepackage[cmintegrals]{newtxmath}
\usepackage{multirow}
\usepackage{enumitem}
\usepackage{pifont}

\usepackage{colortbl} 

\usepackage{amsmath,amsfonts}
\usepackage{algorithmic}
\usepackage{array}
\usepackage{textcomp}
\usepackage{stfloats}
\usepackage{url}
 \def\BibTeX{{\rm B\kern-.05em{\sc i\kern-.025em b}\kern-.08em
     T\kern-.1667em\lower.7ex\hbox{E}\kern-.125emX}}
\usepackage{balance}

\newcommand{\R}{\mathbb{R}}

\newcommand{\xmark}{\ding{55}}
\newcommand{\cmark}{\ding{51}}
\usepackage[dvipsnames]{xcolor}
\usepackage{color,soul}
\colorlet{r1}{red!60}
\colorlet{r2}{Red!20}
\colorlet{r3}{RoyalPurple!20}
\newcommand{\Hl}[2][\empty]{%
    \ifx#1\empty
    \else
    \sethlcolor{#1}%
    \fi
    \hl{#2}}
\soulregister\Hl{7}
\soulregister\cite7
\soulregister\ref7
\soulregister\pageref7
\usepackage{etoolbox}
\makeatletter 
\pretocmd\@bibitem{\csname keycolor#1\endcsname}{}{\fail}
\newcommand\citecolor[3][1]{\@namedef{keycolor#3}{\hspace*{-\labelwidth}\hspace*{-\labelsep}{\color{#2}\rule[-0.3em]{\dimexpr\linewidth+\labelwidth+\labelsep\relax}{#1\baselineskip}}\vspace*{\itemsep}\vspace*{-#1\baselineskip}}}
\makeatother

\begin{document}

\title{Cybersecurity in Critical Infrastructures: A Post-Quantum Cryptography Perspective}

\author{Javier {Oliva delMoral}, Antonio {deMarti iOlius}, Gerard Vidal, Pedro M. Crespo, Josu {Etxezarreta Martinez}
\thanks{J. Oliva delMoral, A. deMarti iOlius, Professor P.M. Crespo and Dr. J. Etxezarreta Martinez are with the Department of Basic Sciences, Tecnun - University of Navarra, 20018 Donostia-San Sebastian, Spain (e-mail: jolivam@unav.es; ademartio@unav.es; pcrespo@unav.es; jetxezarreta@unav.es).}
\thanks{J.Oliva delMoral is with Donostia International Physics Center (DIPC), 20018 Donostia-San Sebastian, Spain.}
\thanks{Dr. G. Vidal is with Opscura, Portuetxe K., 23b, oficina 100, 20018 Donostia-San Sebastian, Spain (e-mail: gerard@opscura.io)}
}

\maketitle

\begin{abstract}

The machinery of industrial environments was connected to the Internet years ago with the scope of increasing their performance. However, this change made such environments vulnerable against cyber-attacks that can compromise their correct functioning resulting in economic or social problems. Moreover, implementing cryptosystems in the communications between operational technology (OT) devices is a more challenging task than for information technology (IT) environments since the OT networks are generally composed of legacy elements, characterized by low-computational capabilities. Consequently, implementing cryptosystems in industrial communication networks faces a trade-off between the security of the communications and the amortization of the industrial infrastructure. Critical Infrastructure (CI) refers to the industries which provide key resources for the daily social and economical development, e.g. electricity. Furthermore, a new threat to cybersecurity has arisen with the theoretical proposal of quantum computers, due to their potential ability of breaking state-of-the-art cryptography protocols, such as RSA or ECC. Many global agents have become aware that transitioning their secure communications to a quantum secure paradigm is a priority that should be established before the arrival of fault-tolerance. In this paper, we aim to describe the problematic of implementing post-quantum cryptography (PQC) to CI environments. For doing so, we describe the requirements for these scenarios and how they differ against IT. We also introduce classical cryptography and how quantum computers pose a threat to such security protocols. Furthermore, we introduce state-of-the-art proposals of PQC protocols and present their characteristics. We conclude by discussing the problematic of integrating PQC in industrial environments.
\end{abstract}

\section{Introduction} \label{Sec: I}
\noindent
The exponential development of communication technologies in the late 20$^{th}$ century and, specially, in the early 21$^{st}$ century has resulted in a contemporary society that exists in a hyperconnected world. In this paradigm, communications do not only refer to the actions of texting, phone (video) calls, social media or news media but also to the control of industrial machines, bank transfers, stock acquisitions, control of unmanned aerial vehicles (UAV) or managing automated houses (domotics), to name a few. Furthermore, strongly tangled concepts such as Smart Cities, Industry 4.0 or the Internet of Things (IoT) are currently being investigated for their convergence with other advanced technologies such as Artificial Intelligence (AI) or Quantum Computing (QC) on a historical inflection point in the form of a Fourth Industrial Revolution \cite{fourthrev} .

In this sense, relying on communications for executing the critical tasks involved in such hyperconnected paradigm requires that those transmissions of information are secure and private. Cyber vulnerabilities in the control systems of a smart city or an automated industry may lead to catastrophic consequences. For example, in a possible future where the transport of people and cargo is exclusively done by autonomous vehicles which rely on the communications among them and some central control stations to move around, the intrusion of a malicious entity on the system to disturb it would lead to fatal consequences economically and socially (casualties). Hence, modern crime and war is heavily based on hacking activities with the scope of manipulating critical infrastructures (CI), to produce economical or social losses by interrupting their production or by decreasing the life-time of their devices,  or obtaining sensitive information regarding state, industrial or personal secrets (banking information or sensitive images, for example). This paradigm of cybercrime and cyberwar is present nowadays with an estimated amount of $2200$ known cyberattacks per day in 2022, posing a threat to the business' infrastructure every $39$ seconds \cite{cyberStats}. Indeed, awareness on cyberattacks among Nation-state actors is increasing due to current geopolitical tensions, as seen recently \cite{CISA_Cyberavengers}. It is due to all these factors that concepts such as the Cyber Apocalypse are being coined to describe the fear that a cyberattack to CI's systems and networks of a country would led to shutting down their capabilities regarding civilian and military services. It is important to state that the possibility of major devastation in the CI of a nation does not have to imply that all the systems consisting it should be attacked, the failure of parts of the structure may lead to a catastrophic propagation of failures through the whole network due to the interconnection among the elements. This effect is known as cascading effect \cite{cascade}.

All of these vulnerabilities make cybersecurity and cryptography to be the pillars to erect the previously described paradigmatic society in a security way. Cybersecurity is defined as the practice of protecting important systems and confidential information from cyberattacks. In this sense, many methods and elements are used for the sake of protecting communication and computer networks, but the algorithms that are employed to cipher sensitive data being communicated in such meshes relate to the field of cryptography. Hence, in this society where the quote ``Information is power'' is getting more and more relevant, the use of such practices is of capital relevance. Importantly, the proposal of the RSA or ECC asymmetric cryptographic systems has maintained the security of communication systems for over 40 years \cite{RSA,eec1,eec2}. The core of those protocols resides in the fact that they are based on hard problems that cannot be solved in a practical time frame by classical computing methods, i.e. thousands of years of computing are required to extract the plain text from the ciphertext if the key is unknown. Unluckily, quantum computers have posed a threat to the security of those asymmetric cryptography protocols. Shor's algorithm is a theoretical quantum algorithm that provides an exponential speedup for solving prime number factorization and the computation of discrete logarithms respectively, which are the hard problems in which the security of the previously commented protocols relies upon \cite{shorFact}. At the current time, quantum computers that can execute such algorithm efficiently and correctly only exists as a theoretical promise. Nonetheless, the past years, quantum computing has proven to be a rapidly evolving field with the achievement of milestones such as the first experimental realizations of quantum advantage \cite{sup1,sup2,sup3,sup4,sup5} or quantum error correction \cite{wallraffSurf,googleSurf}. Such tremendous advancements have made educated voices to estimate the appearance of efficient quantum computers able to make state-of-the-art asymmetric protocols to be deprecated to be within the range of one to two decades \cite{Mosca}. Hence, many have raised the alarm of a possible ``Quantum Apocalypse'' that would result in sensitive data and systems to become completely vulnerable. 

Fortunately, there is hope for making the computer and machine networks of the future to be secure in the fault-tolerant quantum computing era due to the proposals of QKD and PQC. The first refers to using the properties of quantum mechanics in order to secure and transmit information \cite{QKDrev}. This paradigm includes important protocols for QKD such as BB84 \cite{bb84} or E91 \cite{e91}. Although being a very promising candidate for a quantum-safe future, QKD is still a nascent technology posing many challenges that include technical complexity and cost as well as the requirement of sophisticated infrastructure. This comes with the added requirement of still needing to deal with the noise, loss and decoherence that limit the performance of quantum communication systems, e.g. quantum repeaters are being investigated for solving such problem \cite{repeaters}. Hence, PQC has been proposed as the paradigm of classical cryptography schemes that are secure against attackers that have access to fault-tolerant quantum computers \cite{Bernstein2017-zk}. Quantum computers do not provide an exponential speedup to solve every computer science problem \cite{Qcomplexity} and, therefore, the main idea in PQC would be to find hard problems that cannot be efficiently tackled by such technology, even if it is fully operational. Obviously, this should be done in conjunction with security against classical attacks, since PQC protocols would be useless if they were still vulnerable to traditional hacking. The importance of migrating to quantum-secure cryptography has not gone unnoticed for many countries with an open PQC protocol standardization process like the United States National Institute of Standards and Technology (NIST) \cite{NISTstand} or like Europe with the quantum cybersecurity agenda by the European Policy Centre (EPC) \cite{EUagenda}. Several PQC protocols such as hash-, lattice- or code-based cryptography have been proposed as a way of allowing secure communications on the networks of the future. Interestingly, even Google has decided to introduce PQC protocols in their Chrome browser \cite{googlePQC}, announcing that they will admit the use of the X25519Kyber768 protocol to encrypt Transport Level Security (TLS) connections. Such protocol is a combination of a classical ECC based protocol and the lattice-based CRYSTALS-Kyber \cite{Kyber} PQC algorithm, which is one of the algorithms selected at this point by the NIST for standardization. 

Each PQC protocol has its own benefits and disadvantages in terms of security levels, ciphertext size or speed, among other benchmarks. This implies that the selection of PQC protocols is very application-dependent in the sense that as a function of the requirements of a specific system, an approach could be valid or not. Following this logic, PQC protocols are usually proposed for systems in which the cybersecurity is the most critical requirement (IT services), while the latency because of the introduction of those cryptography protocols can be deemed as not too important. However, latency is a key performance parameter in Industrial Control Systems (ICS) and CI, where introducing a delay over the system requirements can imply a failure that cannot be tolerated in such environments \cite{delayICS}. This should obviously be done maintaining a certain level of security on the system. Additionally, it is important to state that implementing cryptography in such networks is done by means of processors that are not powerful enough to manage huge key sizes, mainly because the introduction of such systems should be somehow seamless to the existing communication infrastructure and cheap\footnote{We can speculate that QKD will not be a major player to secure such networks due the fact that they are very costly as well as they require significant infrastructure to be deployed.}. It is in this sense that, the inclusion of PQC in industrial and critical environments poses an interesting trade-off between the benchmarks of those protocols. As mentioned previously, protecting ICS and CI from possible cyberattacks is fundamental due to the immense impact that those systems have in society and industry, making their failure to cause intolerable economic losses and, in the worse scenario, injuries and even casualties. Interestingly, this systems have shown to be vulnerable in the recent times with several hacking proposals \cite{7245330,GooseAttack,PoisonedGOOSE,HackICS}. Therefore, the necessity of transitioning the security of industrial and CI to post-quantum cryptography is central to keep all those systems secure against a possible quantum threat, as it has been recently noted by the United States Cybersecurity and Infrastructure Security Agency (CISA), the National Security Agency (NSA) and the NIST \cite{cisansanist}.

\subsection{Motivation and Related work}
Due to the global security threat posed by the possibility of a fault-tolerant quantum computer, PQC is one of the most important topics in cryptography at the moment. Thus, there are many works pointing out the importance and the lack of cybersecurity in OT environments as well as surveys about PQC cryptosystems. Specifically, there are many theoretical and experimental references regarding PQC cryptosystems, such as \cite{PQCrev1,PQCrev2,PQCrev5,PQCrev6}. However, there is a gap in the literature regarding the merge of both problems. Since OT environments are a clear and critical target for cybercriminals, it is of the most importance to protect those scenarios from quantum attacks as well. In comparison with previous works about PQC cited before, this manuscript aims to provide a perspective on the problem of implementing PQC algorithms in CI and OT environments.  In the literature there are many works regarding the importance of cybersecurity and cryptography in OT \cite{delayICS,7245330,HackICS,IndustrialCyber,OTcyber1,OTcyber3,OTcyber4,OTcyber5,OTcyber6}. As PQC algorithms were traditionally conceived from the point of view of IT communications, there are some requirements in industrial environments that are usually not fulfilled by those. Therefore, we want to emphasize the necessity of more research in PQC algorithms from the point of view of OT communications and, at the same time, more test benches implementing PQC algorithms in industrial environments. This comes with the objective of assessing their reliability for OT communications while providing information for cryptography researchers in order to develop new PQC algorithms that are tailored to fulfill those requirements. Papers such as \cite{NISTPQC,PQCrev2,PQCrev6} provide a good introduction of the problem and a survey for NIST PQC algorithms, while \cite{PQCrev3} gives a good introduction to PQC cybersecurity in OT. In the context of industrial scenarios it is common to underestimate the importance of cybersecurity as well as to consider it as a toll to productivity due to the additional costs. Moreover, each global agent is aiming to standardise PQC protocols in an independent manner, requirements that will be necessary to all vendors to fulfill once established. The NIST standardisation process stands as one of the first efforts for PQC standardisation and, while followed by a considerable amount of occidental countries, it is not the only one in the world, where countries such as France \cite{ANSSI} or Germany \cite{BSI} are also following different processes. Thus, it is very unlikely that a global adoption of the same standard will happen for this new field of cryptography, as it happened for the widespread RSA and ECC cryptographic schemes.

\subsection{Outline and contribution}
In this context, the principal objective of this contribution is to stress out the necessity of the integration of new generation PQC protocols to industrial and CI environments as well as to discuss the state of affairs and challenges regarding such integration. Specifically, we aim to:
\begin{itemize}
    \item Provide an introduction to traditional cryptography in OT environments for industry experts for providing them the basic knowledge of the problem. As mentioned before, many industrial players may be unaware of the importance of integrating cryptography in their environments and, thus, it is our intention to provide them with the basic concepts. Also, this serves to introduce the challenges and requirements of integrating cybersecurity, in general, to CI and industrial environments.
    \item Discuss how quantum computing can pose a threat to OT communications and show how this risk fundamentally differs from the one that IT communications may experience. This is aimed to show industrial players why transitioning to quantum secure cryptography will be critical as well as to show PQC developers how those networks should be protected.
    \item Describe the state-of-the-art PQC families and protocols being considered, not only within the NIST standardization process but also within other processes around the world. This serves as an introduction of PQC cryptography for newcomers. Additionally, we intend to show that due to the uncertainty regarding the PQC algorithms that will be implemented (some strong candidates are being questioned or have been recently broken) and the fact that it seems that many global agents will adopt their own methods, PQC integration in OT environments will require a great degree of flexibility regarding implementation.  
    \item Provide a discussion of the state-of-affairs regarding PQC implementation in OT environments. Specifically, we want to pose the main challenges when integrating those quantum secure protocols in such scenarios. Finally, to show the necessity of more active research on this topic both by cryptographers and industrial players. 
\end{itemize}

The manuscript is organized as follows: in Section \ref{Sec: II} we provide a review of the communication systems in industrial environments as well as of the stringent requirements for integrating cybersecurity in them. We follow, in Section \ref{Sec: III}, with a short review on cryptography and the threat that quantum computers pose to traditional ciphering schemes. Existing proposals of PQC algorithms are surveyed in Section \ref{Sec: IV} presenting several protocols proposed by many worldwide agents. In addition, the performance benchmarks of those PQC candidates are presented. An overview of the state of PQC in industrial and CI environments is finally presented in Section \ref{Sec: V}, were we speculatively discuss which existing protocols may be the ones for integration in those scenarios.

\subsection{Review Methodology}
The methodology rearding the literature review conducted was as follows. An exhaustive online search was conducted in order to identify key works reviewing classical cryptography and PQC algorithms. For such initial search, the traditional databases for classical cryptography were explored, i.e. IEEE Xplore, ACM and the Cryptology ePrint Archive. Many references regarding PQC were found in those sources, but the review by Bernstein and Lange was specially useful to identify many lines and references regarding such field \cite{Bernstein2017-zk}. Furthermore, the NIST standardization process was also an starting point to identify many of the methodologies that present the potential to be implemented. Since the NIST standardization stands as the mainstream process in this line, it was the starting point to identify PQC protocols that go beyond the basic theory for PQC families. Once studied, we followed to other standardization efforts since one of the points of this perspective is to show that many global agents are independently doing such process and we wanted to show the heterogeneous nature of this field. Moreover, (post-quantum) cryptogrpahy is a dynamic field so many related blogs, e.g. Cloudflare\footnote{\url{https://blog.cloudflare.com/}} or Google Bughunters\footnote{\url{https://bughunters.google.com/blog}}, were regularly read in order to follow developments of those fields at the time of writing this article. Regarding industrial cryptography, we based on the knowledge of one of the authors (G. Vidal) in order to identify a preliminary batch of relevant literature regarding this topic. Afterwards, we complemented such literature by means of IEEE Xplore and ACM databases as well as by getting references from such initial set of articles. Last but not least, some missing references were pointed by the referees, which were discussed for completeness of the work.

Once the literature was collected and understood, all the information was used to make discussions regarding the implementation of PQC in industrial networks as well as to identify which are the challenges associated to it. This has been done by comparing the core problem of IT cryptography in contrast to OT cryptography, which is the one tackled in this perspective. In this way, it has been seen that most of the PQC protocols that have been proposed would have difficulties to be integrated in industrial networks as they have been constructed from the IT point of view. Finally, future work has been pointed out in order to make such scenarios quantum secure.

\section{Industrial environments} \label{Sec: II}
\begin{figure}[!t]
    \includegraphics[width=0.9 \columnwidth]{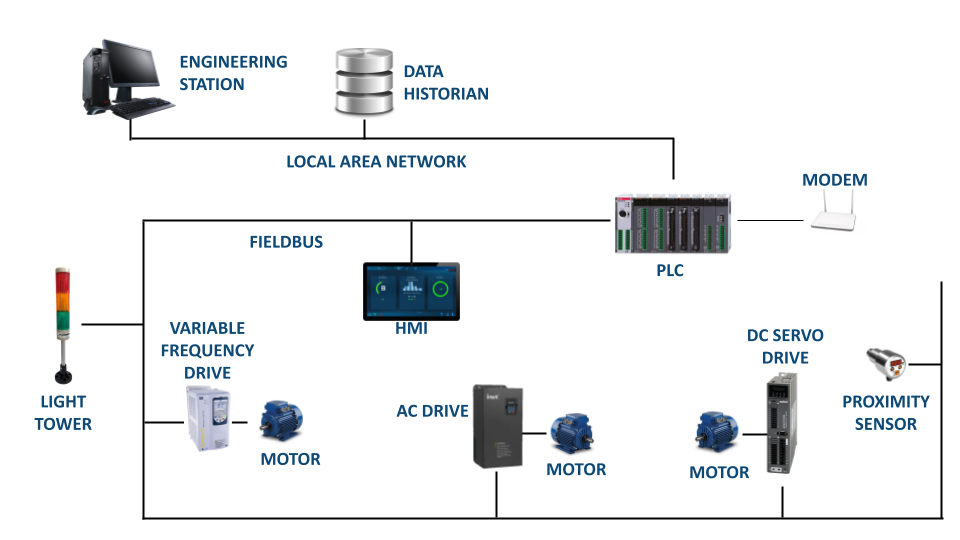}
    \caption{Example of ICS network diagram.}
    \label{fig:ICS}
\end{figure}

\noindent
ICS are the components of the industrial sector and infrastructures, from essential services such as energy, water, transportation systems to manufacturing plants, agricultural systems, building automation systems, etc. In these infrastructures we will always find complex components that share a common denominator: physical processes that are modified by logical computation or viceversa. These components are called Cyber-Physical Systems (CPS). In Fig.~\ref{fig:ICS}, we show an example of ICS. In this example, in short, the instrumentation sensors measure physical variables, programmable logic controllers (PLCs) implement a control loop and send signals to the actuators. All the process is controlled and monitored by the Supervisory Control And Data Acquisition (SCADA) and the operator interacts via the Human-Machine Interface (HMI) or the Engineering Work Station (EWS).

Since these systems control real world processes, any potential cyberattack impact on them could imply a physical effect in the real world. Hence, cyber-risks could turn not only into production downtimes but also physical damage to operators or users. A good survey of different objectives and techniques used to attack ICS networks has been collected in MITRE ATT\&CK \footnote{\url{https://attack.mitre.org/}}.

\begin{figure}[!t] 
    \includegraphics[width=0.9 \columnwidth]{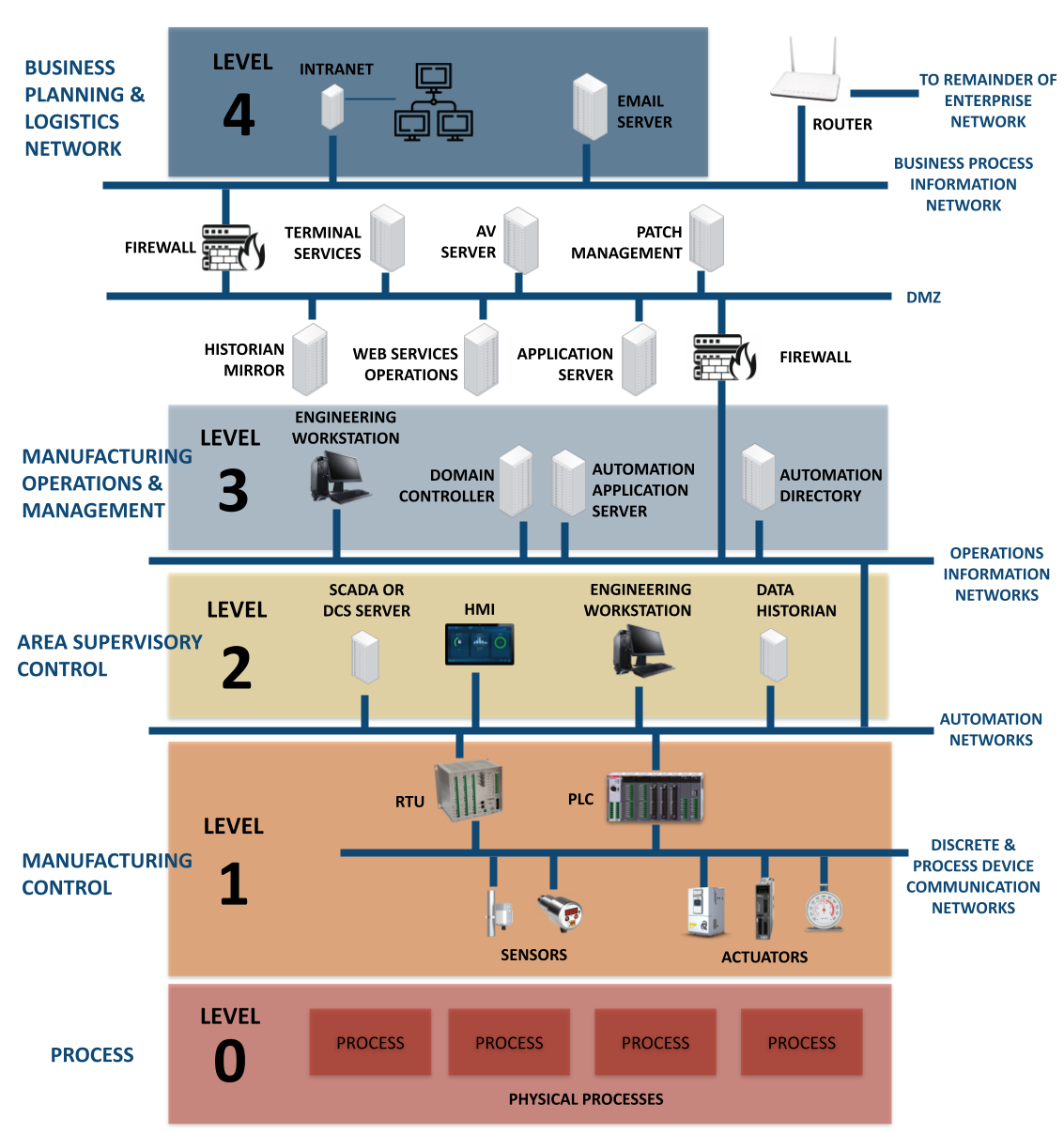}
    \caption{Purdue model of interaction between IT and OT services.}
    \label{fig:Purdue}
\end{figure}

ICS and CPS are a part of larger infrastructures which interact with information technology (IT) systems at certain point. In Fig.~\ref{fig:Purdue}, we show how IT and operational technology (OT) interact in this type of infrastructures according to the Purdue model, even though there are several models such as RIA 4.0 and others.

The mitigation of these risks is challenging since the security mechanisms and techniques that are suitable for IT don't match the needs in OT.
\subsection{Differences between IT and OT} \label{Sec: II A}

Understanding the inner differences between IT and OT communications are essential in the realm of cybersecurity, particularly when safeguarding critical infrastructure like energy systems. Firstly, one of the most prominent distinctions lies in their component lifetimes. OT systems often rely on hardware with a lifespan of up to 20 years, whereas IT systems typically have a significantly shorter lifespan of 3 to 5 years. This variance makes it challenging for OT systems to stay updated with the latest security measures, as their components may become outdated and incompatible with newer cybersecurity technologies over time.

Secondly, availability requirements vary significantly. OT systems demand extremely high levels of availability since any downtime can have severe consequences. In contrast, IT systems usually have more moderate availability requirements. This discrepancy emphasizes the need for robust cybersecurity measures in OT to prevent disruptions that could impact critical operations.

Moreover, real-time requirements diverge between the two domains. OT systems often require real-time responsiveness, with certain elements in the energy sector requiring millisecond-level reactions to commands \cite{IEC62443,delayICS}. This real-time demand can make it challenging to introduce comprehensive cybersecurity measures in OT systems due to the need for rapid response, whereas in IT systems, real-time requirements are typically less stringent, allowing for more deliberate and thorough security implementations.

Additionally, the approach to patching and security standards differs significantly. In IT, security standards are generally more mature, and patching can be executed relatively quickly. In contrast, the OT sector faces slower patching processes often constrained by regulations. This slow pace can leave OT systems vulnerable to emerging threats for extended periods. Lastly, while both domains attend to data integrity, OT systems typically emphasize data integrity as a top priority, while confidentiality is considered a lower to medium priority. In contrast, IT systems prioritize confidentiality as a must, alongside integrity and availability.

Also, we need to remark an additional issue: the cybersecurity generational gap. From a cybersecurity perspective, the age of an industrial plant can significantly impact the cost and complexity of achieving a high level of cybersecurity. Let's delve into this scenario:

When starting a new company or building a modern industrial plant today, you have the advantage of being able to incorporate cybersecurity measures from the very beginning. Many modern components and systems are designed with cybersecurity in mind, often featuring embedded security features and protocols. This not only simplifies the process of implementing cybersecurity but also reduces the overall cost. It is essentially a proactive approach that builds security into the infrastructure from the ground up.

However, the challenge arises when dealing with older industrial plants, where the components and systems were likely not designed with cybersecurity in mind. These legacy systems may lack modern security features, making them vulnerable to cyber threats. Retrofitting these older components with cybersecurity measures can be a complex and costly endeavor. It may involve upgrading or replacing outdated hardware and software, implementing security protocols, and training personnel to operate in a more secure manner.

Furthermore, integrating cybersecurity into an older plant often requires a careful balance between maintaining operational continuity and enhancing security. Downtime can be expensive and disruptive, so the process must be meticulously planned and executed.

In summary, the cost of achieving a high level of cybersecurity can be much higher in older industrial plants due to the need for retrofitting and upgrading legacy systems. In contrast, new companies and modern facilities have the advantage of incorporating cybersecurity measures at a lower cost from the outset, due to the availability of cyber-embedded components and systems. However, it is crucial for all organizations, regardless of age, to prioritize cybersecurity to protect critical infrastructure and assets from evolving cyber threats.

\subsection{Industrial Cybersecurity Standards and Mechanisms} \label{Sec: II b}

Industrial vendors, recognizing the pressing need for enhanced cybersecurity, are taking significant steps to fortify their products and services \cite{IndustrialCyber}. This involves developing and implementing robust security measures throughout their supply chains and lifecycles. Governments worldwide are also taking an active role in formulating regulations and guidelines to address these challenges \cite{ENISAICS,CISAICS}. They are working to create a secure environment for critical infrastructure sectors, including energy, water, and transportation, by establishing cybersecurity frameworks and compliance mandates.

In parallel to the IT sector, where standards like ISO 27001 serve as well-established benchmarks, industrial sectors adhere to their specific standards, with IEC-62443 being the primary global reference \cite{IEC62443}. This standard offers a comprehensive framework for industrial control systems' cybersecurity. It defines guidelines for secure design, deployment, and maintenance, providing a roadmap for organizations to bolster their security posture. Additionally, various countries, such as the United States, the European Union, and China, have crafted regional regulations tailored to their specific requirements, reflecting the nuances of their industrial landscapes \cite{CISAICS,ENISAICS,normaFrancia}.

Furthermore, sector-specific regulations address the unique cybersecurity concerns within industries like water and electricity. These regulations take into account the distinct OT challenges that may not align perfectly with traditional IT standards. By tailoring security measures to the particularities of each sector, these regulations help bridge the gap between the IT and OT worlds, ensuring the protection of critical systems.

\begin{figure}[!t]
    \includegraphics[width=1\columnwidth]{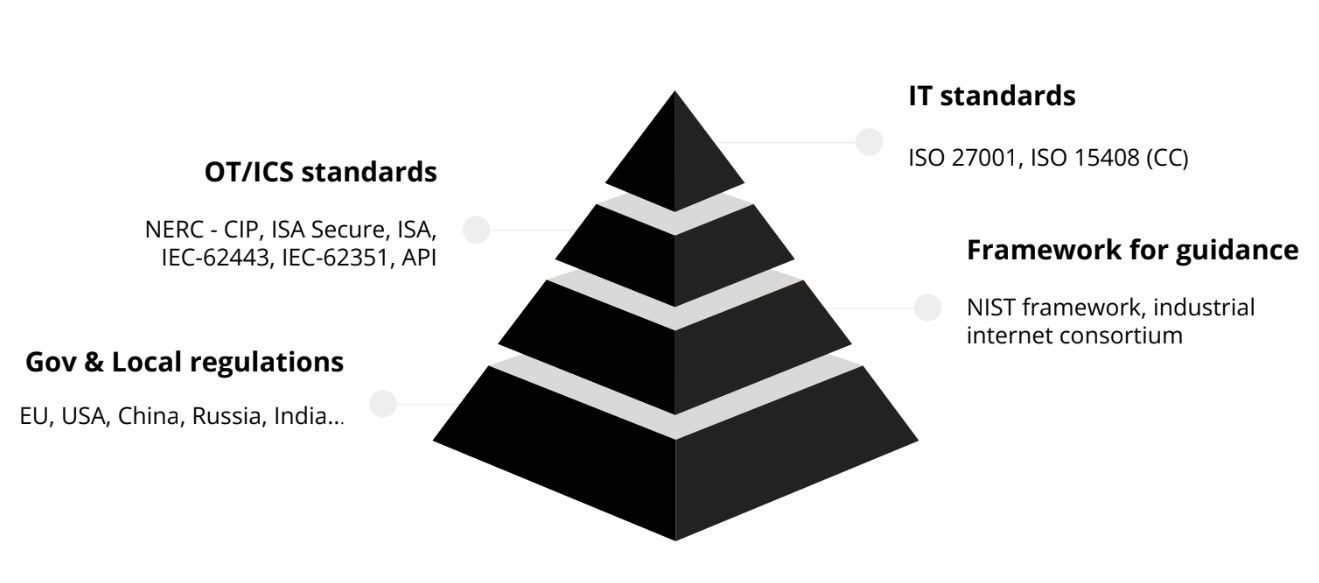}
    \caption{This pyramid shows how standards are organized according to their level of definition in ICS.}
    \label{fig: standards}
\end{figure}

In addition to broader standards and regulations, there are specific standards for components and systems used in industrial environments. These may include industrial communication protocols and HW requirements. By adhering to these standards, organizations can ensure compatibility and security among different components and systems within their infrastructure. The interoperability provided by these specific standards is crucial for maintaining a secure and efficient industrial ecosystem. Fig.~\ref{fig: standards} summarizes these interdependencies.

It is also worth mentioning that many of these legacy protocols were designed without strong security considerations, making them vulnerable to cyberthreats [as shown many times (pipedream \cite{pipedream}, goose attacks \cite{gooseattacks}, rogue7 \cite{rogue7}, etc]. The introduction of Virtual Private Networks (VPNs) into these setups helps create a secure tunnel for data transmission, adding an extra layer of protection. However, this adaptation does come with its own set of latency concerns, further highlighting the ongoing struggle to balance security with operational efficiency.

Additionally to the encryption challenges in industrial cybersecurity, recent regulations have begun mandating the use of VPNs in inherently insecure industrial protocols \cite{normaFrancia}.

Moreover, the emergence of post-quantum encryption technologies has added a new dimension to the regulatory landscape. As quantum computing capabilities continue to advance, traditional encryption methods may become vulnerable to rapid decryption, posing a significant threat to data security as we will discuss in the next section. In response, regulators are intensifying efforts to mandate the adoption of post-quantum encryption techniques in critical infrastructure sectors. This drive for post-quantum encryption standards underscores the necessity for constant adaptation in the industrial cybersecurity field, where maintaining the integrity and confidentiality of data remains paramount, even in the face of evolving threats and technological advancements.

\section{Pre-Quantum Cryptography and Quantum Apocalypses} \label{Sec: III}

\subsection{Overview on cryptography} \label{Sec: III a}

Cryptography is the study of algorithms which are able to make some information unintelligible to a third party (providing confidentiality), to protect it under changes from a third party (providing integrity) and to prevent that a third party masquerades as one of the trusted communication parties (providing authenticity). Mathematical tools for encrypting valuable information have been developed since the dawn of civilizations, with notorious examples such as the Caesar cipher (a shift cipher where each letter is substituted by another letter in a fixed number of positions down the alphabet) used by the Roman Emperor Julius Caesar\footnote{This fact was stated by Roman historian Suetonius in \textit{Vita Divi Julii}, 56.6.}. New encryption algorithms were discovered since those primitive days of cryptography because they had been cracked by other entities trying to obtain the protected information. For example, one of the most important events of World War II was breaking the Enigma code used by Nazi Germany to protect commercial, diplomatic and military communication. The digital revolution brought the possibility of evolving cryptographic techniques by means problems that are harder to solve, but, at the same time, it provided an additional tool for hackers to crack those codes.

Modern cryptography distinguishes two types of cryptosystems depending on how a message is encrypted: symmetric and asymmetric key schemes. Symmetric key schemes make use of the same key for encryption and decryption algorithms. For example, the Advanced Encryption Standard (AES) proposed by J. Daemen and V. Rijmen in 1998, also known as Rijndael scheme, is the most implemented symmetric cryptosystem scheme \cite{AES}. On the other hand, asymmetric key schemes or public key schemes consist of two different algorithms and keys for each of the parties sharing secrets: each of the parties use their public key (generated by the other party) to encrypt messages and their private key (generated by itself) for decrypting the received messages. One of the most famous and most used asymmetric key cryptosystem scheme is the RSA cryptosystem, developed by Rivest, Shamir and Adleman \cite{RSA}. 

In the ensuing paragraphs, we provide a comprehensive explanation of both schemes based on a network composed of three parties: a server, a user who wants to have a secure connection with it and an eavesdropper who wants to obtain information about the message. In cryptography those are usually referred to as Alice, Bob and Eve, respectively.

\subsubsection{Symmetric Key Schemes} \label{Sec: III a 1}

Symmetric key schemes are cryptosystems where both parties, Alice and Bob, share the same key, i.e. they use the same bit string to cipher and to decipher the message. The security of a symmetric scheme relies on the length of the key and on the fact that the key keep its secret to other parties. In this schemes, the key exchange is the most important process since all the security relies on the privacy of such key. If an eavesdropper, Eve, is able to obtain information about the key, the communication is not secure anymore. The most popular symmetric key scheme is the previously mentioned AES \cite{AES}, Blowfish \cite{Blowfish} and its more recent version Twofish \cite{Twofish} are also relatively popular open access schemes. To discuss the basic operation of a symmetric key exchange protocol, we begin by explaining how confidentiality and integrity/authenticity are achieved in the communication between the parties:

\begin{itemize}
    \item Confidentiality: The confidentiality in a symmetric scheme cryptosystem relies on the secrecy of the key. Eve knows how the algorithms which encrypt and decrypt the message work and she could perform a brute force attack by trying all possible key combinations to decipher the message. This, however, is a very inefficient attack. Assuming a key length of $n$, Eve has to try $2^{n-1}$ possibilities in average, which may take several years if the value of $n$ is sufficiently large, even having access to the most powerful computers. Therefore, if the secret key is regularly changed, this type of attack is impossible.
 
    \item Integrity and Authenticity: symmetric key exchange algorithms do not only cipher the plaintext (message), but they also create a Message-Authentication Code (MAC) by means of the message and an authentication key, $k_{auth}$, in order to protect the integrity of the message and verify the identity of Bob. Alice can check this information with the decryption algorithm for deciding if Bob has been the sender or not. 
\end{itemize}
Following our discussion, a symmetric key exchange cryptosystem has three steps for protecting a plaintext message:
\begin{itemize}
    \item Key exchange: Alice and Bob exchange a secret key and an authentication key, ($k_{sym}$ and $k_{auth}$, respectively), which usually are two bit strings, in a secure way. If Eve, an adversary, gets information about the keys, the communication will not be secure. An important challenge is how to do this key exchange using a telecommunication channel certifying that Eve do not get any information about the keys, whose solution will be explained  in the next section.

    \item Encryption Algorithm (Enc) and signature algorithm (Sgn): Bob, by means of the symmetric key ($k_{sym}$) encrypts the plaintext (Msg) generating the ciphertext (Ct). At the same time, he generates the MAC with making use of the authentication key ($k_{auth}$) and the message (Msg). Then, he broadcasts the ciphertext and the MAC, implying that both Alice and Eve have access to them. The broadcasted information would be:
    $$
    \text{(Enc(Msg,$k_{sym}$), Sgn(Msg, $k_{auth}$)) = (Ct,MAC)}.
    $$

    \item Decryption Algorithm (Dec) and verification algorithm (Vry): Alice decrypts the ciphertext using the shared secret key, ($k_{sym}$), and recovers the message. Also, she checks if the MAC corresponds to the $k_{auth}$, she obtains a boolean value, $b$, i.e. $b=1$ if it is correct or $b=0$ if not:
     $$
    \text{(Dec(Ct),Vry(MAC,Msg, $k_{auth}$)) = (Msg,b)}.
    $$
    Note that Eve will not be able to recover the message if she is not able to obtain information about the secret key or makes a brute force attack, which would take a lot of computational time.
\end{itemize}

\subsubsection{Asymmetric Key Schemes} \label{Sec: III a 2}

As defined above, asymmetric key or public key schemes are defined as cryptosystems where each of the parties involved in the communication use their own key to secure the information, the private key and the public key. Each of those are usually employed to share a secret between two parts and, therefore, this encryption schemes may be used for the confidential key exchange required in symmetric schemes, called Key Encapsulation Mechanism (KEM). In this sense, this could refer to the transmission of the bit string used as the input of an algorithm which can be used by Alice and Bob to generate the same symmetric key. Otherwise, if it is used to encrypt the message it is called Public Key Encapsulation (PKE). In asymmetric key schemes confidentiality, integrity and authenticity are achieved in the following ways:

\begin{itemize}
    \item Confidentiality: The confidentiality relies in the hardness of finding the solution of the mathematical problem that Eve, the eavesdropper, has to solve in order to obtain the secret key or recover the plaintext. The problems used in public key schemes are computationally hard problems that cannot be efficiently solved. Thus, with no additional information, Eve cannot obtain the information about the message that has been encrypted or the private key.
    
    \item Integrity and Authenticity: The integrity and authenticity of the message is achieved by means of a digital signature. Digital signatures are based on a hard problem so that the algorithm makes use of a message and a private key as an input for providing a unique output which identifies the party and the message. Any change in the signed ciphertext produces a totally different output providing a way to protect the integrity of the message. The authenticity of the message is determined by the secret key and it can be checked by using the public key, i.e. all parties can check the authenticity of a message but it only can be signed by the owner of the private key.
\end{itemize}

Regarding the general operation of a  public key cryptosystem, those are based on the following three algorithms:

\begin{itemize}
    \item Key Generator Algorithm (Gen): Alice creates her private key (Sk). By means of it and the selected asymmetric cryptography algorithm, she generates the public key (Pk) and sends it to Bob. Note that by sending the key to Bob, Alice broadcasts it as public information and, hence, anyone in the network can access to that information. A potential eavesdropper, Eve, can get and store such key to try to decrypt the message but she will no be able to crack it due to the complexity of the problem she has to solve as we will explain below, she will require an unreasonable amount of time to obtain the plaintext.
    
    \item Encryption Algorithm (Enc): Bob with the public key encrypts the plaintext (the message Msg) generating the ciphertext (Ct). He then proceeds to communicate publicly the protected message:
    $$
    \text{Enc(Msg,Pk) = Ct}
    $$

    \item Decryption Algorithm (Dec): Alice uses her own private key and the decryption algorithm to obtain the message that Bob wanted to provide in a secure manner:
     $$
    \text{Dec(Enc(Msg,Pk),Sk) = Msg}
    $$ 
\end{itemize}

This is the general scheme of an asymmetric key cryptosystem, the security of which relies on the assumption that Eve is not able to recover the original plaintext by just using the public key and the ciphertext. Such thing relies on a mathematical problem that is hard to solve but for which it is easy to prove if a given solution is correct or not. Those are usually referred as one-way functions. They are designed in such a way that Eve can not decipher the message in reasonable time without the private key and she can not get any information about the private key by means of the public information. The most famous, and widely used, scheme is the RSA cryptosystem, developed by R.L. Rivest, A. Shamir and L. Adleman in 1979 \cite{RSA}. Its security is based on the factorization of large numbers in their prime factors. Other used schemes are based on ECC such as the Elliptic Curve Diffie-Hellman (ECDH) scheme \cite{ECDH}.

\subsubsection{Security Notion} \label{Sec: III a 3}

Proving the security of a cybersecurity scheme is not a trivial problem, i.e. mathematically proving that a function is indeed an One-Way function is not a simple task. In fact, one of the ``Millenium Prize Problems'', known as the $P\overset{?}{=}NP$ problem \cite{PNP}, involves to prove if the set of the problems whose solution is hard to find is the same to the set of problems whose solution is easy to check. However, the security of cryptosystems can be studied by means of some security notions. There are three cases which have to be taken into account to study a cryptosystem's security. In all of them, the adversary who wants to break the security is modeled by a probabilistic polynomial time algorithm. The scheme is secure if the algorithm has no advantage for discovering the secret over a random guesser algorithm, i.e. it is said Eve has a negligible advantage if she guess the correct answer with probability $1/2+\epsilon(k)$, where $k$ is a security parameter and $\epsilon(k)$ is a negligible function. Following this logic, the three security notions considered are:

\begin{itemize}
    \item Ciphertext-Indistinguishability under Chosen Plaintext Attacks (IND-CPA): refers to the property where Eve is not able to distinguish a random ciphertext from an actual ciphertext whose plaintext is known by herself.  
    
    \item One-Wayness under Chosen Plaintext Attacks (OW-CPA): refers to the property where Eve is not able to recover the plaintext, even when she has the ability to choose and encrypt any plaintext of her choice and observe the corresponding ciphertext.
    
    \item Key-Indistinguishability under Chosen Ciphertext Attacks (IND-CCA): refers to the property where Eve can call the decryption algorithm as many times she needs for an arbitrary ciphertext but she is not able to guess the key.
\end{itemize}

\subsubsection{Digital Signatures} \label{Sec: III a 4}

Digital signatures allow to sign a message in such a way that the origin of a message can be verified and ensure that the message has not been altered. They are used to verify the authenticity and ensure the integrity of a message in a public key cryptosystem. In order to sign a message, the signer has to generate a private and a public key, the private key is used to sign the message and the public key is needed to verify if a message has been signed by the true party. This process is also done by three polynomial-time algorithms: the key-generation algorithm (Gen), which generates the public (Pk) and the private keys (Sk); the signing algorithm (Sign), which uses the secret key to sign a message; and the verification algorithm which checks by means of the Pk if a message has been signed by the party who generated the public key. 

The security of a Digital Signature resides in the probability that a malicious party is able to sign a message without having access to the Sk and it can be verified satisfactorily. It is worth to say that the security of digital signatures resides on the computational complexity of an algorithm to solve the hard problem in which signature security relies on. Therefore, the same security notions in public key schemes explained above are applied to digital signatures. Nonetheless, in this case the adversary is interested in signing a message. i.e. Eve wants to masquerade as a trusted party, rather than acquiring information about a protected message. One of the most used digital signature scheme is the Elliptic Curve Digital Signature Algorithm (ECDSA) \cite{ECDSA} based on ECC.

\subsection{Quantum Apocalypses} \label{Sec: III b}

Quantum computers promise to be huge step forward in computation as a result of being able to reduce significantly (exponentially for some algorithms) the number of operations an algorithm needs to solve some computational problems. There are two important quantum algorithms which will compromise the security of the current computer network security systems: Grover's algorithm \cite{grover1996fast} and Shor's algorithm \cite{shorFact}.

Grover's algorithm provides a quadratic speedup for searching the secret key in symmetric key cryptosystems \cite{Bernstein2017-zk}. As said before, the security in these schemes relies on the key's secrecy and its length since a brute force attack consists in searching the $n$-bit combination. Therefore, the complexity of a classical brute force attack is bounded by $2^{n-1}$ on average, while a quantum computer could run Grover's algorithm to reduce the maximum number of steps to $2^{n/2}$, on average. This algorithm does not compromise the symmetric cryptography paradigm since the qudaratic boost can be compensated by doubling the key size and increasing the computational cost of the key exchange and encrypt and decrypt algorithms, but as it is just doubling down the key size, the extra costs are rarely noticeable \cite{Bernstein2017-zk}. There are other quantum attack proposals to symmetric cryptography such as Variational Quantum Attack Algorithms (VQAA) \cite{vqaa1,vqaa2}, but they do not compromises its security.

As explained before, the key exchange algorithm for establishing the secret key of a symmetric protocol is done with a public key (asymmetric) cryptosystem. In this sense, doubling the size of the key to be shared increments the complexity of the key generation algorithm of the public key scheme. However, increasing the computational cost is feasible in order to hold the security in almost all the cases and, hence, this is not problematic. Nonetheless, the security of a public key cryptosystem relies on the complexity of the hard problem which has to be solved to get the private key. For example, a brute force attack to RSA consists in trying all primes $p$ and checking whether $p$ is a factor of $N$, requiring $\sqrt{N}$ attempts in the worst case scenario, which is exponential in the digits of $N$ ($d$) and, thus, is an unfeasible task for any classical computer. The most efficient classical algorithm to solve such problem, known as the general number field sieve \cite{NumberFieldSieve}, achieves a complexity of $\mathcal{O}\left(e^{\sqrt{\ln(n)\ln(\ln(n))}}\right)$ asymptotically \cite{complexty}, using this method up to RSA-250 (250 digits) have been factorized satisfactorily \cite{RSA250}.

\begin{figure}[!t]
    \includegraphics[width=1\columnwidth]{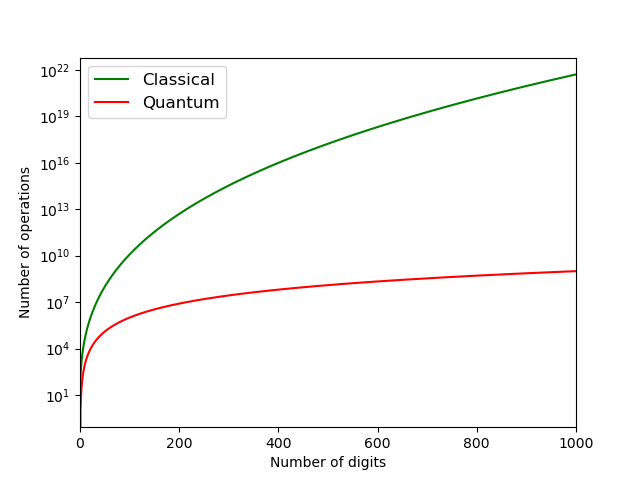}
    \caption{Comparison between the operation's number of the general number field sieve and the Shor's algorithms to break RSA cryptography.}
    \label{fig: QvsC}
\end{figure}

A large enough fault-tolerant quantum computer could run Shor's algorithm, which is able to factor a number in its primes taking advantage of the laws of quantum mechanics. The first large enough quantum computer will be able to solve factorization problem using $10d$ logical qubits\footnote{Note that this number refers to logical or ``noiseless'' qubits \cite{decoders}. For quantum computers to work, quantum error correction is required implying that many more physical or ``noisy'' qubits will be required for an implementation of the Shor algorithm that cracks RSA \cite{rsacrackGid}.}, where $d$ is the number of digits, with a complexity of $\mathcal{O}(d^3)$ \cite{complexty}. A comparison between both algorithms, in terms of their bounds in number of steps, is represented in Fig.~\ref{fig: QvsC}, where the huge difference (exponential speedup) between both algorithms when the number of digits increases can be easily observed.

Therefore, Shor's algorithm is a potential threat to the security of all communications over the world, and will be a real problem in the near future since it is able to break most used public key cryptosystems, RSA and ECC schemes. As a result, governments and private companies, such as technological giants IBM, Google and Microsoft, to cite some of them; are dedicating an immense amount of economical and human resources to develop a large enough functional quantum computer (fault-tolerant) able to compute Shor's algorithm and break the current cybersecurity. Hence, new cryptography schemes are required in order to protect our communications and documents in the post-NISQ\footnote{Noisy intermediate-scale quantum era: refers to time scale in which quantum advantage has been firmly proven, but the quantum computers available are still small and too noisy to offer the full potential of quantum computing \cite{nisq}. NISQ is where we stand right now.} quantum era. Another issue to take into account is the fact that some people speculate about the possibility that some global agents are storing encrypted communications for decrypting it in the future once fault-tolerant quantum computers are available. Thus, finding an alternative to RSA and ECC cryptography resistant to quantum attacks is a very relevant problem nowadays. Moreover, it is estimated that a secret holds its value for 15 years \cite{Mosca} and, thus, it is necessary to consider public key cryptosystems based on hard problems that can not be solved efficiently by quantum computers as soon as possible, i.e. it is necessary to implement quantum resistant algorithms 15 years before the first functional quantum computer is available. Google, IBM, China and Xanadu have shown quantum advantage using their computers and they expect a huge increase in their quantum computer capability during the following years \cite{sup1,sup2,sup3,sup4,sup5}. To sum up, we are at a critical moment for cryptography.

It is in this paradigm that the most promising technologies for the future quantum-safe cryptography are QKD and PQC. QKD takes advantage of quantum superposition and the no-cloning theorem \cite{No-clonning} to exchange a pair of symmetric keys in such a way that Eve is not able to achieve any information of the key, since if she tries to get it Alice and Bob are able to detect such an attempt and discard the exchanged key. Alice and Bob are able to do such things due to the properties of quantum mechanics (no-cloning, entanglement). This proposal is a good candidate for establishing future secure communications, but it is still a nascent technology with many problems as well as requiring a huge investment in infrastructure. Also, QKD is pretty susceptible to DDoS attacks since it can be done just by adding photons to the optical fiber or by measuring the emitted photons, since measuring a quantum state destroys the quantum information inside it. On the other hand, PQC is a family of different asymmetric key schemes which are secure against classical and quantum attacks. This proposals are based on hard problems for which quantum computers do not offer a substantial (not to say any) speedup.  In this paper we will focus on the PQC solution, specially since the objective of the present manuscript is to understand how quantum secure cryptography can be integrated in industrial and critical infrastructure networks. As reviewed before, those networks pose some stringent conditions such as low latencies or adaptability to legacy devices implying that the integration of cryptography to such scenarios must be almost seamless. Hence, it seems straightforward to discard QKD as a realistic candidate for the transition to quantum secure communications in ICS/CI as a results of the high cost and infrastructure need that the technology requires, also because DDoS attacks are very harmful to CI due to the reduction of the availability.

\section{Post-Quantum Cryptography} \label{Sec: IV} 
\noindent
Post Quantum Cryptography (PQC) refers to classical cryptographic methods based on hard problems whose solution cannot be found in polynomial time by neither classical nor quantum computers. The hardness of the problem is defined by the computational complexity of an algorithm capable to solve it. In this sense, a quantum computer should not provide any advantage (or such should be almost negligible) in solving the hard problem that stands at the core of a PQC protocol. Note that, as we commented in Section \ref{Sec: III b} for the case of symmetric cryptography, doubling the key length seems to be enough to keep the security level of classical methods against Grover search attacks. However, CI networks present strong latency requirements implying that doubling the key length could not be an acceptable solution. In this context, there are some proposals for lowering the computational and memory requirements of symmetric key cryptosystems, named Lightweight cryptography (LWC) \cite{LWC}. LWC reduces the required data to achieve secure communication channels and, thus, reduces the needed device computational resources \cite{LWCHW1,LWCHW2}. Furthermore, it is applicable to networks with legacy and computationally limited devices, including CI networks.  Recently, the NIST has finished its standardisation process for LWC \cite{NISTLWC} including some quantum resilient methods such as Ascon-80pq \cite{ASCON}. In this work we will focus on the key agreement processes (asymmetric key schemes), which stand as more problematic in the context of industrial OT communications due to their latency and computational requirements. 

In the following section we will introduce the different hard problems in which PQC methods are based. It is believed that such hard problems are secure against classical and quantum attacks. However, as explained earlier, the security of a cryptosystem is based on some assumption instead of mathematical proofs, hence, it is not possible to assure that any new proposed cryptosystem is secure. Hybrid cryptosystems were proposed in order to keep the current cybersecurity and add quantum resilient protocols, if the PQC protocol is proven to be insecure under classical and quantum attacks in the future the communication protocol would maintain the current security. Hybrid cryptosystems were also conceived to help to the transition from classical to quantum cryptography \cite{Hybrid}. Despite of being a very active field and a good proposal to OT cybersecurity, we will not talk about hybrid solutions. Nonetheless, we encourage the interested reader to read the following papers \cite{Hybrid,HybridImp1,HybridImp2}.
 
\begin{center}

\begin{figure*}[ht!]
    \centering
    \includegraphics[width= 0.8 \textwidth]{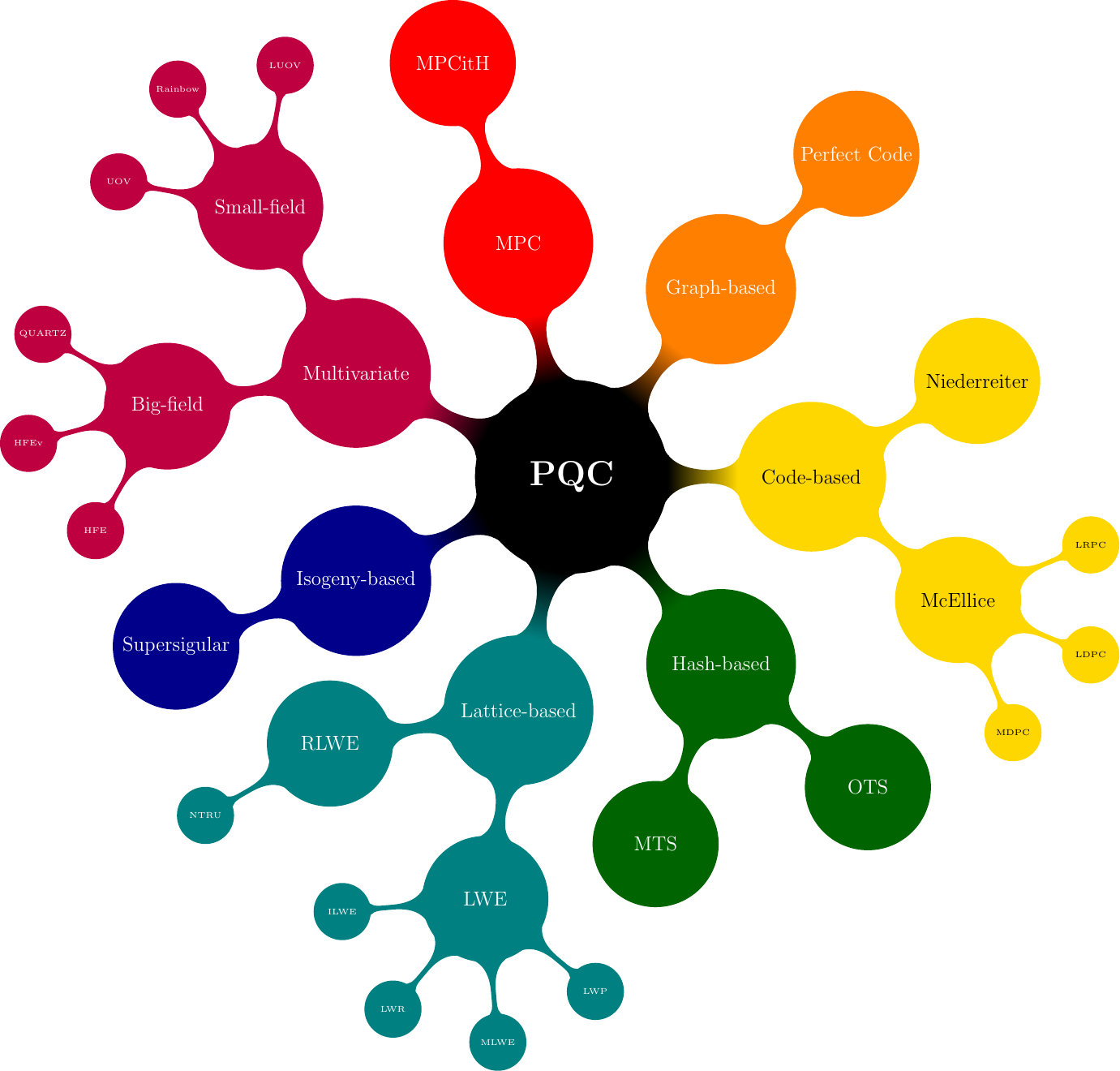}
    \caption{Diagram of the different PQC algorithm families and the most important proposed cryptosystems}
    \label{fig: PQC families}
\end{figure*}
   
\end{center}

PQC algorithms emerge as a consequence of assuming that a possible attacker has, or will have, a large and reliable enough quantum computer to break classical algorithms. Therefore, the new cryptography algorithms have to be hard to solve for classical and quantum computers implying security in the quantum era. These algorithms are usually divided into seven different families based on the hard problem in which their security relies on: Hash-based, Code-based, Lattice-based, Multivariate, Isogeny-based, Multi-Party Computation (MPC) and Graph-based cryptography. A diagram of the different families with the most important proposed schemes is represented in Fig.~\ref{fig: PQC families}. In the following sections we will review the basic operation of such schemes, including an enumeration for each family of the different implementations proposed to be standardized in the future. Several global entities such as the United States, China or the European Union, have started own PQC standardization processes considering different candidates of each family. In this sense, we also provide an overview of those processes around the world. We encourage  the interested reader in this topic to read the review made by D. Bernstein and T. Lange in \cite{Bernstein2017-zk} and by R. Bavdekar et al in \cite{Bavdekar}, based on PQC families proposed to NIST standardization process.

\subsection{Hash-based cryptography} \label{Sec: IV a} 

Hash-based cryptography was first proposed by Ralph. C. Merkel in the 70s \cite{hash}. The security of this cryptosystem relies on hash functions. A hash function ($H_M$) is a mathematical function that compresses an input string of bits of arbitrary length to a string of fixed length, i.e. it maps an input of an undetermined length into an output of fixed length $m$, which appears to be random but is deterministic. Formally, a hash function is defined as:
$$
H_m: \{0,1\}^{*} \xrightarrow{} \{0,1\}^{m}.
$$
Hash functions are usually employed to create digital signatures, deemed as hash-based signatures (HBS), providing authenticity and integrity to the communication. The first signature scheme using hash-functions was introduced by Lamport in 1979 \cite{lamport}. Those functions can be classified into One Way Hash Functions (OWHF), Collision Resistant Hash Functions (CRHF) and Universal One way Hash Functions (UOWHF). The general scheme of a hash function cryptosystem is composed by three algorithms, Gen, Enc and Dec in the case of encryption or Gen, Sign and Vry in the case of a digital signature scheme:

\begin{itemize}
    \item \textbf{Gen}: The key generation algorithm generates a public key (Pk) and private key (Sk). In general, this is done by choosing a private random seed, $\mathrm{seed}(n)$, i.e. a random bit string. A key $(k)$ is derived from the seed by setting it as the input of a hash chain, which is a sequence of hash functions where the output of one hash becomes the input of the next one and it may involve other private functions, $f_i(x_i)$. Usually, the Sk is composed by the key, the seed and the parameters of the hash chain. Finally, the public key (Pk) is a parameter of the entire hash chain, but it does not reveal the individual hash values:
    \begin{align*}
        Sk &  \xleftarrow{} \{H(x_n), n, f(x_i)\}, \\
        Pk &  \xleftarrow{} x'_m | m \leq n.
    \end{align*}

    \item \textbf{Enc/Vry}: Bob can use the public key to encrypt messages, recreating the hash chain to generate the ciphertext (Ct) or to verify the signature of Alice ($b=1$ if Alice is the emitter or $b=0$ if not):
    \begin{align*}
        Ct &  \xleftarrow{}  Enc(H,Pk,msg), \\
        b ={0,1} &  \xleftarrow{} Vry(H,Pk, sig). 
    \end{align*}

    \item \textbf{Dec/Sign}: Alice is able to decrypt the message of Bob with the private key and to sign her messages by means of the private key:
    \begin{align*}
        msg &  \xleftarrow{} Dec(H,Sk,Ct), \\
        Sig &  \xleftarrow{} Sign(H,Sk, msg).
    \end{align*}
   
\end{itemize}

\subsubsection{Security Notion} \label{Sec: IV a 1} 
The security notion of a hash function resides in three characteristics that must have in order to be secure: Preimage resistance, second-preimage resistance and collision resistance. These can be formally defined as:
\begin{itemize}
    \item \textbf{Pre-image problem}: Given the output of the hash function, $H_m(x)$, find the input $x$ (One-wayness function).
    \item \textbf{Second pre-image problem}: Given the output of the hash function, $H_m(x)$, and the input $x$, find another input, $y$, that fulfills $H_m(x)=H_m(y)$ with $y \ne x$ (Weak collision resistance).
    \item \textbf{Collision problem}: Find two inputs, $x$ and $y$, which fulfill $H_m(x)=H_m(y)$ (strong collision resistance). If a Hash-function has collision resistance it implies second pre-image resistance.
\end{itemize}

A hash function takes an input of arbitrary length and gives an output of a fixed length, as we said above. For being a ``good'' hash function it has to give completely different outputs for two random inputs. This characteristic it is guaranteed if the hash function has the following properties:
\begin{itemize}
    \item \textbf{Strong Avalanche effect}: A small change in the input of the Hash function produces a huge change in the output. 
    \item \textbf{Completeness}: Each bit of the input string has an effect on all the output bits.
\end{itemize}

The collision problem is the problem whose solution is found with less computational computational complexity for both classical and quantum computers, hence, its security is bounded by the resilience of a hash function to this attacks, i.e. it has to be collision resistant. A classical algorithm finds the solution with a complexity of $\mathcal{O}(2^n/2)$ and a quantum computer is able to solve it more efficiently with a computational cost of $\mathcal{O}(2^n/3)$ but without compromising its security \cite{CollisionQuantum}.

\subsubsection{PQC protocols} \label{Sec: IV a 2} 

There are two types of hash-based digital signatures (HBS): the One-Time Signature (OTS) scheme and the Multi-Time Signature (MTS) scheme, sometimes called statefull HBS and stateless HBS, respectively. The principal difference between them is the times that a secret key can be used without losing the security assumption. Only Europe, Japan and the United States have a HBS algorithm in their standardization process, SPHINCS+ \cite{Sphincs+}, which is a MTS signature scheme. The parameters of such algorithm is available in Table~\ref{Tab: PQC Hash}. SPHINCS+ has been implemented on an Artix-7 FPGA \cite{SPHINCS+1} and on an ARM CortexM3 \cite{NISTSigperf}.

\subsection{Lattice-based cryptography} \label{Sec: IV b} 

Lattice-based cryptography is based on hard problems involving lattices. It is a special case of the sub-set sum problem based cryptography proposed by Merkle and Hellman \cite{sum-problem}. Generally, a lattice is defined as an infinity grid of points represented by a linear combination of linearly independent vectors, called basis, $B={b_1,b_2,...,b_n}$:
$$
\mathcal{L}(B)=\biggl\{ \sum_i c_i b_i : c_i \in \mathbb{Z} \biggl\}.
$$
Thus, any point of the lattice is represented by an unique combination of the vectors of $B$. It is by definition a discrete subgroup of $\mathbb{R}^n$, since $\mathcal{L}$ spans $\mathbb{Z}^n$ which is a subgroup of $\mathbb{R}^n$.

Hoffstein, Pipher and Silverman proposed a lattice-based cryptosystem based on polynomial rings called 'NTRU' in the 1990s \cite{NTRUOrig}. The polynomial ring is defined as $\mathcal{R}_q=\mathcal{Z}_q[x]/(f(x))$, where $f(x)=x^n-1$ if $n$ is prime, or $f(x)=x^n+1$ if $n$ is a power of two. The  NTRU cryptosystem depends on three integers $(N,p,q)$ and four sets of polynomials ($\mathcal{L}_f,\mathcal{L}_g,\mathcal{L}_\phi,\mathcal{L}_m$), and it is composed by the following algorithms:
\begin{itemize}
    \item \textbf{Gen}: Alice gets two random polynomials $(f,g)$ from $\mathcal{L}_g$, under the condition that $f$ has inverses modulo $q$, $F_q$, and modulo $p$, $F_p$. The secret key (Sk) is the polynomial $f$ and the public key is defined by $Pk \equiv h = F_q \cdot g \hspace{0.1cm} mod \hspace{0.1cm} q$.

    \item \textbf{Enc/Vry}: To encrypt a message $m$ chosen from the polynomial set $\mathcal{L}_m$, Bob has to choose a random polynomial $\phi$ from $\mathcal{L}_\phi$:
    $$
    Ct \xleftarrow{} Enc(Pk,\phi,m)  \equiv p\phi \cdot h + m \hspace{0.1cm} mod \hspace{0.1cm} q.
    $$
    \item \textbf{Dec/Sign}: Alice uses the Sk to decrypt the message computing
    \begin{align*}
        a &\equiv f \cdot Ct \mod q \\
        msg &\xleftarrow{} F_p \cdot a \mod  p,
    \end{align*}
        
    where the coefficients of $a$ are selected between $-q/2$ and $q/2$.

\end{itemize}

Another lattice-based cryptosystem is based on the Learning With Errors (LWE) problem, which was proposed by Regev \cite{Regev-scheme} and which will be later described. The three algorithms involving such cryptography scheme are:
\begin{itemize}
    \item \textbf{Gen}: Given the dimension, $n$, of the lattice, the key generator algorithm generates the public and the private key as:
    \begin{align*}
        Sk & \equiv s^t \xleftarrow{} \mathbb{Z}^n_q, \\
        Pk & \equiv b^t = s^t A + e^t, 
    \end{align*}
    where $A$ is a $n \times m$ modulo $q$ random matrix and $e^t$ is a random error vector, both of them selected from a probabilistic distribution.
    \item \textbf{Enc/Vry}: Bob encodes the message, a bit string ($msg$),  by using a secret vector $x \xleftarrow{} \{0,1\}^m$ and the public key:
    \begin{align*}
        Ct & \equiv (u,u')  \xleftarrow{}  (Ax, b^t x + msg\cdot q/2). \\
    \end{align*}    
    \item \textbf{Dec/Sign}: Alice is able to decrypt the Ct using the Sk by computing:
    \begin{align*}
        msg \cdot q/2 & \approx u'-s^tu.  
    \end{align*}    
   
\end{itemize}

\subsubsection{Security Notion} \label{Sec: IV b 1} 
The security of Lattice-based cryptography relies on the following worst-case problems:
\begin{itemize}

    \item \textbf{Shortest Vector Problem (SVP)}: Given a lattice $\mathcal{L}$ find a non-zero vector $\Vec{v}\in \mathcal{L}$, whose norm $|\Vec{v}|$ is minimized. 
    \item \textbf{Closest Vector Problem (CVP)}: Given a lattice $\mathcal{L}$ and a vector $\Vec{u}$ find  vector $\Vec{v} \in \mathcal{L}$ such that the distance between $\Vec{u}$ and $\Vec{v}$ is shorter or equal to the distance between $\Vec{u}$ and the lattice $\mathcal{L}$, i.e. the problem involves minimizing the norm $|\Vec{v}-\Vec{u}|$.

\end{itemize}

Despite of the existence of an algorithm which finds the non-zero vector of the SVP in time $\mathcal{O}(2^{n})$ \cite{Lll1982}, there is no quantum algorithm providing an exponential speedup. Indeed, it is worth noting that the SVP problem can be reduced to the CVP problem. In 1996, Ajtai et al. showed that there is a connection between the worst-case problems and the average-case problems in lattice-based cryptography and explain how to construct hard lattice instances from random instances \cite{Ajtai}. This connection implies that if there is a probabilistic algorithm that solves the hard problems in the average-case,  then there exists a solution for the worst-case scenario. However, it has been proven that there is no probabilistic algorithm in polynomial time for the worst-case scenario. Finally, they conclude that a lattice-based cryptography scheme based on average-case problems can be designed with the security of the worst-case problems. The average-case problems related with lattices are:

\begin{itemize}
    \item \textbf{Short Integer Solution (SIS)}: Given a set of $m$ vectors $\Vec{a_i}\in \mathbb{Z}_q^n$ as the columns of a matrix $A_{n \times m}$, where $q$ is a prime number and $q$ define the modulus of the lattice, find a non-zero integer vector $\Vec{v} \in \mathbb{Z}^m$ which fulfills $A\cdot \Vec{v}=0 \in \mathbb{Z}_q^n$. 
    \item \textbf{Learning With Errors (LWE)}: Given a set of pairs $(\Vec{a_i},b_i)$, where $\Vec{a_i}$ and $b$ are sampled from a certain distribution, find a secret vector $\Vec{s}$ such that $\Vec{s} \cdot a_i \hspace{0.2cm} \text{mod} \hspace{0.2cm} e_i = b_i$ and $e_i$ is sampled from a Gaussian distribution over $\mathbb{Z}$.
\end{itemize}

Both problems are considered to be computationally hard for classical and quantum computers giving the security notion to lattice-based cryptography schemes. Specifically, the security of NTRU cryptography relies on the Ring-LWE (RLWE) problem, a variant of LWE problem where the secret vector is an unknown polynomial $s(x)$ in $\R_q$ and is faster than LWE-based cryptography \cite{RLWE}.

\subsubsection{PQC protocols} \label{Sec: IV b 2} 

Lattice-based cryptography can be splitted in two diffferent algorithms: NTRU, developed by mathematicians Jeffrey Hoffstein, Jill Pipher, and Joseph H. Silverman in 1996 \cite{NTRUOrig}; and Learning With Errors (LWE) first introduced by Oded Regev in 2009 \cite{Regev-scheme}. Lattice-based cryptography is the most promising quantum resistant cryptosystem, for encryption and signatures. This can be seen as different countries are proposing a large amount of lattice-based protocols to be be standardized. The NTRU encryption protocols that have been proposed to be standardized are: NTRU \cite{NTRU}, NTRU-HRSS \cite{NTRU-HRSS}, NTRU-Prime \cite{NTRU-prime} and NTRU+ \cite{NTRU+}. The following NTRU protocols have been also proposed to generate signatures: Falcon \cite{Falcon}, FatSeal \cite{Fatseal}, Peregrine \cite{Peregrine} and SOLMAE \cite{Solmae}. On the other hand, the LWE encryption protocols proposed to be standardized are: CRYSTALS-Kyber \cite{Kyber}, Saber \cite{Saber}, FrodoKEM \cite{FrodoKEMLW}, LAC PKE \cite{LAC}, Aigis-Enc \cite{Aigis-Enc} , AKCN-MLWE \cite{AKCN-MLWE}, TALE \cite{TALE} , AKCN-E8 \cite{AKCN-E8} SCloud \cite{Scloud} and SMAUG-T, which is a merge of SMAUG \cite{SMAUG} and \cite{Tiger}; and the following LWE protocols have been also proposed to generate signatures: CRYSTALS-Dilithium \cite{Dilithium}, Aigis-Sig \cite{Aigis-sig}, Mulan \cite{Mulan}, NCC-Sign \cite{NCC-sign} and HAETAE \cite{HAETAE}.

The following lattice-based PQC protocols have been implemented in HW devices, some of them in devices with low computational resources: NTRU, NTRU-HRSS and NTRU-Prime on an Artix-7 and on aZynq UltraScale+ \cite{NTRUCrystalSaber}; NTRU+ on a Xilinx Zynq-7000 \cite{NTRU+1}; Falcon on an ARM Cortex-A53 \cite{Falcon1} and on an ARM CortexM3\cite{NISTSigperf}; CRYSTALS-Kyber on a Xilinx Artix-7 \cite{CrystalKyber1}, on a Xilinx Artix-7 and on a Virtex-7 FPGAs \cite{CrystalKyber2} and on 64-bit ARM Cortex-A processors \cite{kyberARM64} using number-theoretic transform (NTT) optimization \cite{NTT2,NTT}, Saber on a Xilinx UltraScale+ \cite{Saber1} and on an Artix-7 and on a Zynq UltraScale+\cite{NTRUCrystalSaber} ; LAC on a Xilinx Zynq-7000 \cite{LAC1} and  CRYSTALS-Dilithium in \cite{CRYSTAL_Dilithium1} on Virtex UltraScale+ and on an ARM Cortex-M4 \cite{CRYSTAL_Dilithium1}.

The parameters of those algorithms are available in Table~\ref{Tab: PQC Lattice}. In \cite{LatticeSurvey} they give a good survey explaining different aspects of lattice-based cryptography as their theory, security and performance. They include NIST and Chinese standardisation processes.

\subsection{Code-based cryptography} \label{Sec: IV c} 

The first instance of a code-based cryptosystem was conceived by Robert McEliece in 1978 \cite{eliece}, which consisted in using the complexity of decoding a syndrome within code theory in order to encrypt messages with a high level of security. While proving a fine level of security, the McEliece cryptosystem usually suffers from excess in memory since it precises large ciphertexts and key pairs. Given this backdrops, a second version introduced by Harald Niederreiter in 1986 \cite{niederreiter} proposed a variation that allowed faster key generation and message sending while preserving the security given by the McEliece cryptosystem \cite{equiv}.

The McEliece cryptosystem requires being able to generate a random $t$-correctable ($n$,$k$)-code, that is, a $k$-dimensional code composed of $n$ bits able to correct all errors of weight equal or less than $t$. Given this condition the key generation goes as follows:

\begin{itemize}
    \item \textbf{Gen}: a $t$-correctable ($n$,$k$)-code with generator matrix $G$ and parity check matrix $H$ is randomly generated altogether with a random non-singular binary matrix $S$ of size $k\times k$ and a random permutation matrix $P$ of size $n \times n$. Finally, the matrix product $G' = SGP$ is computed, which is itself a new code. The key pair can now be introduced:
    \begin{align*}
        Sk & \equiv G \in \mathbb{F}_{2}^{k \times n}, S \in  \mathbb{F}_{2}^{k \times k}, P \in  \mathbb{F}_{2}^{n \times n}, \\
        Pk & \equiv G' \in \mathbb{F}_{2}^{k \times n}.
    \end{align*}
    
    \item \textbf{Enc}: the plaintext that is to be encrypted must be presented as a $k$-dimensional binary vector $\textbf{m}\in\mathbb{F}_{2}^{k}$. The message is encoded within the public key code, $G'$, as $ \textbf{m}G'$. Additionally, an $n$-dimensional random error $\textbf{e} \in \mathbb{F}_{2}^{n}$ of Hamming weight equal or lower than $t$ must be introduced. The encryption process consists in adding the error $\textbf{e}$ to the encoded message as: $\textbf{c} = \textbf{m}G' + \textbf{e}$. 
    
    \item \textbf{Dec}: The private key owner can decipher a ciphertext by first computing $CP^{-1} = \textbf{m}SGPP^{-1} + \textbf{e}P^{-1} =  \textbf{m}SG + \textbf{e}P^{-1}$. Afterwards, one can follow by computing $\textbf{e}P^{-1}H^T = \textbf{m}SGH^T + \textbf{e}P^{-1}H^T = \textbf{e}P^{-1}H^T = \textbf{z}$, where $\textbf{z} \in \mathbb{F}_{2}^{n-k}$ is the syndrome. Given that one knows $H$ and $\textbf{z}$, the error $\textbf{e}P^{-1}$ can be obtained through decoding, which allows the private key holder to find $\textbf{m}SG$. Furthermore, $\textbf{m}SGG^T = \textbf{m}S$ and, finally, by computing $\textbf{m}SS^{-1} = \textbf{m}$, one can reach the plaintext.
   
\end{itemize}

The Niederreiter cryptosystem builds upon the McEliece one by encrypting the plaintext through the error $\textbf{e}$. In other words, the cryptosystem is slightly modified into:

\begin{itemize}
    \item \textbf{Gen}: a $t$-correctable ($n$,$k$)-code with parity check matrix $H$ is randomly generated altogether with a random non-singular binary matrix $S$ of size $n-k\times n-k$ and a random permutation matrix $P$ of size $n \times n$. Finally, the matrix product $K = SHP$ is computed. The key pair can now be introduced:
    \begin{align*}
        Sk & \equiv H \in \mathbb{F}_{2}^{n-k \times n}, S \in  \mathbb{F}_{2}^{n-k \times n-k}, P \in  \mathbb{F}_{2}^{n \times n}, \\
        Pk & \equiv K \in \mathbb{F}_{2}^{k \times n}.
    \end{align*}
    
    \item \textbf{Enc}: the plaintext which is to be encrypted must be presented as an $n$-dimensional error vector $\textbf{e}\in\mathbb{F}_{2}^{n}$ of Hamming weight less or equal to $t$. Now, the message is encrypted through the product $\textbf{z} = \textbf{e}K^T$, where $\textbf{z}\in \mathbb{F}_{2}^{n-k}$ is the outcoming syndrome, which is the ciphertext.
    
    \item \textbf{Dec}: The ciphertext can be decrypted by the key pair owner by first operating $\textbf{z}(S^{-1})^T = \textbf{e}P^TH^TS^T(S^{-1})^T = \textbf{e}P^TH^T$, the result is a decodable syndrome given that we know the parity check matrix $H$. After decoding, one recovers $\textbf{e}P^T$, upon applying $\textbf{e}P^T(P^T)^{-1} = \textbf{e}P^T P = \textbf{e}$, which returns the plaintext. Note that permutation matrices are always orthogonal.
   
\end{itemize}

The Niederreiter cryptosystem allows for a code based-cryptosystem to happen without the requirement of storing the plain text within the code, which yields a time improvement. Nevertheless, this is done at the expense of the dimensionality change within the two cryptosystems. While the dimension of the plaintext of a McEliece cryptosystem using a $t$-correctable ($n$,$k$)-code is $k$, for a Niederreiter cryptosystem it is correspondent to the dimension of the space of $n$-dimensional binary error vectors of Hamming weight less or equal to $t$, which may be lower. This translates to the fact that less possible plaintexts can be ciphered.

\subsubsection{Security Notion} \label{Sec: IV c 1} 

Attacking either code-based cryptosystems may be done in two different ways. The first consists in attempting to separate the public key into the private key, which has been proved to be unfeasible \cite{francessigunature}. The second one consists in attempting to decode the syndrome given $K$\footnote{For the McEliece case, $K$ can be obtained through $G'$.}. This problem is defined as the computation syndrome decoding problem, which is proven to be NP-complete \cite{reviewjavi,complexitydecoding}:

\begin{itemize}
    \item \textbf{Computing the syndrome decoding problem}: given a binary linear $t$-correctable code of parity check matrix $K$ and a syndrome $\textbf{z}\in\mathbb{F}_{2}^{n-k}$ produced by the summation of a code word $\textbf{x}\in\mathbb{F}_{2}^{n}$ with and an error $\textbf{e}\mathbb{F}_{2}^{n}$ of weight equal or less than $t$ by $\textbf{z} = (\textbf{x}+\textbf{e})K^T$, find the error $\textbf{e}$.
\end{itemize}

 The best known generic attack on both cryptosystems is the Lee-Brickell attack \cite{lee-brickell}, which sets the security of the code and proves that both the McEliece and the Niederreiter cryptosystem have the same level of security \cite{equiv}.

\subsubsection{PQC protocols} \label{Sec: IV c 2} 

Two Niederreiter protocols have been proposed to be standardized: BIKE \cite{BIKE}, proposed to the NIST standardization process \cite{NIST3}: and PALOMA \cite{PALOMA}, proposed to the Korea PQC standardization process \cite{KpqC}. In the case of McEllice protocols, NIST process is considering Classical McEllice \cite{McEliece} and Hamming Quasi-Cyclic (HQC)protocols \cite{HQC} and South Korea is considering REDOG \cite{REDOG}.

The parameters of those algorithms are available in Table~\ref{Tab: PQC Code}.The following code-based PQC protocols have been implemented on a Xilinx Artix-7: BIKE \cite{BIKE1}, the classic McEliece \cite{McEliece1} and HQC \cite{HQC1}.

\subsection{Multivariate cryptography} \label{Sec: IV d} 

Multivariate cryptography is a public-key cryptography scheme based on a multivariate and non-linear polynomial map of a field $\mathbb{F}$. It was firstly proposed by Tsutomu Matsumoto and Hideki Imai in 1988 \cite{MultivariateOrig}.  In this scheme, a vector $x \in \mathbb{F}^q$ is mapped to a vector $x' \in \mathrm{F}^q $ through a map $P$ composed of a set of non-linear polynomials $p_1$, $p_2$, $\dots$, $p_m$:
\begin{align}
    P: & \hspace{1cm} \mathbb{F}^q \hspace{1.37cm} \xrightarrow{} \hspace{1.5cm} \mathbb{F}^q, \\
    &x=  (x_1,\dots, x_n) \hspace{0.5cm} \xrightarrow{} \hspace{0.5cm} x'= (p_1(x),\dots, p_m(x)).
\end{align}

The secret key is a non-linear map $P$, and two affine maps $S$ and $T$. An affine map is a linear map which connected two different affine spaces. The public key is the composition of the map with the affine maps $S$ and $T$: $P_k = T \circ P \circ S$, which looks like a random map. Multivariate cryptosystems are composed by the following algorithms:

\begin{itemize}
    \item \textbf{Gen}: Alice generates an easily invertible quadratic map $P: \mathbb{F}^q \xrightarrow{} \mathbb{F}^p$ and composes it with two affine maps, $S: \mathbb{F}^q \xrightarrow{} \mathbb{F}^q$ and $T: F^p \xrightarrow{} \mathbb{F}^p$:
    \begin{align*}
        Sk &  \xleftarrow{} P,S,T, \\
        Pk & = P' \xleftarrow{} T \circ P \circ S.   
    \end{align*}
                   
    \item \textbf{Enc}: Bob uses the Pk to encrypt a message $m \in \mathbb{F}^q$ getting the ciphertext (Ct) and sends it to Alice:
    \begin{align*}
       Ct \xleftarrow{} P'(m) \in  \mathbb{F}^q.
    \end{align*}
    \item \textbf{Dec}: Alice, by means of the Sk, decrypts the ciphertex obtaining the message: 
    \begin{align*}
       Ct \xleftarrow{} P(m) \in  \mathbb{F}^q.
    \end{align*}    
\end{itemize}
\subsubsection{Security Notion} \label{Sec: IV d 1} 
The security notion in multivariate cryptography relies on the NP-hard problem of finding preimages of multivariate polynomial maps. This is defined as: 

\begin{itemize}
    \item \textbf{The Multivariate Quadratic (MQ) problem:} Given a system of polynomials $p_1$, $p_2$, $\dots$, $p_m$, where each $p_i$ is a non-linear polynomial in $n$ variables whose coefficients and variables are defined over $\mathrm{F}^q$, find a solution $x = (x_1,\dots, x_n)$ that satisfies $p_1(x)=p_2(x) = \dots = p_m(x)=0$.
\end{itemize}

The computational complexity to solve the MQ problem depends on the degree of the polynomials, the number of variables $n$ and polynomials $m$ and the field $\mathbb{F}$. The best classical algorithm to solve this problem is based on  Gröbner bases, known as F5 \cite{F5}, and there is not a known quantum algorithm which solves MQ problem faster than F5 algorithm \cite{MQQuantum}.

\subsubsection{PQC protocols} \label{Sec: IV d 2}
There are two multivariate cryptosystems proposed to be standardized: Rainbow \cite{Rainbow} and GeMSS \cite{GeMSS}.An implementation on an ARM CortexM3 of Rainbow and GeMSS is done in \cite{NISTSigperf}. Both of them have been proposed in the NIST standardization process. However, recently, a new paper has been published claiming that the Rainbow protocol can be broken in just one weekend by means of a laptop \cite{Rainbowbroken}. On the other hand, South Korea has proposed a multivariate cryptography protocol called MQ-sign \cite{MQ-sign}.

The parameters of those algorithms are available in Table~\ref{Tab: PQC Multi}.Rainbow has been implemented in a HW device in \cite{Rainbow1}.

\subsection{Isogeny-based cryptography} \label{Sec: IV e}

Elliptic curve cryptography (ECC) is a public key cryptosystem developed in the 1980s by Miller \cite{Miller} and Koblitz \cite{KoblitzECC}, who suggested separately in the same year to use elliptic curves in the Diffie-Hellman cryptography protocol. Hasse (1936) discovered supersingular elliptic curves during his work on the Riemann hypothesis for elliptic curves \cite{Hasse1936}. Since quantum attacks based on Shor's algorithm break ECC relying on the discrete-logarithm problem \cite{shorFact}, this field was deemed as not secure. However, the interest in elliptic-curves for PQC was recovered due to the work in Supersingular-Isogeny Diffie-Hellman (SIDH)  published by L. de Feo \cite{SIDH2} and the work of A. Rostovtsev and A. Stolbunov \cite{SIDH1}; and in Hard Homogeneous Spaces \cite{HHS} protocols, both of them are based on random walks in graphs of horizontal isogenies. 

An elliptic curve is a projective curve defined over a field $k$. Specifically, it composed by the set of points that fulfill the following equation:

$$
E: y^2 z = x^3 +axz^2+b^3 \quad a,b \in k \quad  \text{ and } 4a^3 +27b^2 \ne 0,
$$
where the point at infinity is $(0:1:0)$, when $z=0$. In the affine space it is defined as:
$$
y^2 = x^3 +ax + b,
$$
with $\mathcal{O}=(0:1:0)$ as the point at infinity.


Following such definition, an isogeny is a map between two elliptic curves $\phi: E \xrightarrow{} E'$, such as $\phi$ is a surjective group morphism that preserves its identity $\phi(\mathbb{O})= \mathbb{O}'$. Two curves are called isogenous if there exists an isogeny between them and, there is a isogeny if and only if $ \# E(k) = \# E'(k)$, where $ \# $ is defined as the cardinality of the elliptic curve. An elliptic curve defined over a field $\mathbb{F}_p$ has an invariant, called the j-invariant of a Montgomery curve, which determines the isomorphism class. It is defined by:
$$
j(E_{a,b})=  \frac{256(a^2-3)^3}{a^2-4}.
$$

Two isogenous elliptic curves have different j-invariants, $j(E) \neq j(E')$ if they can be mapped by an isogeny $\phi$, hence, an isogeny maps one isomorphism to another.

\begin{figure}[!t]
    \centering
    \includegraphics[width= 0.8 \textwidth]{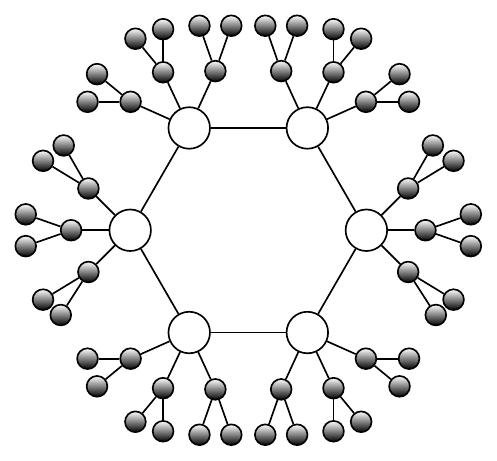}
    \caption{Graphical representation of a graph of isogenies, where each node is a elliptic curve and each edge connect two elliptic curves if there exits an isogeny between them}
    \label{fig: Isogeny}
\end{figure}

A graph of isogenies is a collection of elliptic curves, i.e. isomorphisms, connected by isogenies, where the elliptic curves are the nodes and the isogenies are the edges. An example is represented in Fig.~\ref{fig: Isogeny}.

Isogeny-based PQC protocols are based on random walks in isogeny graphs, obtaining a shared secret to be used as symmetric key between Alice and Bob. Isogeny-based protocols are only composed by the key exchange algorithm: 

\begin{itemize}
    \item Key exchange: The key exchange algorithm consists on random walks taken by Alice and Bob from the same Elliptic Curve $E_0$ along the graph. They publish the EC they have reached, $E_A$ and $E_B$, and they repeat exactly the same random walk they followed before but from the EC the opposite party have published, reaching the same secret EC, $E_S$. It is easy to see that both paths have to commute: $P_A(P_B(E_0))$ has to be equal to $P_B(P_A(E_0)$, where $P_i$ is the secret path they followed. Alice and Bod end the algorithm in the same elliptic curve with the same j-invariant, i.e. in the same isomorphism.
\end{itemize}

A singular EC is an EC with singular points, which are defined as points within the EC and in the curves defined by the two partial derivatives of the EC, known as the Jacobi criterion. Supersingular Isogeny Diffie-Hellman (SSIDH) cryptography is a special case of Isogeny-based cryptography whose security against classical and quantum attacks has been proved \cite{Tani_2009}.

\subsubsection{Security Notion} \label{Sec: IV e 1}

The security of Isogeny-based cryptography against classical and quantum attacks relies on the Supersingular Isogeny Problem (SSIP):
\begin{itemize}
    \item \textbf{Supersingular Isogeny problem}: Given a prime $p$ and two supersingular elliptic curves over $\mathbb{F}_{p^2}$, $E$ and $E'$ find an isogeny $\phi: E \xrightarrow{} E'$.
\end{itemize}

The best algorithm to solve the SSIP and, hence, that breaks isogeny-based cryptography, is based on meet-in-the-middle attack and its complexity is $\mathcal{O}(p^{1/4})$, which requires $\mathcal{O}(p^{1/4})$ storage capability \cite{Tani_2009}, or, in the case of Supersingular Isogeny Key Encapsulation (SIKE), its security is based on van Oorschot-Wiener (vOW) golden collision finding algorithm \cite{AttackSIKE}. The best quantum algorithm is able to find the secret with $\mathcal{O}(p^{1/6})$, which is not a notorious advantage and does not compromise isogeny-based cryptography \cite{Tani_2009}.

\subsubsection{PQC protocols} \label{Sec: IV e 2}

Supersingular Isogeny Key Encapsulation (SIKE) \cite{SIKE} is an encryption scheme based on Supersingular Isogeny Diffie-Hellman protocol proposed for the NIST standardization process. On the other hand, FIBS is a signature scheme based on the same protocol proposed in the South-Korean standardization process \cite{FIBS}. A compressed alternative of SIKE was proposed in 2016 by Axzarderakhsh et al. in \cite{CompressedSIKE}, requiring to transmit half of the data. This compressed SIKE was implemented in an ARM Cortex-M4 processor \cite{SIKEARM} and on a Xilinx Virtex-7 \cite{SIKEFPGA}, the authors claimed that it is the algorithm which introduces the lowest latency to communications due to its low computational requirements and the extremely compact key sizes. There are also works that try to speed-up the algorithm proposed by L. de Feo \cite{SIDH2}, such as the work of B. Koziel et al. \cite{NEON-SIDH}.

The parameters of those algorithms are available in Table~\ref{Tab: PQC Isogeny}.

\subsection{Multiparty Computation protocol and Graph-based cryptography} \label{Sec: IV f}
In the PQC standardization processes around the word only South Korea is still considering a protocol based on one of these problems as we will see in the next section \cite{KpqC}.

\subsubsection{Multiparty Computation protocol}
Multiparty Computation Protocol (MPC) cryptography is used in scenarios where several parties $P_i$ want to make some data available keeping its confidentiality. As an example, we could imagine two different countries which wants to keep the trajectory of its spy satellite secret but they want to be sure that they have different trajectory in order to avoid collisions. Therefore, each party has a secret $x_i$, the trajectory of their satellite, which has to remain secret but they want to share some confidential information making sure they will not collide. This is known as the MPC problem. In this sense, the MPC paradigm relates with the interest of allowing some party to do some computations to extract some conclusions using some protected data without actually having access to the raw data. The zero-knowledge proof allows the entities to convince the other ones about something without making the data public. The first MPC scheme was proposed by Y. Ishai, E. Kushilevitz, R. Ostrovsky and A. Sahai in 2007 \cite{MPC}, known as MPC-in-the-head paradigm. The first application proof of this paradigm was presented by Giacomelli et al. at 2016 in \cite{MPCapplication} and protocols based on this paradigm can be used to generate signatures. Picnic \cite{Picnic} is a PQC signature scheme that has been proposed for the NIST standardization process, but that was rejected in the third round \cite{NIST3}. However, AIMer\cite{AIMer} PQC algorithm based on MPC-in-the-head paradigm had been selected for the second round in the South Korea PQC standardization process.

\subsubsection{Graph-based Cryptography}

Graph-based cryptography refers to Perfect Code Cryptosystems (PCC) proposed by Koblitz et al. in 1993 \cite{Koblitz}. The study of Perfect Codes (PC) emerged in the field of information theory, since PCs over graphs corresponds to PCs over structured alphabets \cite{PCC-HP}. Despite of having its origin in the study of codes, we separate this cryptography from code-based cryptography explained in Section \ref{Sec: IV c} since the hard problem to be solved in order to break each of the cryptosystems is different. A graph is a mathematical object formed by a set of edges and vertices, where each vertex is connected to other vertices through edges. A Perfect Code (PC) in a graph is defined as the set of vertices $A$ such that every vertex not included in $A$ is connected to only one element in $A$. A graph is defined as $G={V,E}$, where $V$ is a set of vertices and $E$ is a set of edges. Therefore, $A \subseteq V$  and for every $v \in V$, $N[v]$ contains only one element of $A$, where the set $N[v]$ is formed by the vertices connected by an edge to $v$. The security of this cryptosystem relies on the hardness of knowing if a graph has PCs or not, which is a hard problem by itself (NP-complete problem) \cite{PCC-HP}, as well as on the hardness of finding the vertices which form the PC, which is conjectured to be a hard problem (NP-complete problem). IPCC \cite{IPCC} is a graph-based post-quantum cryptography protocol proposed to be standardized in the South-Korean standardization process, which improves the original ideas proposed by Koblitz.

\subsection{Performance of PQC algorithms}
\label{Sec: IV g}

\begin{table}[t]
\resizebox{\textwidth}{!}{%
\begin{tabular}{|c||c | c | c |} 
 \hline
 Lattice-based &  Public Key &  Private key   & Ct/Signature      \\
 \hline\hline
 \textbf{NTRU} \cite{NTRUECCperf,NIST3} &  930 & 1234 &  930  \\
 \hline 
 \textbf{NTRU-HRSS} \cite{NTRUHRSSPerf, NIST3} &1 138 & 1 450 & 1 138  \\
 \hline 
 \textbf{NTRU-Prime} \cite{NTRUHRSSPerf,NTRUPrimePerf,NIST3}  & 994 &15 158 & 897  \\
 \hline 
 \textbf{NTRU+} \cite{NTRU+,KorPerf} & 864 & 1 728 & 864 \\
 \hline 
 \textbf{Falcon} cite{FalconPerf,NIST3} & 897 & 7 553 &  666  \\
 \hline
 \textbf{FatSeal} \cite{Fatseal} & 2 321 & 385 & 2 048     \\
 \hline
 \textbf{Peregrine} \cite{Peregrine,KorPerf} & 897 & 7 553 & 666  \\
 \hline
 \textbf{SOLMAE} \cite{Solmae,KorPerf} & 897 & 7 553 & 666    \\
 \hline
 \textbf{CRYSTALS-Kyber} \cite{SPHINCSKyberPerf,NIST3}  & 800 & 1 632 & 768  \\
 \hline
 \textbf{Saber} \cite{NTRUHRSSPerf,NIST3} & 672 & 832 & 736  \\
 \hline
 \textbf{FrodoKEM} \cite{FrodoKEMLW,NIST3} & 9 619 & 19 888 & 9 720  \\
 \hline
 \textbf{LAC PKE} \cite{LAC} & 544 & 1 056 & 704    \\
 \hline
 \textbf{Aigis-Enc} \cite{Aigis-Enc,Aigis-Enc2} & 672  & 1 568 & 672     \\
 \hline
 \textbf{AKCN-MLWE (AES256)} \cite{AKCN-MLWE} & 10 560 & 10 560 & 8 610    \\
 \hline
 \textbf{TALE} \cite{TALE,TALE2} & 736 & 1 504 & 704    \\
 \hline
 \textbf{AKCN-E8 (AES256)} \cite{AKCN-E8} & 928 & 928 & 896   \\
 \hline
 \textbf{SCloud (AES256)} \cite{Scloud} & 98 800 & 19 800 & 82 300   \\
 \hline
 \textbf{TiGER}  \cite{KorPerf,Tiger} &  480 & 177 & 640  \\
 \hline
 \textbf{SMAUG} \cite{KorPerf,SMAUG}& 174 & 672 & 768   \\
 \hline
 \textbf{CRYSTALS-Dilithium(AES192)} \cite{NIST3} & 1 312 & 2 528 & 2 420 \\
 \hline
 \textbf{Aigis-Sig} \cite{Aigis-sig} & 672  & 1 568 & 672   \\
 \hline
 \textbf{Mulan} \cite{Mulan}& 1 312 & 2 528 & 2 420    \\
 \hline
 \textbf{GCKSign} \cite{GCKSign,KorPerf} & 1 760 & 288 & 1 952    \\
 \hline
 \textbf{NCC-Sign} \cite{NCC-sign, KorPerf} & 1 440 & 2 400 & 2 529    \\
 \hline
 \textbf{HAETAE} \cite{HAETAE,KorPerf} & 992 & 1 376 & 1 463    \\
 \hline
 
\hline 

\end{tabular}}
\caption{Performance of Lattice-based PQC protocols. Ciphertext (Ct) and keys length are expressed in Bytes. All values are taken from the level I security defined by NIST: "Any attack that breaks the relevant security definition must require computational
resources comparable to or greater than those required for key search on a block
cipher with a 128-bit key (e.g. AES128)" \cite{Levels}}
\label{Tab: PQC Lattice}
\end{table} 
\begin{table}[h]
\resizebox{\textwidth}{!}{%
\begin{tabular}{|c||c | c | c |} 
 \hline
  Hash-based &  Public Key &  Private key   & Ct/signature   \\
 \hline\hline
 \textbf{SPHINCS+} \cite{Sphincs+,SPHINCSKyberPerf,NIST3} & 32 & 64 & 7 856  \\
 \hline

\end{tabular}}
\caption{Performance of Isogeny-based PQC protocols. Ciphertext (Ct) and keys length are expressed in Bytes. All values are taken from the level I security defined by NIST: "Any attack that breaks the relevant security definition must require computational
resources comparable to or greater than those required for key search on a block
cipher with a 128-bit key (e.g. AES128)" \cite{Levels}}
\label{Tab: PQC Hash}
\end{table}

\begin{table}[t]
\resizebox{\textwidth}{!}{%
\begin{tabular}{|c||c | c | c |} 
 \hline
 Code-based &  Public Key &  Private key   & Ct/Signature     \\
 \hline\hline
 \textbf{BIKE} \cite{NIST3} & 1 540 & 280 &  1 572   \\
 \hline 
 \textbf{PALOMA} \cite{PALOMA,KorPerf} & 319 488 & 7 808 & 32    \\
 \hline 
 \textbf{Cl. McEllice} \cite{NIST3} & 261 129 & 6 492 & 128  \\
 \hline
 \textbf{HQC}  \cite{NIST3} & 2 249 & 40 & 4 481      \\
 \hline 
 \textbf{Piglet-1} \cite{Piglet} & 1 212 & 32 & 1 801    \\
 \hline
 \textbf{ROLLO-I} \cite{Rollo,KorPerf} & 1 240  & 120 & 620    \\
 \hline
 \textbf{REDOG} \cite{REDOG,KorPerf} & 14 250 & 1 450 & 830    \\
 \hline
 \textbf{Enhanced PQsigRm} \cite{pqsigRM,KorPerf} & 2 000 000 & \xmark  & 1 032  \\
 \hline

\end{tabular}}
\caption{Performance of Code-based PQC protocols. Ciphertext (Ct) and keys length are expressed in Bytes. All values are taken from the level I security defined by NIST: "Any attack that breaks the relevant security definition must require computational
resources comparable to or greater than those required for key search on a block
cipher with a 128-bit key (e.g. AES128)" \cite{Levels}}
\label{Tab: PQC Code}
\end{table}

\begin{table}[t]
\resizebox{\textwidth}{!}{%
\begin{tabular}{|c||c | c | c |} 
 \hline
  Multivariate &  Public Key &  Private key   & Ct/Signature  \\ 
 \hline\hline
 \textbf{GeMSS} \cite{SPHINCSKyberPerf,NIST3} & 352 168 &  16 & 33  \\
 \hline 
 \textbf{Rainbow} \cite{Rainbow, NIST3} & 157 800 & 611 300 &  164  \\
 \hline 
 \textbf{MQ-Sign} \cite{MQ-sign,KorPerf} & 328 441 & 15 561 & 134  \\
 \hline 
 
\end{tabular}}
\caption{Performance of Multivariate PQC protocols. Ciphertext (Ct) and keys length are expressed in Bytes. All values are taken from the level I security defined by NIST: "Any attack that breaks the relevant security definition must require computational
resources comparable to or greater than those required for key search on a block
cipher with a 128-bit key (e.g. AES128)" \cite{Levels}}
\label{Tab: PQC Multi}
\end{table}

\begin{table}[t]
\resizebox{\textwidth}{!}{%
\begin{tabular}{|c||c | c | c |} 
 \hline
  Isogeny-based &  Public Key &  Private key   & Ct/Signature \\  
 \hline\hline
 \textbf{SIKE} \cite{NIST3} & 330 & 374 &  346  \\
 \hline 
 \textbf{FIBS} \cite{FIBS} & 32 & 64 & 17 088  \\
 \hline 
  
\end{tabular}}
\caption{Performance of Isogeny-based PQC protocols. Ciphertext (Ct) and keys length are expressed in Bytes. All values are taken from the level I security defined by NIST: "Any attack that breaks the relevant security definition must require computational
resources comparable to or greater than those required for key search on a block
cipher with a 128-bit key (e.g. AES128)" \cite{Levels}}
\label{Tab: PQC Isogeny}
\end{table} 

\begin{table}[]
\resizebox{\textwidth}{!}{%
\begin{tabular}{|c||c | c | c |} 
 \hline
  Others &  Public Key &  Private key   & Ct/Signature  \\  
 \hline\hline
 \textbf{IPCC (AES80)}\cite{IPCC} & 4 800 & 400 &  92 000 \\
 \hline 
 \textbf{AIMer} \cite{AIMer}  & 32 & 16 & 5 904  \\
 \hline 
  
\end{tabular}}
\caption{Performance of Isogeny-based PQC protocols. Ciphertext (Ct) and keys length are expressed in Bytes. All values are taken from the level I security defined by NIST: "Any attack that breaks the relevant security definition must require computational
resources comparable to or greater than those required for key search on a block
cipher with a 128-bit key (e.g. AES128)" \cite{Levels}}
\label{Tab: PQC Others}
\end{table} 
The NIST has defined 5 security levels to compare the performance of the different PQC algorithms when the security they provide is the same \cite{Levels}. Some metrics on the performance of the different PQC algorithms proposed to be standardized to achieve first level of security are presented in: Table~\ref{Tab: PQC Lattice} (Lattice-based), Table~\ref{Tab: PQC Hash} (Hash-based), Table~\ref{Tab: PQC Code} (Code-based), Table~\ref{Tab: PQC Multi} (Multivariate cryptography), Table~\ref{Tab: PQC Isogeny} (Isogeny-based) and Table~\ref{Tab: PQC Others} (MPC and Graph-based). In general terms their performance in OT communications can be briefly discussed:
\begin{table*}[t!]
\resizebox{\textwidth}{!}{%
\begin{tabular}{|c||c | c | c | c | c |} 
 \hline
  PQC family &  Key lengths &  Ct/Signature length   & Number of PQC algorithms & Time* & Appropriate for CI  \\  
 \hline\hline
 Hash-based & \cellcolor{green!60} A & \cellcolor{red!60} D & \cellcolor{orange!60} C  & \cellcolor{red!60} D & \cellcolor{red!60} D \\
 \hline 
 Lattice-based & \cellcolor{yellow!60} B  & \cellcolor{yellow!60} B & \cellcolor{green!60} A & \cellcolor{green!60} A & \cellcolor{green!60} A  \\
 \hline 
 Code-based & \cellcolor{orange!60} C  & \cellcolor{yellow!60} B & \cellcolor{yellow!60} B & \cellcolor{red!60} D & \cellcolor{red!60}  D  \\
 \hline 
 Multivariate & \cellcolor{red!60} D  & \cellcolor{green!60} A & \cellcolor{yellow!60} B & \cellcolor{red!60} D & \cellcolor{orange!60} C\\
 \hline 
 Isogeny-based & \cellcolor{green!60} A & \cellcolor{yellow!60} B & \cellcolor{orange!60} C & \cellcolor{orange!60} C & \cellcolor{yellow!60} B \\
 \hline 
 Multiparty  & \cellcolor{green!60} A  & \cellcolor{orange!60} C & \cellcolor{red!60} D & \cellcolor{red!60} D & \cellcolor{red!60} D \\
 \hline  
 Graph-based & \cellcolor{orange!60} C  & \cellcolor{red!60} D & \cellcolor{red!60} D & \cellcolor{orange!60} C & \cellcolor{red!60}  D \\
 \hline 
\end{tabular}}
\caption{Adequacy of PQC protocols to CI in terms of their key and ciphertext/Signature lengths, the number of PQC algorithms proposed to be standardized and the average computational time. The letters and the colors define the suitability of the PQC family, being the most suitable (A/green), moderately suitable (B/yellow), hardly suitable (C/orange) and not suitable (D/red).}
\label{Tab: PQC performance}
\end{table*} 
\begin{itemize}

    \item \textbf{Lattice-based} (Table~\ref{Tab: PQC Lattice}): Lattice-based cryptography requires relatively small public and private keys as well as ciphertexts while showing a lower computational cost compared to other PQC families \cite{Perform-lattice-multivariate}. Hence, it is the most promising candidate to be implementable in ICS/CI scenarios.

    \item \textbf{Hash-based} (Table~\ref{Tab: PQC Hash}): There is only one hash-based PQC protocol, Sphincs+, proposed along the world, and its problems are the large ciphertext and the lengthy signing time \cite{SPHINCSKyberPerf}, which could make it not suitable for ICS/CI.
    
    \item \textbf{Code-based} (Table~\ref{Tab: PQC Code}): Their public and private keys' length are longer than for other PQC families implying that more memory is required. Also, the fact that deciphering the ciphertext requires a high computational cost makes code-based not to be the best option to be deployed in an ICS/CI environment a priori \cite{NIST3}.

    \item \textbf{Multivariate-based} (Table \ref{Tab: PQC Multi}): Multivariate-based cryptosystems require large keys while their ciphertext lengths are the same as for other PQC families, or even shorter. Nonetheless, due to the high computational and memory resources requirements \cite{Perform-lattice-multivariate} multivariate cryptography is not the best option to be used in CI.

    \item \textbf{Isogeny-based} (Table~\ref{Tab: PQC Isogeny}): Isogeny based cryptography presents keys of moderate length, however, FIBS has a large signature, making it unappealing to be implemented in CI. Although isogeny-based protocols have small key sizes, the computational cost is a great disadvantage due to the low computational requirements in ICS/CI.

    \item \textbf{Graph-based and multiparty computation cryptography} (Table~\ref{Tab: PQC Others}):  These are the least used PQC algorithms along the world. They are not receiving a lot of attention and, thus, their security against classical and quantum attacks are not as studied as for other PQC families. Due to the nature of ICS/CI is not in principle recommended to use protocols based on these hard problems for such contexts.
\end{itemize}

The result of this high level comparison among PQC algorithms implies that, a priori, the most suitable protocols for its integration in ICS/CI networks are those that belong to the lattice-based cryptography family. It is important to note that during the writing of this perspective, a new preprint was posted proposing a quantum algorithm that solves LWE problem with a polynomial complexity \cite{LWEQuantum}. While the method proposed does not break NIST PQC candidates, developments in this field may lead to such outcome. Thus, even if out speculative comparison leans towards them, this advances in quantum algorithms should be taken into account. The comparison is summarized in Table~\ref{Tab: PQC performance}. However, and as discussed before, this conclusion is rather speculative and, thus, an actual comparison in the conditions of those scenarios should be done for obtaining accurate results and conclusions because it is probably that any of them satisfies all industrial infrastructure communication's constrains.

\begin{table*}[ht!]
\resizebox{\textwidth}{!}{%
\begin{tabular}{|c||c | c | c | c | c | c | c |} 
 \hline
 PQC Algo. & \textbf{Europe (ETSI)} & \textbf{China (CACR)} & \textbf{USA (NIST)}   &  \textbf{Germany (BSI)}  &  \textbf{Japan (CRYPTREC)} & \textbf{France (ANSSI)}  & \textbf{South Korea (KpqC)}  \\ [0.5ex] 
 \hline\hline
 \textbf{Code-based} & \cmark & \cmark & \cmark &  \cmark & \cmark & \xmark & \cmark  \\
 \hline 
 Niederreiter & BIKE \cite{BIKE} & - & BIKE \cite{BIKE} &  -  & BIKE \cite{BIKE} & - & PALOMA  \cite{PALOMA}\\
 \hline
 McEllice & \begin{tabular}{@{}c@{}}Cl. McEllice \cite{McEliece}\\ HQC \cite{HQC} \end{tabular}  & Piglet-1 \cite{Piglet}& \begin{tabular}{@{}c@{}}Cl. McEllice \cite{McEliece} \\ HQC \cite{HQC} \end{tabular}  & Cl. McEllice \cite{McEliece} & \begin{tabular}{@{}c@{}}Cl. McEllice \cite{McEliece} \\ HQC \cite{HQC} \end{tabular} & - & \begin{tabular}{@{}c@{}} \\ REDOG  \cite{REDOG} \end{tabular}  \\
 \hline
 Signature & - & - & - &  -  & - & - &  \\
 \hline \hline
 \textbf{Lattice-based} & \cmark & \cmark & \cmark  & \cmark & \cmark & \cmark & \cmark \\
 \hline
 LWE & \begin{tabular}{@{}c@{}}CRYSTALS-Kyber \cite{Kyber}\\ Saber \cite{Saber} \end{tabular} & \begin{tabular}{@{}c@{}} Aigis-Enc \cite{Aigis-Enc} \\ AKCN-MLWE \cite{AKCN-MLWE} \\  SCloud \cite{Scloud}\end{tabular}   & \begin{tabular}{@{}c@{}}CRYSTALS-Kyber \cite{Kyber} \\ Saber \cite{Saber} \end{tabular}  & FrodoKEM \cite{FrodoKEMLW}& \begin{tabular}{@{}c@{}}CRYSTALS-Kyber \cite{Kyber} \\ Saber \cite{Saber} \end{tabular} & \begin{tabular}{@{}c@{}}CRYSTALS-Kyber \cite{Kyber} \\ FrodoKEM \cite{FrodoKEMLW} \end{tabular} & - \\
 \hline
 LWE Signature & CRYSTALS-Dilithium \cite{Dilithium} & \begin{tabular}{@{}c@{}} Aigis-Sig    \cite{Aigis-sig}\\ Mulan \cite{Mulan}  \end{tabular} & CRYSTALS-Dilithium \cite{Dilithium}& - & CRYSTALS-Dilithium \cite{Dilithium} & CRYSTALS-Dilithium \cite{Dilithium} & \begin{tabular}{@{}c@{}}  HAETAE \cite{HAETAE} \end{tabular} \\
 \hline 
 RLWE & \begin{tabular}{@{}c@{}}NTRU \cite{NTRU} \\ NTRU-HRSS \cite{NTRU-HRSS}\\ NTRU-Prime \cite{NTRU-prime}\end{tabular} &  \begin{tabular}{@{}c@{}}TALE \cite{TALE} \\ LAC PKE \cite{LAC} \\ AKCN-E8 \cite{AKCN-E8} \\ \end{tabular} & \begin{tabular}{@{}c@{}}NTRU \cite{NTRU} \\ NTRU-HRSS \cite{NTRU-HRSS}\\ NTRU-Prime \cite{NTRU-prime} \end{tabular} &  - & \begin{tabular}{@{}c@{}}NTRU \cite{NTRU} \\ NTRU-HRSS \cite{NTRU-HRSS}\\ NTRU-Prime \cite{NTRU-prime} \end{tabular} & NTRU+ \cite{NTRU+} & \begin{tabular}{@{}c@{}}NTRU+ \cite{NTRU+} \\ SMAUG+TiGER(merged) \cite{SMAUG,Tiger} \end{tabular}  \\
 
 \hline
 RLWE Signature & Falcon \cite{Falcon} & FatSeal \cite{Fatseal} & Falcon \cite{Falcon} & - & Falcon \cite{Falcon} & Falcon \cite{Falcon} & \begin{tabular}{@{}c@{}} NCC-Sign \cite{NCC-sign}\end{tabular} \\
 \hline \hline
 \textbf{Hash-Based} & \cmark & \xmark & \cmark &  \xmark & \cmark & \xmark & \xmark \\
 \hline
 MTS & SPHINCS+ \cite{Sphincs+}& - & SPHINCS+ \cite{Sphincs+} & -  & SPHINCS+ \cite{Sphincs+} & - & - \\
 \hline \hline
 \textbf{Multivariate} & \cmark & \xmark & \cmark &  \xmark & \cmark & \xmark & \cmark  \\
 \hline 
 Small-Field  & - & - & - & - & - & - & -  \\
 \hline 
  Big-Field  & GeMSS \cite{GeMSS} & - & GeMSS \cite{GeMSS} & - & GeMSS \cite{GeMSS} & - & - \\
 \hline
 Signatures & - & - & - & - & - & - & MQ-Sign \cite{MQ-sign}\\
 \hline 

\hline \hline
\textbf{Isogeny} & \cmark & \xmark & \cmark &  \xmark & \cmark & \xmark & \cmark \\
\hline 
Supersingular & SIKE \cite{SIKE} & - & SIKE \cite{SIKE} & - & SIKE \cite{SIKE} & - & - \\
\hline
Signatures & - & - & - & - & - & - & - \\
\hline \hline
\textbf{MPC} & \xmark & \xmark & \xmark & \xmark & \xmark & \xmark & \cmark  \\
\hline
Signatures & - & - & - & - & - & - & AIMer \cite{AIMer}\\
\hline \hline 
\textbf{Graph-based} & \xmark & \xmark & \xmark & \xmark & \xmark & \xmark & \cmark  \\
\hline
Perfect Code & - & - & - &  -  & - & - & - \\
\hline

\end{tabular}}
\caption{The data have been obtained from ETSI (European Telecommunications Standards Institute) in the case of Europe \cite{ETSI}; from CACR (Chinese Association for Cryptologic Research) in the case of china \cite{CACR}; from NIST (National Institute of Standards and Technology) in the case of United States of America \cite{NIST3}; from BSI (Bundesamt für Sicherheit in der Informationstechnik) in the case of Germany\cite{BSI}; from CRYPTREC (Cryptography Research and Evaluation Committees) in the case of Japan \cite{cryptrec}; from ANSSI (Agence Nationale de la Sécurité des Systèmes d'Information) in France \cite{ANSSI} and from KpqC (Korean post-quantum Cryptography) in the case of South Korea \cite{KpqC}. A discussion comparing their performances is given in Section \ref{Sec: IV g}, while a set of tables presnting the performance is given in Section \ref{Sec: IV}.}
\label{Tab: PQC countries}
\end{table*} 
\begin{table*}[]
\begin{tabular}{|c||c | c | c |} 
 \hline
  Country &  Quantum Public Spend 2023 (Million \euro) & (Million \$)  & Quantum Public Spend 2022 (Million \euro)    \\
 \hline\hline
 China & 13 500 ($+0\%$) & 15 000 & 13 500  \\
 \hline
 UK & 3 600 ($+200\%$) & 4 300 &1 200  \\
 \hline
 USA & 3 000 ($+172.73\%$) & 3 750 & 1 100 \\
 \hline
 Germany & 3 000  ($+13.33\%$) & 3 300 & 2 600  \\
 \hline
 South Korea & 2 000 ($+56143 \%$) & 2 350 & 35  \\
 \hline
 France &1 800 ($+0\%$) & 2 200 & 1 800  \\
 \hline
 Russia & 1 250 ($+119.3 \%$) & 1 450 & 570  \\
 \hline
 Europe & 1 000 ($+0\%$) & 1 100 & 1 000  \\ 
 \hline
 Canada & 1 000 ($+0\%$) & 1 100 & 1 000 \\
 \hline
 India & 630 ($-30 \%$) & 735 & 900  \\
 \hline
 Japan & 600 ($+0\%$) & 700 & 600  \\
 \hline 
 Others & 4 620 ($+71.1\%$) & 5 100 & 2 700 \\
 \hline \hline
 Total & 36 000 ($+33\%$) & 40 000 & 27 000 \\
\hline
  
\end{tabular}
\caption{Global distribution of the public investment in quantum technologies in 2023 \cite{QURECA23} and in 2022 \cite{QURECA22}. The total investment was around 27 billion euros in 2022 and increased a $33\%$ in 2023, to 36 billion euros.} 
\label{Tab: Investments}
\end{table*}

\subsection{PQC standardization processes around the world} \label{Sec: IV h}

Despite of the global concern regarding the cybersecurity threat posed by the possibility of constructing fault-tolerant quantum computers, not every country/entity has started a standardization process by their own, or if they has started they do not publish this information. In this context, the United States is the pioneer by means of the NIST post-quantum cryptography standardization process, which started in 2017 and is yet unfinished (at the moment they are conducting the fourth round \cite{NISTstand}). Even if it stands as the largest standardization process, not all countries will adopt NIST recommendation and standardized PQC algorithms as a result of economical and political differences. Countries not aligned with the North Atlantic Treaty Organization (NATO) and the USA foreign policy are developing their own algorithms such as, for example, China. However, even EU countries that belong to such alliances are developing PQC protocols and requirements by their own, e.g. Germany and France. There are other countries which are investing in quantum technologies but they do not have published a PQC standardisation process. However, by comparing the Table~\ref{Tab: PQC countries} with Table~\ref{Tab: Investments} it can be inferred that it is quite probable that they are making their own efforts to be quantum secure.

The main issue with the adoption of PQC schemes for cryptography tasks resides on the youth of most of the proposals. More concretely, there is not a wide experience in the integration of the schemes in real systems implying that the security of the experimental implementation of PQC schemes is still a question to be deeply explored. Due to the criticality of cybersecurity regarding the security of national secrets, most governments are still skeptical of completely relying on those novel methods. This is also a result of the fact that many of the cryptography protocols proposed in the past were proved to be insecure years later of their proposal. This is somehow result of the fact that some of the hard problems in which cryptography is based on rely on mathematical assumptions regarding their hardness, which is a evolving science. Obviously, this problem is exacerbated when PQC security proofs are considered, mainly because the class of problems that are solvable by quantum computers and its relationship to other complexity classes is still a question under research. Also, even with formal proofs of the hardness, the actual implementations of the protocols are not guaranteed to be safe. Therefore, due to the early stage of PQC proposals, many countries are exploring different possibilities as a function of their own analyses and interests rather than relying on the recommendations by a single entity such as the NIST. In Table~\ref{Tab: PQC countries} and Table~\ref{Tab: Investments}, we present the global efforts regarding PQC integration in their communications and public investments, respectively, in quantum technologies as a way of showing a picture of the state of affairs at the time of writing.

Recently, South Korea has finished round 1 of their PQC Standardisation process \cite{KpqC}. After the whole process that has taken almost one year, they have decided to discard ROLLO-1 \cite{Rollo}, Enhanced pqsigRm \cite{pqsigRM}, SMAUG \cite{SMAUG}, GCKSign \cite{GCKSign}, TiGER \cite{Tiger}, Peregrine \cite{Peregrine}, SOLMAE \cite{Solmae}, FIBS \cite{FIBS} and IPCC \cite{IPCC}. On the contrary, the signature algorithms AIMer \cite{AIMer} as a MPC signature; HAETAE \cite{HAETAE} and NCC-Sign \cite{NCC-sign} as lattice-based signatures; and MQ-Sign \cite{MQ-sign} as Multivariate signature have been maintained as candidates for the second round. Regarding PKE/KEM algorithms they have selected  NTRU+ and a merge of SMAUG \cite{SMAUG} and TiGER \cite{Tiger} as lattice-based; and PALOMA \cite{PALOMA} and REDOG \cite{REDOG} as code-based for the second round.

\section{Post-Quantum cryptography in Critical Infrastructure} \label{Sec: V}
\noindent

\begin{figure*}[h!]
    \centering
    \captionsetup{justification=centering}
    \includegraphics[width= 0.9 \textwidth]{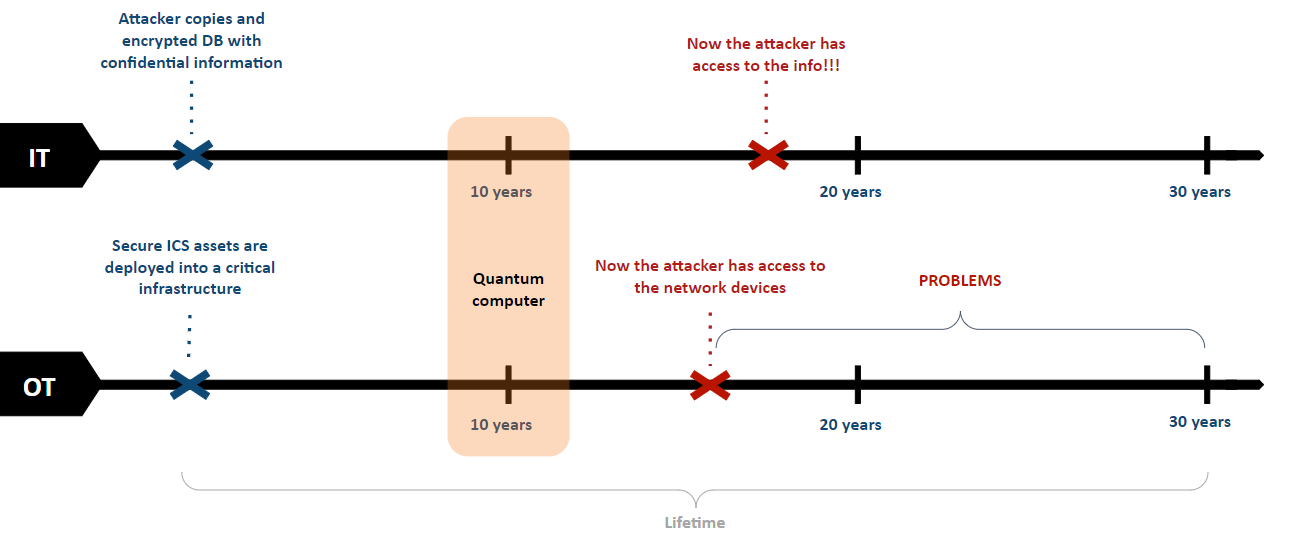}
    \caption{A comparison of the lifespan of secrets in IT environments and the lifespan of devices in OT environments reveals significant differences. In IT communications, security must withstand quantum attacks several years prior to the advent of fault-tolerant quantum computers. This preemptive security measure is essential due to the risk of intercepting and storing communications today for decryption at a later time, known as 'Harvest now, decrypt later' attack. Conversely, in OT communications, this particular issue is less pressing. However, cybersecurity in OT environments must still be quantum resilient, primarily because OT devices have long lifespans.}
    \label{fig: OTvsIT}
\end{figure*}

Providing cybersecurity resources to CIs is an indispensable task due to the fact that their constituting elements are interconnected among them and with other CI industries, implying that a weakness in any point of the network could produce a cascading effect that would result in a large economical, social and human cost, as we have explained in the Section \ref{Sec: II}. An adversary with the computational power of a quantum computer could take advantage of weakness in a concrete part of CI's communications to launch a fatal quantum attack that could potentially affect several urban centers, industrial and state infrastructures. Therefore, integrating classical cryptographic methods in CI networks will provide security nowadays, but since the life-time span of OT devices is longer than the time scale estimated for the construction of the first fault tolerant quantum computer\footnote{Note that it is projected that, at the current pace, IBM quantum processors could crack RSA by 2040 \cite{RSAcrackIBM}}, such solution can be deemed as a patch in the goal of protecting CI networks. Fig.~\ref{fig: OTvsIT}, shows the essential difference of the problems of securing IT and OT environments. The main point in trying to accelerate the integration of PQC in IT networks is the ``harvest now, decrypt later" paradigm, i.e. possible attackers store encrypted data for decrypting it once a quantum processor is available. Therefore, a fast integration is required for IT from the point of view of confidentiality, since the attackers are interested in the actual content of the encrypted data. On the other hand, in OT environments a possible attacker is not interested in reading the content of the information being communicated, but it is interested in being able to violate the system, i.e. to break data authenticity and integrity to attack the critical infrastructure. Thus, the ``harvest now, decrypt later" paradigm is not very relevant in this scenario. The big problem lays in the lifespan of the devices. As discussed before, industrial networks are assumed to last many decades and consist of legacy equipment, so integrating a secure solution against classical attacks may not be useful once quantum attacks can be realized, putting the entire system into a high degree of vulnerability. Protecting CI networks with quantum resilient solutions should aim to make them secure over the whole lifespan of the system. Consequently, PQC solutions for CI are a crucial necessity and, hence, it is necessary to target PQC from the point of view of CIs, fulfilling the required stringent communications requirements with the low computational resources/legacy devices and testing the proposed solutions in real environments.

One of the main issues with implementing PQC in communication networks, both in OT and IT, is the increased duration of the handshake between parties. This is primarily due to the larger key length of PQC cryptosystems compared to traditional protocols. The handshake between parties typically occurs by means of the Transport Layer Security (TLS) protocol. There are experiments documented in the literature that compare the handshake times between PQC cryptosystems and classical ones, as noted in \cite{Cloudfare}. The latency increase in IT communications poses a significant challenge, especially when there is a high volume of communication relying on this protocol. However, in the case of OT communications, an increase in latency is not merely an optimization concern; it could potentially result in fatal errors. Therefore, the most important characteristic a PQC algorithm should present in order to be suitable for its implementation in CI, in combination with security (data integrity and authenticity), is a low computational time (as this relates to the latency added to the system). This comes from the fact that some controlling operations in OT require latencies of the order of milliseconds, as in IEC-62443 \cite{IEC62443}, failing to satisfy such latency constraints may cause failures on the system. 

Another important concern in OT cybersecurity are the Side-Channel Attacks. A side-channel attack is a type of security breach that involves analyzing patterns of information leakage from a system to gain unauthorized access to sensitive data. Instead of directly attacking the cryptographic algorithm itself, side-channel attacks exploit unintended side channels such as power consumption, electromagnetic emissions, acoustic emanation, or timing information to infer secret information, such as encryption keys. A comprehensive analysis and definition of each type of Side-Channel Attack (SCA) is provided in \cite{SCA}.

By observing these side channels, attackers can deduce valuable information about the internal operations of a system and potentially compromise its security. While side-channel attacks are a crucial consideration in IT cybersecurity, they are less of a concern in Operational Technology (OT) environments. This is because conducting a side-channel attack in OT environments typically requires physical access to the devices or prior infection with malware. However, nowadays physical access is not always necessary to conduct a side-channel attack. Some attacks can be executed remotely or with limited physical proximity to the target system, assuming that the device has been priory infected by a malware which send enough information to the attacker for doing a SCA. However, the feasibility and effectiveness of a side-channel attack may vary depending on the specific type of attack and the level of access to the target device. Although physical access may facilitate certain types of side-channel attacks on ICS, it is not always a strict requirement for successful attacks. For example, in \cite{SCA-PLC} they perform a SCA to a infected PLC knowing its cache behaviour. This paper contributes to a better understanding of the risks posed by SCA in industrial control environments and emphasize the need for robust countermeasures to protect CI against  this kind of attacks. Another important issue related to SCA is error and fault detection \cite{Pomaranch}. An error in the PQC algorithm could leak enough information to enable a SCA attack. For instance, some research focuses on enhancing the NTT \cite{NTT,NTT1}, which can reduce the computation time of lattice-based algorithms \cite{DVFS}, but it may also leak information. A good review of NTT and its applications in PQC is given in \cite{NTT2}. Error and fault detection present a significant challenge not only in lattice-based cryptography but also in other PQC families, such as hash-based algorithms \cite{Faultdet,Faultdet1}.

Many research efforts aiming the implementation of PQC algorithms in HW are being conducted by the community, as pointed out in Section~\ref{Sec: IV g}, and examples of these implementations are given in the subsection 'PQC protocols' for each PQC family (Sections~\ref{Sec: IV a 2},~\ref{Sec: IV b 2},~\ref{Sec: IV c 2},~\ref{Sec: IV d 2},~\ref{Sec: IV e 2} and ~\ref{Sec: IV f}. However, since each of the discussed PQC protocols has not been tested under the same conditions (processor, benchmark, security level ...), performing a high-level comparison by means of the provided latencies would be inaccurate. Thus, the conclusions would not be relevant for the application of the protocols in ICS/CI. This also comes in hand with the fact that since such implementations have not been realized from the point of view of ICS/CI, i.e. trying to fulfill the stringent conditions imposed by such systems, the obtained conclusions would only be partially true. Recently, a study on implementing CRYSTALS-Kyber and CRYSTALS-Dilithium in IoT environments, meaning environments with limited computational resources, has been published \cite{PQCIoT}. These PQC algorithms were selected in the NIST standardization process \cite{NISTstand}. The study pointed out the challenges of implementing PQC algorithms in IoT environments and presented an efficient and innovative lattice-based cryptography processor to make them suitable for IoT environments. However, more studies and implementations of PQC algorithms in IoT infrastructures are needed, particularly concerning SCA and the potential delay they could introduce to communication processes. Despite of the lack of such benchmark, it is possible to somehow bound the performance of PQC families in order to select which of them could be a potentially good option to be deployed in an ICS system. We have done this by using the tables in section \ref{Sec: IV g}, where we compare the keys and ciphertext length of each PQC algorithm and their computational cost, characteristics that, in the end, are related with the latency introduced to the communications and the requirements of the hardware used in the network. However, the lack of actual fair comparative metrics for PQC in industrial and critical infrastructure networks urges for performing such comparison in the same conditions and from the point of view of the necessities of such scenarios. In this way, the selection of PQC protocols for deployment in ICS/CI will be actually possible, feature that has recently been pointed out by the CISA, NSA and NIST to be of critical importance \cite{cisansanist}.

Another challenge in implementing any new cryptosystem into a standardized communication protocol is that it may require changes across all systems \cite{PQCrev1}. All systems must adopt the same cryptosystem for a successful handshake initiation, necessitating a migration of all systems from the classical TLS protocol to a quantum-secure protocol. However, not all systems are prepared to incorporate PQC in their current state. Some are old and lack the resources, while others may not be designed to accommodate this type of cryptography, even in IT communication\cite{ETSI-Whitepaper15}. In the context of OT communications, the fact that all devices have to adopt the same PQC protocol is less problematic due to the confined nature of communications within an industry. Communication between devices typically occurs only within the same industrial setting, necessitating standardization only within that specific industry. For instance, a PLC primarily communicates with other devices within the manufacturing control layer or the Area Supervisory Control layer (refer to Fig.~\ref{fig:Purdue}), rather than with devices outside of this network. In order to communicate with elements outside this network, the protocols used are IT standards and, thus, do not concern the discussions presented here.

Another important issue for integrating PQC in ICS/CI infrastructure is that it is also necessary to think about a resiliency solution, with the capability to adapt for different cryptography algorithms, since it is not exactly known if they would be secure against quantum attacks \footnote{Recall the new proposal for solving LWE with quantum computers in polynomial time \cite{LWEQuantum}.} or specific cybersecurity requirements could be imposed by different countries and OT environments. Note also the recent controversy regarding the calculations of the security level of Kyber-512 posed by Daniel Bernstein \cite{KyberControversy} and the recently proposed method for breaking the Rainbow PQC cryptosystem in one weekend requiring a single laptop \cite{Rainbowbroken}. In principle, it seems rather difficult that such flexibility can be achieved by implementing those cryptographic protocols with a hardware solution, which is the most explored one for low latency solutions as shown in Section \ref{Sec: IV}. As an example, if the network of a electricity provider is secured by integrating a specific PQC protocol in hardware, were the case that such method is not reliable anymore, then all those chips introduced in the network should be substituted by new ones that implement the alternative. It is straightforward to see that such scenario would lead to a high cost in terms of money and man power. Hybrid cryptography could be a good solution for maintaining classical cybersecurity even if the implemented PQC algorithm is proven to be insecure, but it will not be secure against quantum computers until the industry upgrade its PQC communication protocol. Furthermore, as we exposed before, there are many PQC standardisation processes along the word. The generalized increment in the public investment in quantum  technologies, Table~\ref{Tab: Investments}, is a signal that countries consider their development a priority. Moreover, this table refers only to public expend and it does not show private and/or military investment in quantum technology. It is a signal that in the near future each country could adopt their own PQC cryptosystem protocols and will require to fulfill their cybersecurity requirements to companies that operate/sell in their territory. Note that France has recently established a normative requiring cybersecurity for the communication protocols within critical infrastructures \cite{normaFrancia} and, even if no specifics on PQC are required yet, it is a matter of time that they will. Hence, having flexible PQC solutions would allow fulfilling the specific requirements imposed by these global agents.

Moreover, it is important to note that the PQC solutions integrated in this networks should be low-power and autonomous in terms of energy. This is related with the previous discussion on the required flexibility. Note that if the power of the PQC solutions require to be changed regularly (for example by using batteries), such requirement would also result in huge costs for the industry. Therefore, it is essential that the solutions are powered by in the same way as the other elements of the network. This is why they should be low-power as not to make the power system to be saturated, i.e. they should not be a problem for the power system of the network \cite{lowpowermidori}. Also, some environments are more challenging than other (e.g. an oil extraction plant), so self powering methods may also be required as a function of the ICS to secure. For example, energy harvesting methods could be required to power some of the devices as in some IoT sensor networks \cite{iotSensEnergy}, implying low-power consumption requirements.

Regarding current PQC algorithms, those mainly focus on achieving quantum security to provide confidentiality, integrity and authenticity, being the first one the most important characteristic; due to the fact that they are thought mainly from the point of view of IT. For those to be implemented in OT, they should assure high availability and adaptability as discussed in Section \ref{Sec: II b}. Several countries are trying to standardize PQC algorithms taking only into account the perspective of IT systems, nonetheless, considering also the OT context is of pivotal importance for the security and safety of the industry.

Subsequently, since ICS/CIs have to be secure under quantum attacks as well, it is necessary to study how those algorithms work on OT networks and see which of them achieve all the requirements. Despite of the fact that there are many PQC families and many different protocols for each family, it is probable that none of them fulfill all the demanding requirements in CI environments. So, there could not be only a lack of PQC experimental work in OT environments, but also a lack of theoretical framework. If such were the case, cryptography should urge cryptographers to find other possible families or protocols that fit those conditions before a future comes in which operational technologies have no protection to quantum attacks.

\section{Conclusion and a guidelines}
\noindent
In this paper we have provided a comprehensive review of the state-of-the-art of post-quantum cryptography from the perspective of industrial and critical infrastructure networks. For doing so, we defined what are ICSs specifying its different layers and their communication protocols. Among the ICSs and different industries, we have focused on critical infrastructure networks, which provide goods and services that are indispensable for providing social and economical necessities on a day-to-day basis. In this sense, the stringent conditions that the communication network of a CI should meet have been presented. Therefore, the integration of cybersecurity in such OT systems is much more difficult than for IT services, but protecting them is of vital importance as cyberattacks on CI may lead to unbearable economical and social losses. Thus, we have provided a comprehensive overview of cryptography, the mathematical tool to maintain information secure, and discussed why quantum technologies can make state-of-the-art classical cryptography methods deprecated in a timescale of around 20 years, imposing a threat to ICS/CI components whose lifetimes are deemed to be of around 40 years. Hence, the paradigm of post-quantum cryptography has arised as the possible solution to such quantum apocalypse, and it consists in designing cryptographic methods that rely on hard problems for which quantum computing does not provide exponential speedups. Thus, we have given a review of the state-of-the-art of PQC families and protocols. We have also discussed that PQC development is being done by different global agents in an somehow independent way, implying that the near future will probably see many protocols operating at distinct industries or countries around the world, different to what happened with the widespread RSA and ECC protocols. Finally, we have discussed the current state of affairs regarding the integration of such families in ICS/CI networks. 

\begin{table*}[t!]
\resizebox{\textwidth}{!}{%
\begin{tabular}{|c||c | c |} 
 \hline
  Problems &  Importance &  Section   \\  
 \hline\hline
 Latency & OT protocols require low latency communications \cite{IEC62443,delayICS,Cloudfare}  & \ref{Sec: II A}   \\
 \hline 
 Legacy devices & PQC require a great amount of memory and computational resources \cite{pipedream,rogue7,gooseattacks}   & \ref{Sec: II A}   \\
\hline 
 Data integrity and availability & Data integrity and availability are the most important characteristics in OT cryptography   & \ref{Sec: II A}  \\
 \hline 
 Regulation & The solution has to be resilient to fulfill the regulations of the different countries \cite{ENISAICS,CISAICS,normaFrancia}  & \ref{Sec: IV h}    \\
 \hline 
 Agility & PQC primitives security are not proved and HW solutions are not flexible    & \ref{Sec: IV}   \\
 \hline
 Theory & There are not theoretical PQC specific proposals for OT protocols   & \ref{Sec: V}  \\
 \hline
 Experiments & More tests of PQC algorithms in CI environments have to be done  & \ref{Sec: V}   \\
 \hline  
\end{tabular}}
\caption{In this table we present the most valuable insights and problems for implementing PQC security in CI}
\label{Tab: Conclusion}
\end{table*} 
We have concluded that although there are many different PQC alternatives that seem to provide good security against quantum attacks for IT services in the near-future, their implementation in CI is not a trivial problem. The lack of security notions for the PQC families, the long lifetime span of OT devices, the fact that the communications within ICS have stringent requirements and that those are mainly composed of legacy elements of little computational capabilities imply that there is a current gap in terms of PQC protocols that can be seamlessly adapted to such scenarios. Moreover, the absence of a general benchmark of PQC algorithms under the same conditions (e.g. same processor for latency tests) makes it hard to make a top view comparison among them to conclude which could be well suited for implementation in CI networks. This is really important since cryptosystems that introduce too much latency reduce the availability of communication protocols, which could produce fatal consequences not only to the specific industry which suffers the shutdown, but also to all the interconnected industrial chain due to cascade effects. Therefore,we consider that the following points stand as some of the most relevant future research lines regarding this topic:
\begin{itemize}
    \item Conduct experimental studies comparing different PQC families under the same conditions. As explained through the article, the latency of PQC protocols is provided for specific scenarios and implementations (e.g. different processors), implying that a straight comparison by means of the literature data would not be accurate. This type of studies would clarify which protocols could be more suitable for integration in industrial networks as well as providing information to the community.
    \item Optimized PQC implementations for OT networks should be investigated in order to understand the capabilities of current proposals. As stated before, there are not many fair comparisons of state-of-the-art PQC protocols to understand which could be potentially implementable in industrial networks. However, the optimization of those existing protocols to be dedicated to such scenarios should also be investigated. Those could be candidates for OT communication systems would there be any successful protocols to fill all the conditions required.
    \item Propose PQC protocols from the point-of-view of the stringent conditions of OT sevices. As discussed, the problem of securing industrial networks is fundamentally different to the one of protecting IT systems. For example, in OT the aim should be authenticity/integrity rather than confidentiality. This with the very important requirement of low latency. In this sense, researcher on PQC proposals could think of their methods to target this problematic instead of protecting confidentiality, which is the usual target.
    \item Propose flexible solutions for PQC integration in ICS/CI networks. In Section \ref{Sec: V}, we discussed that PQC solutions for ICS/CI networks should be flexible in order to avoid huge economic and manpower costs if the implemented protocol results to be deprecated or if new governmental requirements are imposed since, for example, hardware solutions would imply the substitution of a humongous amounts of elements introduced in different points of a network that could be enormous in terms of space, i.e. in of the order of hundred of kilometers. Thus, proposing, for example, programmable methods that are flexible enough to change protocols if required is important for this post-quantum transition.
    \item Standardization processes for PQC implementations in industrial environments should be conducted. Through this document, many PQC standardization efforts around the world have been discussed, but those are oriented towards IT communications. As commented before, both communication scenarios are very different in the requirements for cryptography, indicating that the IT and OT implementations will diverge. Therefore, efforts regarding PQC implementations for OT networks should be pushed worldwide. Importantly, this should be done with a fast pace since, as discussed before, equipment that is not quantum secure could be vulnerable through their lifetime.
\end{itemize}

\section*{Competing Interests}
\noindent
The authors declare no competing interests.

\section*{Acknowledgements}
\noindent
We want to acknowledge Reza Dastbasteh for fruitful discussions regarding code-based cryptography. 

This work was supported by the Spanish Ministry of Economy and Competitiveness through the MADDIE project (Grant No. PID2022-137099NB-C44) and by the Gipuzkoako Foru Aldundia through the ``Post-Quantum Cryptographic Strategies for Critical Infrastructures'' project (IS172551022).

\bibliographystyle{IEEEtran}

\bibliography{Bibliography}

\begin{thebibliography}{100}
\providecommand{\url}[1]{#1}
\csname url@samestyle\endcsname
\providecommand{\newblock}{\relax}
\providecommand{\bibinfo}[2]{#2}
\providecommand{\BIBentrySTDinterwordspacing}{\spaceskip=0pt\relax}
\providecommand{\BIBentryALTinterwordstretchfactor}{4}
\providecommand{\BIBentryALTinterwordspacing}{\spaceskip=\fontdimen2\font plus
\BIBentryALTinterwordstretchfactor\fontdimen3\font minus \fontdimen4\font\relax}
\providecommand{\BIBforeignlanguage}[2]{{%
\expandafter\ifx\csname l@#1\endcsname\relax
\typeout{** WARNING: IEEEtran.bst: No hyphenation pattern has been}%
\typeout{** loaded for the language `#1'. Using the pattern for}%
\typeout{** the default language instead.}%
\else
\language=\csname l@#1\endcsname
\fi
#2}}
\providecommand{\BIBdecl}{\relax}
\BIBdecl

\bibitem{fourthrev}
\BIBentryALTinterwordspacing
T.~Philbeck and N.~Davis, ``The fourth industrial revolution: Shaping a new era,'' \emph{Journal of International Affairs}, vol.~72, no.~1, pp. 17--22, 2018. [Online]. Available: \url{https://www.jstor.org/stable/26588339}
\BIBentrySTDinterwordspacing

\bibitem{cyberStats}
\BIBentryALTinterwordspacing
Packetlabs. (2023) Cybersecurity statistics (2023). (Accessed on 03/2024). [Online]. Available: \url{https://www.packetlabs.net/posts/239-cybersecurity-statistics-2023/}
\BIBentrySTDinterwordspacing

\bibitem{CISA_Cyberavengers}
\BIBentryALTinterwordspacing
C.~. I. S.~A. (CISA), ``Irgc-affiliated cyber actors exploit plcs in multiple sectors, including u.s. water and wastewater systems facilities.'' [Online]. Available: \url{https://www.cisa.gov/news-events/cybersecurity-advisories/aa23-335a}
\BIBentrySTDinterwordspacing

\bibitem{cascade}
\BIBentryALTinterwordspacing
V.~R. Palleti, S.~Adepu, V.~K. Mishra, and A.~Mathur, ``Cascading effects of cyber-attacks on interconnected critical infrastructure,'' \emph{Cybersecurity}, vol.~4, no.~1, p.~8, Mar 2021. [Online]. Available: \url{https://doi.org/10.1186/s42400-021-00071-z}
\BIBentrySTDinterwordspacing

\bibitem{RSA}
\BIBentryALTinterwordspacing
R.~L. Rivest, A.~Shamir, and L.~Adleman, ``A method for obtaining digital signatures and public-key cryptosystems,'' \emph{Commun. ACM}, vol.~21, no.~2, p. 120–126, feb 1978. [Online]. Available: \url{https://doi.org/10.1145/359340.359342}
\BIBentrySTDinterwordspacing

\bibitem{eec1}
\BIBentryALTinterwordspacing
N.~Koblitz, ``Elliptic curve cryptosystems,'' \emph{Mathematics of Computation}, vol.~48, no. 177, pp. 203--209, 1987. [Online]. Available: \url{http://www.jstor.org/stable/2007884}
\BIBentrySTDinterwordspacing

\bibitem{eec2}
V.~S. Miller, ``Use of elliptic curves in cryptography,'' in \emph{Advances in Cryptology --- CRYPTO '85 Proceedings}, H.~C. Williams, Ed.\hskip 1em plus 0.5em minus 0.4em\relax Berlin, Heidelberg: Springer Berlin Heidelberg, 1986, pp. 417--426.

\bibitem{shorFact}
P.~Shor, ``Algorithms for quantum computation: discrete logarithms and factoring,'' in \emph{Proceedings 35th Annual Symposium on Foundations of Computer Science}, 1994, pp. 124--134.

\bibitem{sup1}
\BIBentryALTinterwordspacing
F.~Arute, K.~Arya, R.~Babbush, D.~Bacon, J.~C. Bardin, R.~Barends, R.~Biswas, S.~Boixo, F.~G. S.~L. Brandao, D.~A. Buell \emph{et~al.}, ``Quantum supremacy using a programmable superconducting processor,'' \emph{Nature}, vol. 574, no. 7779, pp. 505--510, Oct 2019. [Online]. Available: \url{https://doi.org/10.1038/s41586-019-1666-5}
\BIBentrySTDinterwordspacing

\bibitem{sup2}
\BIBentryALTinterwordspacing
H.-S. Zhong, H.~Wang, Y.-H. Deng, M.-C. Chen, L.-C. Peng, Y.-H. Luo, J.~Qin, D.~Wu, X.~Ding, Y.~Hu, P.~Hu \emph{et~al.}, ``Quantum computational advantage using photons,'' \emph{Science}, vol. 370, no. 6523, pp. 1460--1463, 2020. [Online]. Available: \url{https://www.science.org/doi/abs/10.1126/science.abe8770}
\BIBentrySTDinterwordspacing

\bibitem{sup3}
\BIBentryALTinterwordspacing
Y.~Wu, W.-S. Bao, S.~Cao, F.~Chen, M.-C. Chen, X.~Chen, T.-H. Chung, H.~Deng, Y.~Du, D.~Fan, M.~Gong \emph{et~al.}, ``Strong quantum computational advantage using a superconducting quantum processor,'' \emph{Phys. Rev. Lett.}, vol. 127, p. 180501, Oct 2021. [Online]. Available: \url{https://link.aps.org/doi/10.1103/PhysRevLett.127.180501}
\BIBentrySTDinterwordspacing

\bibitem{sup4}
\BIBentryALTinterwordspacing
H.-Y. Huang, M.~Broughton, J.~Cotler, S.~Chen, J.~Li, M.~Mohseni, H.~Neven, R.~Babbush, R.~Kueng, J.~Preskill, and J.~R. McClean, ``Quantum advantage in learning from experiments,'' \emph{Science}, vol. 376, no. 6598, pp. 1182--1186, 2022. [Online]. Available: \url{https://www.science.org/doi/abs/10.1126/science.abn7293}
\BIBentrySTDinterwordspacing

\bibitem{sup5}
\BIBentryALTinterwordspacing
L.~S. Madsen, F.~Laudenbach, M.~F. Askarani, F.~Rortais, T.~Vincent, J.~F.~F. Bulmer, F.~M. Miatto, L.~Neuhaus, L.~G. Helt, M.~J. Collins, A.~E. Lita \emph{et~al.}, ``Quantum computational advantage with a programmable photonic processor,'' \emph{Nature}, vol. 606, no. 7912, pp. 75--81, Jun 2022. [Online]. Available: \url{https://doi.org/10.1038/s41586-022-04725-x}
\BIBentrySTDinterwordspacing

\bibitem{wallraffSurf}
\BIBentryALTinterwordspacing
S.~Krinner, N.~Lacroix, A.~Remm, A.~Di~Paolo, E.~Genois, C.~Leroux, C.~Hellings, S.~Lazar, F.~Swiadek, J.~Herrmann, G.~J. Norris \emph{et~al.}, ``Realizing repeated quantum error correction in a distance-three surface code,'' \emph{Nature}, vol. 605, no. 7911, pp. 669--674, May 2022. [Online]. Available: \url{https://doi.org/10.1038/s41586-022-04566-8}
\BIBentrySTDinterwordspacing

\bibitem{googleSurf}
\BIBentryALTinterwordspacing
R.~Acharya, I.~Aleiner, R.~Allen, T.~I. Andersen, M.~Ansmann, F.~Arute, K.~Arya, A.~Asfaw, J.~Atalaya, R.~Babbush, D.~Bacon \emph{et~al.}, ``Suppressing quantum errors by scaling a surface code logical qubit,'' \emph{Nature}, vol. 614, no. 7949, pp. 676--681, Feb 2023. [Online]. Available: \url{https://doi.org/10.1038/s41586-022-05434-1}
\BIBentrySTDinterwordspacing

\bibitem{Mosca}
\BIBentryALTinterwordspacing
M.~Mosca and M.~Piani. Global risk institute: Quantum threat timeline report 2022. (Accessed on 03/2024). [Online]. Available: \url{https://globalriskinstitute.org/publication/2022-quantum-threat-timeline-report/}
\BIBentrySTDinterwordspacing

\bibitem{QKDrev}
\BIBentryALTinterwordspacing
N.~Gisin, G.~Ribordy, W.~Tittel, and H.~Zbinden, ``Quantum cryptography,'' \emph{Rev. Mod. Phys.}, vol.~74, pp. 145--195, Mar 2002. [Online]. Available: \url{https://link.aps.org/doi/10.1103/RevModPhys.74.145}
\BIBentrySTDinterwordspacing

\bibitem{bb84}
\BIBentryALTinterwordspacing
C.~H. Bennett and G.~Brassard, ``Quantum cryptography: Public key distribution and coin tossing,'' \emph{Theoretical Computer Science}, vol. 560, pp. 7--11, 2014, theoretical Aspects of Quantum Cryptography – celebrating 30 years of BB84. [Online]. Available: \url{https://www.sciencedirect.com/science/article/pii/S0304397514004241}
\BIBentrySTDinterwordspacing

\bibitem{e91}
\BIBentryALTinterwordspacing
A.~K. Ekert, ``Quantum cryptography based on bell's theorem,'' \emph{Phys. Rev. Lett.}, vol.~67, pp. 661--663, Aug 1991. [Online]. Available: \url{https://link.aps.org/doi/10.1103/PhysRevLett.67.661}
\BIBentrySTDinterwordspacing

\bibitem{repeaters}
K.~{Azuma}, S.~E. {Economou}, D.~{Elkouss}, P.~{Hilaire}, L.~{Jiang}, H.-K. {Lo}, and I.~{Tzitrin}, ``{Quantum repeaters: From quantum networks to the quantum internet},'' \emph{arXiv e-prints}, p. arXiv:2212.10820, Dec. 2022.

\bibitem{Bernstein2017-zk}
D.~J. Bernstein and T.~Lange, ``\BIBforeignlanguage{en}{Post-quantum cryptography},'' \emph{\BIBforeignlanguage{en}{Nature}}, vol. 549, no. 7671, pp. 188--194, Sep. 2017.

\bibitem{Qcomplexity}
\BIBentryALTinterwordspacing
T.~Vidick and J.~Watrous, ``Quantum proofs,'' \emph{Foundations and Trends® in Theoretical Computer Science}, vol.~11, no. 1-2, pp. 1--215, 2016. [Online]. Available: \url{http://dx.doi.org/10.1561/0400000068}
\BIBentrySTDinterwordspacing

\bibitem{NISTstand}
\BIBentryALTinterwordspacing
NIST, ``Post-quantum cryptography standarization,'' 2017. [Online]. Available: \url{https://csrc.nist.gov/Projects/post-quantum-cryptography/post-quantum-cryptography-standardization}
\BIBentrySTDinterwordspacing

\bibitem{EUagenda}
\BIBentryALTinterwordspacing
A.~Rodriguez, ``A quantum cybersecurity agenda for europe,'' 2023. [Online]. Available: \url{https://www.epc.eu/en/publications/A-quantum-cybersecurity-agenda-for-Europe~526b9c}
\BIBentrySTDinterwordspacing

\bibitem{googlePQC}
\BIBentryALTinterwordspacing
D.~O'Brien, ``Protecting chrome traffic with hybrid kyber kem,'' 2023. [Online]. Available: \url{https://blog.chromium.org/2023/08/protecting-chrome-traffic-with-hybrid.html}
\BIBentrySTDinterwordspacing

\bibitem{Kyber}
J.~Bos, L.~Ducas, E.~Kiltz, T.~Lepoint, V.~Lyubashevsky, J.~M. Schanck, P.~Schwabe, G.~Seiler, and D.~Stehle, ``Crystals - kyber: A cca-secure module-lattice-based kem,'' in \emph{2018 IEEE European Symposium on Security and Privacy (EuroS \& P)}, 2018, pp. 353--367.

\bibitem{delayICS}
E.~Korkmaz, M.~Davis, A.~Dolgikh, and V.~Skormin, ``Detection and mitigation of time delay injection attacks on industrial control systems with plcs,'' in \emph{Computer Network Security}, J.~Rak, J.~Bay, I.~Kotenko, L.~Popyack, V.~Skormin, and K.~Szczypiorski, Eds.\hskip 1em plus 0.5em minus 0.4em\relax Cham: Springer International Publishing, 2017, pp. 62--74.

\bibitem{7245330}
Z.~Drias, A.~Serhrouchni, and O.~Vogel, ``Analysis of cyber security for industrial control systems,'' in \emph{2015 International Conference on Cyber Security of Smart Cities, Industrial Control System and Communications (SSIC)}, 2015, pp. 1--8.

\bibitem{GooseAttack}
J.~Hoyos, M.~Dehus, and T.~X. Brown, ``Exploiting the goose protocol: A practical attack on cyber-infrastructure,'' in \emph{2012 IEEE Globecom Workshops}, 2012, pp. 1508--1513.

\bibitem{PoisonedGOOSE}
N.~Kush, M.~Branagan, E.~Foo, and E.~Ahmed, ``Poisoned goose : exploiting the goose protocol,'' vol. 149, 01 2014.

\bibitem{HackICS}
\BIBentryALTinterwordspacing
E.~D. Knapp and J.~T. Langill, ``Chapter 7 - hacking industrial control systems,'' in \emph{Industrial Network Security (Second Edition)}, E.~D. Knapp and J.~T. Langill, Eds.\hskip 1em plus 0.5em minus 0.4em\relax Boston: Syngress, 2015, pp. 171--207. [Online]. Available: \url{https://www.sciencedirect.com/science/article/pii/B9780124201149000071}
\BIBentrySTDinterwordspacing

\bibitem{cisansanist}
\BIBentryALTinterwordspacing
NISA, NSA, and NIST, ``Quantum-readiness: Migration to post-quantum cryptography,'' 2023. [Online]. Available: \url{https://www.cisa.gov/sites/default/files/2023-08/Quantum%20Readiness_Final_CLEAR_508c%20%283%29.pdf}
\BIBentrySTDinterwordspacing

\bibitem{PQCrev1}
E.~Zeydan, Y.~Turk, B.~Aksoy, and S.~B. Ozturk, ``Recent advances in post-quantum cryptography for networks: A survey,'' in \emph{2022 Seventh International Conference On Mobile And Secure Services (MobiSecServ)}, 2022, pp. 1--8.

\bibitem{PQCrev2}
\BIBentryALTinterwordspacing
K.~Kan and M.~Une, ``{Recent Trends on Research and Development of Quantum Computers and Standardization of Post-Quantum Cryptography},'' \emph{Monetary and Economic Studies}, vol.~39, pp. 77--108, November 2021. [Online]. Available: \url{https://ideas.repec.org/a/ime/imemes/v39y2021p77-108.html}
\BIBentrySTDinterwordspacing

\bibitem{PQCrev5}
M.~Alvarado, L.~Gayler, A.~Seals, T.~Wang, and T.~Hou, ``A survey on post-quantum cryptography: State-of-the-art and challenges,'' 2023.

\bibitem{PQCrev6}
\BIBentryALTinterwordspacing
D.-T. Dam, T.-H. Tran, V.-P. Hoang, C.-K. Pham, and T.-T. Hoang, ``A survey of post-quantum cryptography: Start of a new race,'' \emph{Cryptography}, vol.~7, no.~3, 2023. [Online]. Available: \url{https://www.mdpi.com/2410-387X/7/3/40}
\BIBentrySTDinterwordspacing

\bibitem{IndustrialCyber}
\BIBentryALTinterwordspacing
(2022) The state of industrial security in 2022. (Accessed on 03/2024). [Online]. Available: \url{https://www.barracuda.com/reports/iiot-2022-report}
\BIBentrySTDinterwordspacing

\bibitem{OTcyber1}
\BIBentryALTinterwordspacing
M.~Lezzi, M.~Lazoi, and A.~Corallo, ``Cybersecurity for industry 4.0 in the current literature: A reference framework,'' \emph{Computers in Industry}, vol. 103, pp. 97--110, 2018. [Online]. Available: \url{https://www.sciencedirect.com/science/article/pii/S0166361518303658}
\BIBentrySTDinterwordspacing

\bibitem{OTcyber3}
\BIBentryALTinterwordspacing
D.~Bhamare, M.~Zolanvari, A.~Erbad, R.~Jain, K.~Khan, and N.~Meskin, ``Cybersecurity for industrial control systems: A survey,'' \emph{Computers \& Security}, vol.~89, p. 101677, 2020. [Online]. Available: \url{https://www.sciencedirect.com/science/article/pii/S0167404819302172}
\BIBentrySTDinterwordspacing

\bibitem{OTcyber4}
\BIBentryALTinterwordspacing
H.~M.~H. Uchenna P. Daniel~Ani and A.~Tiwari, ``Review of cybersecurity issues in industrial critical infrastructure: manufacturing in perspective,'' \emph{Journal of Cyber Security Technology}, vol.~1, no.~1, pp. 32--74, 2017. [Online]. Available: \url{https://doi.org/10.1080/23742917.2016.1252211}
\BIBentrySTDinterwordspacing

\bibitem{OTcyber5}
M.~Krotofil and D.~Gollmann, ``Industrial control systems security: What is happening?'' in \emph{2013 11th IEEE International Conference on Industrial Informatics (INDIN)}, 2013, pp. 664--669.

\bibitem{OTcyber6}
K.-K.~R. Choo, S.~Gritzalis, and J.~H. Park, ``Cryptographic solutions for industrial internet-of-things: Research challenges and opportunities,'' \emph{IEEE Transactions on Industrial Informatics}, vol.~14, no.~8, pp. 3567--3569, 2018.

\bibitem{NISTPQC}
A.~K. Pandey, A.~Banati, B.~Rajendran, S.~D. Sudarsan, and K.~K.~S. Pandian, ``Cryptographic challenges and security in post quantum cryptography migration: A prospective approach,'' in \emph{2023 IEEE International Conference on Public Key Infrastructure and its Applications (PKIA)}, 2023, pp. 1--8.

\bibitem{PQCrev3}
\BIBentryALTinterwordspacing
S.~Paul, ``\BIBforeignlanguage{en}{On the transition to post-quantum cryptography in the industrial internet of things},'' Ph.D. dissertation, Technische Universit{\"a}t Darmstadt, Darmstadt, 2022. [Online]. Available: \url{http://tuprints.ulb.tu-darmstadt.de/21368/}
\BIBentrySTDinterwordspacing

\bibitem{ANSSI}
\BIBentryALTinterwordspacing
French national agency for the security of information systems. (Accessed on 03/2024). [Online]. Available: \url{https://www.ssi.gouv.fr/en/}
\BIBentrySTDinterwordspacing

\bibitem{BSI}
\BIBentryALTinterwordspacing
Federal office for information security. (Accessed on 03/2024). [Online]. Available: \url{https://www.bsi.bund.de/EN/Home/home_node.html}
\BIBentrySTDinterwordspacing

\bibitem{IEC62443}
\BIBentryALTinterwordspacing
Isa/iec 62443 series of standards. (Accessed on 03/2024). [Online]. Available: \url{https://www.isa.org/standards-and-publications/isa-standards/isa-iec-62443-series-of-standards}
\BIBentrySTDinterwordspacing

\bibitem{ENISAICS}
\BIBentryALTinterwordspacing
R.~Mattioli and K.~Moulinos, ``Analysis of ics-scada cyber security maturity levels in critical sectors,'' 2015. [Online]. Available: \url{https://www.enisa.europa.eu/publications/maturity-levels}
\BIBentrySTDinterwordspacing

\bibitem{CISAICS}
\BIBentryALTinterwordspacing
Cisa: Industrial control systems. (Accessed on 03/2024). [Online]. Available: \url{https://www.cisa.gov/topics/industrial-control-systems}
\BIBentrySTDinterwordspacing

\bibitem{normaFrancia}
\BIBentryALTinterwordspacing
{Agence nationale de la s\'ecurit\'e des syst\`emes d'information}. La cybers\'ecurit\'e des syst\`emes industriels. (Accessed on 03/2024). [Online]. Available: \url{https://cyber.gouv.fr/publications/la-cybersecurite-des-systemes-industriels}
\BIBentrySTDinterwordspacing

\bibitem{pipedream}
R.~Ramirez, C.-K. Chang, and S.-H. Liang, ``Plc cyber-security challenges in industrial networks,'' in \emph{2022 18th IEEE/ASME International Conference on Mechatronic and Embedded Systems and Applications (MESA)}, 2022, pp. 1--6.

\bibitem{gooseattacks}
M.~T.~A. Rashid, S.~Yussof, Y.~Yusoff, and R.~Ismail, ``A review of security attacks on iec61850 substation automation system network,'' in \emph{Proceedings of the 6th International Conference on Information Technology and Multimedia}, 2014, pp. 5--10.

\bibitem{rogue7}
E.~Biham, S.~Bitan, A.~Carmel, A.~Dankner, U.~Malin, and A.~Wool, ``Rogue7: Rogue engineering-station attacks on s7 simatic plcs,'' \emph{Black Hat USA}, vol. 2019, 2019.

\bibitem{AES}
J.~Daor, J.~Daemen, and V.~Rijmen, ``Aes proposal: rijndael,'' 10 1999.

\bibitem{Blowfish}
B.~Schneier, ``Description of a new variable-length key, 64-bit block cipher (blowfish),'' in \emph{Fast Software Encryption}, R.~Anderson, Ed.\hskip 1em plus 0.5em minus 0.4em\relax Berlin, Heidelberg: Springer Berlin Heidelberg, 1994, pp. 191--204.

\bibitem{Twofish}
B.~Kelsey, D.~Whiting, D.~Wagner, C.~Hall, and N.~Ferguson, ``Twofish: A 128bit block cipher,'' 01 1998.

\bibitem{ECDH}
W.~Diffie and M.~Hellman, ``New directions in cryptography,'' \emph{IEEE Transactions on Information Theory}, vol.~22, no.~6, pp. 644--654, 1976.

\bibitem{PNP}
\BIBentryALTinterwordspacing
Millenium prize problems. (Accessed on 03/2024). [Online]. Available: \url{https://www.claymath.org/millennium-problems/}
\BIBentrySTDinterwordspacing

\bibitem{ECDSA}
\BIBentryALTinterwordspacing
A.~Menezes, U.~of~Waterloo. Department~of Combinatorics, Optimization, J.~D, and U.~of~Waterloo. Faculty~of Mathematics, \emph{The Elliptic Curve Digital Signature Algorithm (ECDSA)}, ser. Technical report (University of Waterloo. Faculty of Mathematics).\hskip 1em plus 0.5em minus 0.4em\relax Faculty of Mathematics, University of Waterloo, 1999. [Online]. Available: \url{https://books.google.es/books?id=gAPeMwEACAAJ}
\BIBentrySTDinterwordspacing

\bibitem{grover1996fast}
L.~K. Grover, ``A fast quantum mechanical algorithm for database search,'' 1996.

\bibitem{vqaa1}
\BIBentryALTinterwordspacing
Z.~Wang, S.~Wei, G.-L. Long, and L.~Hanzo, ``Variational quantum attacks threaten advanced encryption standard based symmetric cryptography,'' \emph{Science China Information Sciences}, vol.~65, no.~10, p. 200503, Jul 2022. [Online]. Available: \url{https://doi.org/10.1007/s11432-022-3511-5}
\BIBentrySTDinterwordspacing

\bibitem{vqaa2}
B.~{Aizpurua}, P.~{Bermejo}, J.~{Etxezarreta Martinez}, and R.~{Orus}, ``{Hacking Cryptographic Protocols with Advanced Variational Quantum Attacks},'' \emph{arXiv e-prints}, p. arXiv:2311.02986, Nov. 2023.

\bibitem{NumberFieldSieve}
J.~P. Buhler, H.~W. Lenstra, and C.~Pomerance, ``Factoring integers with the number field sieve,'' in \emph{The development of the number field sieve}, A.~K. Lenstra and H.~W. Lenstra, Eds.\hskip 1em plus 0.5em minus 0.4em\relax Berlin, Heidelberg: Springer Berlin Heidelberg, 1993, pp. 50--94.

\bibitem{complexty}
K.~K. Soni and A.~Rasool, ``Cryptographic attack possibilities over rsa algorithm through classical and quantum computation,'' in \emph{2018 International Conference on Smart Systems and Inventive Technology (ICSSIT)}, 2018, pp. 11--15.

\bibitem{RSA250}
\BIBentryALTinterwordspacing
Factorization of rsa-250. (Accessed on 03/2024). [Online]. Available: \url{https://listserv.nodak.edu/cgi-bin/wa.exe?A2=NMBRTHRY;dc42ccd1.2002}
\BIBentrySTDinterwordspacing

\bibitem{decoders}
\BIBentryALTinterwordspacing
A.~{deMarti iOlius}, P.~{Fuentes}, R.~{Or{\'u}s}, P.~M. {Crespo}, and J.~{Etxezarreta Martinez}, ``{Decoding algorithms for surface codes},'' \emph{arXiv e-prints}, p. arXiv:2307.14989, Jul. 2023. [Online]. Available: \url{https://doi.org/10.48550/arXiv.2307.14989}
\BIBentrySTDinterwordspacing

\bibitem{rsacrackGid}
\BIBentryALTinterwordspacing
C.~Gidney and M.~Eker{\aa{}}, ``How to factor 2048 bit {RSA} integers in 8 hours using 20 million noisy qubits,'' \emph{{Quantum}}, vol.~5, p. 433, Apr. 2021. [Online]. Available: \url{https://doi.org/10.22331/q-2021-04-15-433}
\BIBentrySTDinterwordspacing

\bibitem{nisq}
\BIBentryALTinterwordspacing
J.~Preskill, ``Quantum {C}omputing in the {NISQ} era and beyond,'' \emph{{Quantum}}, vol.~2, p.~79, Aug. 2018. [Online]. Available: \url{https://doi.org/10.22331/q-2018-08-06-79}
\BIBentrySTDinterwordspacing

\bibitem{No-clonning}
W.~K. {Wootters} and W.~H. {Zurek}, ``{A single quantum cannot be cloned},'' vol. 299, no. 5886, pp. 802--803, Oct. 1982.

\bibitem{LWC}
W.~J. Buchanan, S.~Li, and R.~Asif, ``Lightweight cryptography methods,'' \emph{Journal of Cyber Security Technology}, vol.~1, no. 3-4, pp. 187--201, 2017.

\bibitem{LWCHW1}
J.~Kaur, A.~Sarker, M.~M. Kermani, and R.~Azarderakhsh, ``Hardware constructions for error detection in lightweight welch-gong (wg)-oriented streamcipher wage benchmarked on fpga,'' \emph{IEEE Transactions on Emerging Topics in Computing}, vol.~10, no.~2, pp. 1208--1215, 2022.

\bibitem{LWCHW2}
J.~Kaur, M.~M. Kermani, and R.~Azarderakhsh, ``Hardware constructions for lightweight cryptographic block cipher qarma with error detection mechanisms,'' \emph{IEEE Transactions on Emerging Topics in Computing}, vol.~10, no.~1, pp. 514--519, 2022.

\bibitem{NISTLWC}
M.~S. Turan, M.~S. Turan, K.~McKay, D.~Chang, L.~E. Bassham, J.~Kang, N.~D. Waller, J.~M. Kelsey, and D.~Hong, \emph{Status report on the final round of the NIST lightweight cryptography standardization process}.\hskip 1em plus 0.5em minus 0.4em\relax US Department of Commerce, National Institute of Standards and Technology, 2023.

\bibitem{ASCON}
\BIBentryALTinterwordspacing
C.~Dobraunig, M.~Eichlseder, F.~Mendel, and M.~Schl{\"{a}}ffer, ``Ascon v1.2: Lightweight authenticated encryption and hashing,'' \emph{J. Cryptol.}, vol.~34, no.~3, p.~33, 2021. [Online]. Available: \url{https://doi.org/10.1007/s00145-021-09398-9}
\BIBentrySTDinterwordspacing

\bibitem{Hybrid}
N.~Bindel, U.~Herath, M.~McKague, and D.~Stebila, ``Transitioning to a quantum-resistant public key infrastructure,'' in \emph{Proc. 8th International Conference on Post-Quantum Cryptography (PQCrypto) 2017}, ser. LNCS, T.~Lange and T.~Takagi, Eds., vol. 10346.\hskip 1em plus 0.5em minus 0.4em\relax Springer, June 2017, pp. 384--405.

\bibitem{HybridImp1}
\BIBentryALTinterwordspacing
H.~Seo and R.~Azarderakhsh, ``Curve448 on 32-bit arm cortex-m4,'' Cryptology ePrint Archive, Paper 2021/1355, 2021, \url{https://eprint.iacr.org/2021/1355}. [Online]. Available: \url{https://eprint.iacr.org/2021/1355}
\BIBentrySTDinterwordspacing

\bibitem{HybridImp2}
M.~Bisheh-Niasar, R.~Azarderakhsh, and M.~Mozaffari-Kermani, ``Cryptographic accelerators for digital signature based on ed25519,'' \emph{IEEE Transactions on Very Large Scale Integration (VLSI) Systems}, vol.~29, no.~7, pp. 1297--1305, 2021.

\bibitem{Bavdekar}
R.~Bavdekar, E.~Jayant~Chopde, A.~Agrawal, A.~Bhatia, and K.~Tiwari, ``Post quantum cryptography: A review of techniques, challenges and standardizations,'' in \emph{2023 International Conference on Information Networking (ICOIN)}, 2023, pp. 146--151.

\bibitem{hash}
\BIBentryALTinterwordspacing
R.~C. MERKLE, ``\BIBforeignlanguage{English}{Secrecy, authentication, and public key systems},'' Ph.D. dissertation, 1979, copyright - Database copyright ProQuest LLC; ProQuest does not claim copyright in the individual underlying works; Última actualización - 2023-02-23. [Online]. Available: \url{https://www.proquest.com/dissertations-theses/secrecy-authentication-public-key-systems/docview/302984000/se-2}
\BIBentrySTDinterwordspacing

\bibitem{lamport}
\BIBentryALTinterwordspacing
L.~Lamport, ``Constructing digital signatures from a one way function,'' Tech. Rep. CSL-98, October 1979, this paper was published by IEEE in the Proceedings of HICSS-43 in January, 2010. [Online]. Available: \url{https://www.microsoft.com/en-us/research/publication/constructing-digital-signatures-one-way-function/}
\BIBentrySTDinterwordspacing

\bibitem{CollisionQuantum}
\BIBentryALTinterwordspacing
G.~Brassard, P.~H{\O}yer, and A.~Tapp, ``Quantum cryptanalysis of hash and claw-free functions,'' in \emph{{LATIN}{\textquotesingle}98: Theoretical Informatics}.\hskip 1em plus 0.5em minus 0.4em\relax Springer Berlin Heidelberg, 1998, pp. 163--169. [Online]. Available: \url{https://doi.org/10.1007%2Fbfb0054319}
\BIBentrySTDinterwordspacing

\bibitem{Sphincs+}
\BIBentryALTinterwordspacing
D.~J. Bernstein, A.~H\"{u}lsing, S.~K\"{o}lbl, R.~Niederhagen, J.~Rijneveld, and P.~Schwabe, ``The sphincs+ signature framework,'' in \emph{Proceedings of the 2019 ACM SIGSAC Conference on Computer and Communications Security}, ser. CCS '19.\hskip 1em plus 0.5em minus 0.4em\relax New York, NY, USA: Association for Computing Machinery, 2019, p. 2129–2146. [Online]. Available: \url{https://doi.org/10.1145/3319535.3363229}
\BIBentrySTDinterwordspacing

\bibitem{SPHINCS+1}
D.~Amiet, L.~Leuenberger, A.~Curiger, and P.~Zbinden, ``Fpga-based sphincs+ implementations: Mind the glitch,'' in \emph{2020 23rd Euromicro Conference on Digital System Design (DSD)}, 2020, pp. 229--237.

\bibitem{NISTSigperf}
R.~Gonzalez, A.~H{\"u}lsing, M.~J. Kannwischer, J.~Kr{\"a}mer, T.~Lange, M.~St{\"o}ttinger, E.~Waitz, T.~Wiggers, and B.-Y. Yang, ``Verifying post-quantum signatures in 8 kb of ram,'' in \emph{Post-Quantum Cryptography}, J.~H. Cheon and J.-P. Tillich, Eds.\hskip 1em plus 0.5em minus 0.4em\relax Cham: Springer International Publishing, 2021, pp. 215--233.

\bibitem{sum-problem}
R.~Merkle and M.~Hellman, ``Hiding information and signatures in trapdoor knapsacks,'' \emph{IEEE Transactions on Information Theory}, vol.~24, no.~5, pp. 525--530, 1978.

\bibitem{NTRUOrig}
J.~Hoffstein, J.~Pipher, and J.~H. Silverman, ``Ntru: A ring-based public key cryptosystem,'' in \emph{Algorithmic Number Theory}, J.~P. Buhler, Ed.\hskip 1em plus 0.5em minus 0.4em\relax Berlin, Heidelberg: Springer Berlin Heidelberg, 1998, pp. 267--288.

\bibitem{Regev-scheme}
\BIBentryALTinterwordspacing
O.~Regev, ``On lattices, learning with errors, random linear codes, and cryptography,'' \emph{J. ACM}, vol.~56, no.~6, sep 2009. [Online]. Available: \url{https://doi.org/10.1145/1568318.1568324}
\BIBentrySTDinterwordspacing

\bibitem{Lll1982}
A.~K. Lenstra, H.~W. Lenstra, and L.~Lov{\'a}sz, ``Factoring polynomials with rational coefficients,'' \emph{Mathematische Annalen}, vol. 261, no.~4, pp. 515--534, Dec. 1982.

\bibitem{Ajtai}
\BIBentryALTinterwordspacing
M.~Ajtai, ``Generating hard instances of lattice problems (extended abstract),'' in \emph{Proceedings of the Twenty-Eighth Annual ACM Symposium on Theory of Computing}, ser. STOC '96.\hskip 1em plus 0.5em minus 0.4em\relax New York, NY, USA: Association for Computing Machinery, 1996, p. 99–108. [Online]. Available: \url{https://doi.org/10.1145/237814.237838}
\BIBentrySTDinterwordspacing

\bibitem{RLWE}
V.~Lyubashevsky, C.~Peikert, and O.~Regev, ``On ideal lattices and learning with errors over rings,'' in \emph{Advances in Cryptology -- EUROCRYPT 2010}, H.~Gilbert, Ed.\hskip 1em plus 0.5em minus 0.4em\relax Berlin, Heidelberg: Springer Berlin Heidelberg, 2010, pp. 1--23.

\bibitem{NTRU}
\BIBentryALTinterwordspacing
W.~Whyte and J.~Hoffstein, \emph{NTRU}.\hskip 1em plus 0.5em minus 0.4em\relax Boston, MA: Springer US, 2011, pp. 858--861. [Online]. Available: \url{https://doi.org/10.1007/978-1-4419-5906-5_464}
\BIBentrySTDinterwordspacing

\bibitem{NTRU-HRSS}
\BIBentryALTinterwordspacing
A.~H{\"u}lsing, J.~Rijneveld, J.~Schanck, and P.~Schwabe, ``\BIBforeignlanguage{English}{Ntru-hrss-kem - submission to the nist post-quantum cryptography project},'' 2017. [Online]. Available: \url{https://csrc.nist.gov/CSRC/media/Projects/Post-Quantum-Cryptography/documents/round-1/submissions/NTRU_HRSS_KEM.zip}
\BIBentrySTDinterwordspacing

\bibitem{NTRU-prime}
\BIBentryALTinterwordspacing
D.~J. Bernstein, C.~Chuengsatiansup, T.~Lange, and C.~van Vredendaal, ``Ntru prime: reducing attack surface at low cost,'' Cryptology ePrint Archive, Paper 2016/461, 2016. [Online]. Available: \url{https://eprint.iacr.org/2016/461}
\BIBentrySTDinterwordspacing

\bibitem{NTRU+}
\BIBentryALTinterwordspacing
J.~Kim and J.~H. Park, ``Ntru+: Compact construction of ntru using simple encoding method,'' Cryptology ePrint Archive, Paper 2022/1664, 2022. [Online]. Available: \url{https://eprint.iacr.org/2022/1664}
\BIBentrySTDinterwordspacing

\bibitem{Falcon}
P.-A. Fouque, J.~Hoffstein, P.~Kirchner, V.~Lyubashevsky, T.~Pornin, T.~Prest, T.~Ricosset, G.~Seiler, W.~Whyte, and Z.~Zhang, ``Falcon: Fast-fourier lattice-based compact signatures over ntru.''

\bibitem{Fatseal}
\BIBentryALTinterwordspacing
T.~Xie, H.~Li, Y.~Zhu, Y.~Pan, Z.~Liu, and Z.~Yang, ``Fatseal: An efficient lattice-based signature algorithm,'' \emph{电子与信息学报}, vol.~42, no.~2, pp. 333--340, 2020. [Online]. Available: \url{https://jeit.ac.cn/cn/article/doi/10.11999/JEIT190678}
\BIBentrySTDinterwordspacing

\bibitem{Peregrine}
\BIBentryALTinterwordspacing
E.-Y. Seo, Y.-S. Kim, J.-W. Lee, and J.-S. No, ``Peregrine: Toward fastest falcon based on gpv framework,'' Cryptology ePrint Archive, Paper 2022/1495, 2022. [Online]. Available: \url{https://eprint.iacr.org/2022/1495}
\BIBentrySTDinterwordspacing

\bibitem{Solmae}
\BIBentryALTinterwordspacing
K.~Kim, ``How solmae was designed.'' [Online]. Available: \url{https://ircs.re.kr/wp-content/uploads/2023/06/40CISC_S23_2col_final.pdf}
\BIBentrySTDinterwordspacing

\bibitem{Saber}
\BIBentryALTinterwordspacing
J.-P. D’Anvers, A.~Karmakar, S.~S. Roy, and F.~Vercauteren, ``Saber: Module-lwr based key exchange, cpa-secure encryption and cca-secure kem,'' Cryptology ePrint Archive, Paper 2018/230, 2018. [Online]. Available: \url{https://eprint.iacr.org/2018/230}
\BIBentrySTDinterwordspacing

\bibitem{FrodoKEMLW}
\BIBentryALTinterwordspacing
E.~Alkım, J.~W. Bos, L.~Ducas, P.~Longa, I.~Mironov, M.~Naehrig, V.~Nikolaenko, C.~Peikert, A.~Raghunathan, and D.~Stebila, ``Frodokem learning with errors key encapsulation algorithm specifications and supporting documentation,'' 2019. [Online]. Available: \url{https://frodokem.org/}
\BIBentrySTDinterwordspacing

\bibitem{LAC}
X.~Lu, Y.~Liu, Z.~Zhang, D.~Jia, H.~Xue, J.~He, and B.~Li, ``Lac: Practical ring-lwe based public-key encryption with byte-level modulus,'' \emph{IACR Cryptol. ePrint Arch.}, vol. 2018, p. 1009, 2018.

\bibitem{Aigis-Enc}
\BIBentryALTinterwordspacing
J.~Zhang, Y.~Yu, S.~Fan, Z.~Zhang, and K.~Yang, ``Tweaking the asymmetry of asymmetric-key cryptography on lattices: Kems and signatures of smaller sizes,'' Cryptology ePrint Archive, Paper 2019/510, 2019. [Online]. Available: \url{https://eprint.iacr.org/2019/510}
\BIBentrySTDinterwordspacing

\bibitem{AKCN-MLWE}
Z.~Jin and Y.~Zhao, ``Optimal key consensus in presence of noise,'' 2017.

\bibitem{TALE}
Y.~Zhu, Z.~Liu, and Y.~Pan, ``When ntt meets karatsuba: Preprocess-then-ntt technique revisited,'' in \emph{Information and Communications Security}, D.~Gao, Q.~Li, X.~Guan, and X.~Liao, Eds.\hskip 1em plus 0.5em minus 0.4em\relax Cham: Springer International Publishing, 2021, pp. 249--264.

\bibitem{AKCN-E8}
\BIBentryALTinterwordspacing
Z.~JIn and Y.~Zhao, ``Akcn-e8: Compact and flexible kem from ideal lattice,'' Cryptology ePrint Archive, Paper 2020/056, 2020. [Online]. Available: \url{https://eprint.iacr.org/2020/056}
\BIBentrySTDinterwordspacing

\bibitem{Scloud}
\BIBentryALTinterwordspacing
Z.~Zheng, A.~Wang, H.~Fan, C.~Zhao, C.~Liu, and X.~Zhang, ``Scloud: Public key encryption and key encapsulation mechanism based on learning with errors,'' Cryptology ePrint Archive, Paper 2020/095, 2020. [Online]. Available: \url{https://eprint.iacr.org/2020/095}
\BIBentrySTDinterwordspacing

\bibitem{SMAUG}
\BIBentryALTinterwordspacing
J.~H. Cheon, H.~Choe, D.~Hong, and M.~Yi, ``Smaug: Pushing lattice-based key encapsulation mechanisms to the limits,'' Cryptology ePrint Archive, Paper 2023/739, 2023. [Online]. Available: \url{https://eprint.iacr.org/2023/739}
\BIBentrySTDinterwordspacing

\bibitem{Tiger}
\BIBentryALTinterwordspacing
S.~Park, C.-G. Jung, A.~Park, J.~Choi, and H.~Kang, ``Tiger: Tiny bandwidth key encapsulation mechanism for easy migration based on rlwe(r),'' Cryptology ePrint Archive, Paper 2022/1651, 2022. [Online]. Available: \url{https://eprint.iacr.org/2022/1651}
\BIBentrySTDinterwordspacing

\bibitem{Dilithium}
\BIBentryALTinterwordspacing
L.~Ducas, T.~Lepoint, V.~Lyubashevsky, P.~Schwabe, G.~Seiler, and D.~Stehle, ``Crystals -- dilithium: Digital signatures from module lattices,'' Cryptology ePrint Archive, Paper 2017/633, 2017. [Online]. Available: \url{https://eprint.iacr.org/2017/633}
\BIBentrySTDinterwordspacing

\bibitem{Aigis-sig}
\BIBentryALTinterwordspacing
J.~Zhang, Y.~Yu, S.~Fan, Z.~Zhang, and K.~Yang, ``Tweaking the asymmetry of asymmetric-key cryptography on lattices: Kems and signatures of smaller sizes,'' Cryptology ePrint Archive, Paper 2019/510, 2019. [Online]. Available: \url{https://eprint.iacr.org/2019/510}
\BIBentrySTDinterwordspacing

\bibitem{Mulan}
\BIBentryALTinterwordspacing
J.~Zheng, F.~He, S.~Shen, C.~Xue, and Y.~Zhao, ``Parallel small polynomial multiplication for dilithium: A faster design and implementation,'' in \emph{Proceedings of the 38th Annual Computer Security Applications Conference}, ser. ACSAC '22.\hskip 1em plus 0.5em minus 0.4em\relax New York, NY, USA: Association for Computing Machinery, 2022, p. 304–317. [Online]. Available: \url{https://doi.org/10.1145/3564625.3564629}
\BIBentrySTDinterwordspacing

\bibitem{NCC-sign}
\BIBentryALTinterwordspacing
K.-A. Shim, J.~Kim, and Y.~An. Ncc-sign: A new lattice-based signature scheme using non-cyclotomic polynomials. (Accessed on 03/2024). [Online]. Available: \url{https://www.kpqc.or.kr/images/pdf/NCC-Sign.pdf}
\BIBentrySTDinterwordspacing

\bibitem{HAETAE}
\BIBentryALTinterwordspacing
J.~H. Cheon, H.~Choe, J.~Devevey, T.~Güneysu, D.~Hong, M.~Krausz, G.~Land, M.~Möller, D.~Stehlé, and M.~Yi, ``Haetae: Shorter lattice-based fiat-shamir signatures,'' Cryptology ePrint Archive, Paper 2023/624, 2023. [Online]. Available: \url{https://eprint.iacr.org/2023/624}
\BIBentrySTDinterwordspacing

\bibitem{NTRUCrystalSaber}
V.~B. Dang, K.~Mohajerani, and K.~Gaj, ``High-speed hardware architectures and fpga benchmarking of crystals-kyber, ntru, and saber,'' \emph{IEEE Transactions on Computers}, vol.~72, no.~2, pp. 306--320, 2023.

\bibitem{NTRU+1}
Z.~Qin, R.~Tong, X.~Wu, G.~Bai, L.~Wu, and L.~Su, ``A compact full hardware implementation of pqc algorithm ntru,'' in \emph{2021 International Conference on Communications, Information System and Computer Engineering (CISCE)}, 2021, pp. 792--797.

\bibitem{Falcon1}
S.~Paul and P.~Scheible, ``Towards post-quantum security for cyber-physical systems: Integrating pqc into industrial m2m communication,'' in \emph{Computer Security -- ESORICS 2020}, L.~Chen, N.~Li, K.~Liang, and S.~Schneider, Eds.\hskip 1em plus 0.5em minus 0.4em\relax Cham: Springer International Publishing, 2020, pp. 295--316.

\bibitem{CrystalKyber1}
W.~Guo, S.~Li, and L.~Kong, ``An efficient implementation of kyber,'' \emph{IEEE Transactions on Circuits and Systems II: Express Briefs}, vol.~69, pp. 1562--1566, 2022.

\bibitem{CrystalKyber2}
Y.~Huang, M.~Huang, Z.~Lei, and J.~Wu, ``A pure hardware implementation of crystals-kyber pqc algorithm through resource reuse,'' \emph{IEICE Electron. Express}, vol.~17, p. 20200234, 2020.

\bibitem{kyberARM64}
P.~Sanal, E.~Karagoz, H.~Seo, R.~Azarderakhsh, and M.~Mozaffari-Kermani, ``Kyber on arm64: Compact implementations of kyber on 64-bit arm cortex-a processors,'' in \emph{Security and Privacy in Communication Networks}, J.~Garcia-Alfaro, S.~Li, R.~Poovendran, H.~Debar, and M.~Yung, Eds.\hskip 1em plus 0.5em minus 0.4em\relax Cham: Springer International Publishing, 2021, pp. 424--440.

\bibitem{NTT2}
A.~Satriawan, I.~Syafalni, R.~Mareta, I.~Anshori, W.~Shalannanda, and A.~Barra, ``Conceptual review on number theoretic transform and comprehensive review on its implementations,'' \emph{IEEE Access}, vol.~11, pp. 70\,288--70\,316, 2023.

\bibitem{NTT}
A.~Sarker, M.~Mozaffari-Kermani, and R.~Azarderakhsh, ``Hardware constructions for error detection of number-theoretic transform utilized in secure cryptographic architectures,'' \emph{IEEE Transactions on Very Large Scale Integration (VLSI) Systems}, vol.~27, no.~3, pp. 738--741, 2019.

\bibitem{Saber1}
\BIBentryALTinterwordspacing
S.~Sinha~Roy and A.~Basso, ``High-speed instruction-set coprocessor for lattice-based key encapsulation mechanism: Saber in hardware,'' \emph{IACR Transactions on Cryptographic Hardware and Embedded Systems}, vol. 2020, no.~4, p. 443–466, Aug. 2020. [Online]. Available: \url{https://tches.iacr.org/index.php/TCHES/article/view/8690}
\BIBentrySTDinterwordspacing

\bibitem{LAC1}
R.~Tong, Y.~Yin, L.~Wu, X.~Zhang, Z.~Qin, X.~Wu, and L.~Su, ``High-speed hardware implementation of pqc algorithm lac,'' in \emph{2020 IEEE 14th International Conference on Anti-counterfeiting, Security, and Identification (ASID)}, 2020, pp. 104--108.

\bibitem{CRYSTAL_Dilithium1}
L.~Beckwith, D.~Nguyen, and K.~Gaj, ``High-performance hardware implementation of crystals-dilithium,'' \emph{2021 International Conference on Field-Programmable Technology (ICFPT)}, pp. 1--10, 2021.

\bibitem{LatticeSurvey}
\BIBentryALTinterwordspacing
A.~Wang, D.~Xiao, and Y.~Yu, ``Lattice-based cryptosystems in standardisation processes: A survey,'' \emph{IET Information Security}, vol.~17, no.~2, pp. 227--243, 2023. [Online]. Available: \url{https://ietresearch.onlinelibrary.wiley.com/doi/abs/10.1049/ise2.12101}
\BIBentrySTDinterwordspacing

\bibitem{eliece}
R.~J. {McEliece}, ``{A Public-Key Cryptosystem Based On Algebraic Coding Theory},'' \emph{Deep Space Network Progress Report}, vol.~44, pp. 114--116, Jan. 1978.

\bibitem{niederreiter}
H.~Niederreiter, ``{Knapsack-type cryptosystems and algebraic coding theory},'' \emph{Prob. Contr. Inform. Theory}, vol.~15, pp. 157--166, 1986.

\bibitem{equiv}
Y.~X. Li, R.~Deng, and X.~M. Wang, ``On the equivalence of mceliece's and niederreiter's public-key cryptosystems,'' \emph{IEEE Transactions on Information Theory}, vol.~40, no.~1, pp. 271--273, 1994.

\bibitem{francessigunature}
\BIBentryALTinterwordspacing
N.~Courtois, M.~Finiasz, and N.~Sendrier, ``How to achieve a mceliece-based digital signature scheme,'' Cryptology ePrint Archive, Paper 2001/010, 2001. [Online]. Available: \url{https://eprint.iacr.org/2001/010}
\BIBentrySTDinterwordspacing

\bibitem{reviewjavi}
\BIBentryALTinterwordspacing
R.~Overbeck and N.~Sendrier, \emph{Code-based cryptography}.\hskip 1em plus 0.5em minus 0.4em\relax Berlin, Heidelberg: Springer Berlin Heidelberg, 2009, pp. 95--145. [Online]. Available: \url{https://doi.org/10.1007/978-3-540-88702-7_4}
\BIBentrySTDinterwordspacing

\bibitem{complexitydecoding}
E.~Berlekamp, R.~McEliece, and H.~van Tilborg, ``On the inherent intractability of certain coding problems (corresp.),'' \emph{IEEE Transactions on Information Theory}, vol.~24, no.~3, pp. 384--386, 1978.

\bibitem{lee-brickell}
P.~J. Lee and E.~F. Brickell, ``An observation on the security of mceliece's public-key cryptosystem,'' in \emph{Advances in Cryptology --- EUROCRYPT '88}, D.~Barstow, W.~Brauer, P.~Brinch~Hansen, D.~Gries, D.~Luckham, C.~Moler, A.~Pnueli, G.~Seegm{\"u}ller, J.~Stoer, N.~Wirth, and C.~G. G{\"u}nther, Eds.\hskip 1em plus 0.5em minus 0.4em\relax Berlin, Heidelberg: Springer Berlin Heidelberg, 1988, pp. 275--280.

\bibitem{BIKE}
\BIBentryALTinterwordspacing
N.~Aragon, P.~S. Barreto, S.~Bettaieb, L.~Bidoux, O.~Blazy, J.-C. Deneuville, P.~Gaborit, S.~Gueron, T.~Guneysu, C.~A. Melchor \emph{et~al.} (2017) Bike: bit flipping key encapsulation. (Accessed on 03/2024). [Online]. Available: \url{https://bikesuite.org/}
\BIBentrySTDinterwordspacing

\bibitem{NIST3}
\BIBentryALTinterwordspacing
G.~Alagic, D.~Cooper, Q.~Dang, T.~Dang, J.~M. Kelsey, J.~Lichtinger, Y.-K. Liu, C.~A. Miller, D.~Moody, R.~Peralta, R.~Perlner, A.~Robinson, D.~Smith-Tone, and D.~Apon, ``\BIBforeignlanguage{en}{Status report on the third round of the nist post-quantum cryptography standardization process},'' 2022-07-05 04:07:00 2022. [Online]. Available: \url{https://tsapps.nist.gov/publication/get_pdf.cfm?pub_id=934458}
\BIBentrySTDinterwordspacing

\bibitem{PALOMA}
\BIBentryALTinterwordspacing
D.-C. Kim, C.-Y. Jeon, Y.~Kim, and M.~Kim. Paloma: Binary separable goppa-based kem. (Accessed on 03/2024). [Online]. Available: \url{https://www.kpqc.or.kr/images/pdf/PALOMA.pdf}
\BIBentrySTDinterwordspacing

\bibitem{KpqC}
\BIBentryALTinterwordspacing
Korean post-quantum cryptography. (Accessed on 03/2024). [Online]. Available: \url{https://www.kpqc.or.kr/}
\BIBentrySTDinterwordspacing

\bibitem{McEliece}
R.~J. {McEliece}, ``{A Public-Key Cryptosystem Based On Algebraic Coding Theory},'' \emph{Deep Space Network Progress Report}, vol.~44, pp. 114--116, Jan. 1978.

\bibitem{HQC}
\BIBentryALTinterwordspacing
C.~Aguilar-Melchor, N.~Aragon, S.~Bettaieb, L.~Bidoux, O.~Blazy, J.-C. Deneuville, P.~Gaborit, E.~Persichetti, and G.~Z{\'e}mor, ``Hamming quasi-cyclic (hqc),'' 2017. [Online]. Available: \url{https://api.semanticscholar.org/CorpusID:127090340}
\BIBentrySTDinterwordspacing

\bibitem{REDOG}
\BIBentryALTinterwordspacing
J.-L. Kim, J.~Hong, T.~S.~C. Lau, Y.~Lim, and B.-S. Won, ``Redog and its performance analysis,'' Cryptology ePrint Archive, Paper 2022/1663, 2022. [Online]. Available: \url{https://eprint.iacr.org/2022/1663}
\BIBentrySTDinterwordspacing

\bibitem{BIKE1}
J.~Richter-Brockmann, J.~Mono, and T.~Güneysu, ``Folding bike: Scalable hardware implementation for reconfigurable devices,'' \emph{IEEE Transactions on Computers}, vol.~71, no.~5, pp. 1204--1215, 2022.

\bibitem{McEliece1}
\BIBentryALTinterwordspacing
``Complete and improved fpga implementation of classic mceliece,'' vol. 2022, p. 71–113, Jun. 2022. [Online]. Available: \url{https://tches.iacr.org/index.php/TCHES/article/view/9695}
\BIBentrySTDinterwordspacing

\bibitem{HQC1}
\BIBentryALTinterwordspacing
S.~Deshpande, C.~Xu, M.~Nawan, K.~Nawaz, and J.~Szefer, ``Fast and efficient hardware implementation of hqc,'' Cryptology ePrint Archive, Paper 2022/1183, 2022, \url{https://eprint.iacr.org/2022/1183}. [Online]. Available: \url{https://eprint.iacr.org/2022/1183}
\BIBentrySTDinterwordspacing

\bibitem{MultivariateOrig}
T.~Matsumoto and H.~Imai, ``Public quadratic polynomial-tuples for efficient signature-verification and message-encryption,'' in \emph{Advances in Cryptology --- EUROCRYPT '88}, D.~Barstow, W.~Brauer, P.~Brinch~Hansen, D.~Gries, D.~Luckham, C.~Moler, A.~Pnueli, G.~Seegm{\"u}ller, J.~Stoer, N.~Wirth, and C.~G. G{\"u}nther, Eds.\hskip 1em plus 0.5em minus 0.4em\relax Berlin, Heidelberg: Springer Berlin Heidelberg, 1988, pp. 419--453.

\bibitem{F5}
\BIBentryALTinterwordspacing
J.~C. Faug\`{e}re, ``A new efficient algorithm for computing gr\"{o}bner bases without reduction to zero (f5),'' in \emph{Proceedings of the 2002 International Symposium on Symbolic and Algebraic Computation}, ser. ISSAC '02.\hskip 1em plus 0.5em minus 0.4em\relax New York, NY, USA: Association for Computing Machinery, 2002, p. 75–83. [Online]. Available: \url{https://doi.org/10.1145/780506.780516}
\BIBentrySTDinterwordspacing

\bibitem{MQQuantum}
\BIBentryALTinterwordspacing
T.~Yasuda, X.~Dahan, Y.-J. Huang, T.~Takagi, and K.~Sakurai, ``A multivariate quadratic challenge toward post-quantum generation cryptography,'' \emph{ACM Commun. Comput. Algebra}, vol.~49, no.~3, p. 105–107, nov 2015. [Online]. Available: \url{https://doi.org/10.1145/2850449.2850462}
\BIBentrySTDinterwordspacing

\bibitem{Rainbow}
J.~Ding and D.~Schmidt, ``Rainbow, a new multivariable polynomial signature scheme,'' in \emph{Applied Cryptography and Network Security}, J.~Ioannidis, A.~Keromytis, and M.~Yung, Eds.\hskip 1em plus 0.5em minus 0.4em\relax Berlin, Heidelberg: Springer Berlin Heidelberg, 2005, pp. 164--175.

\bibitem{GeMSS}
\BIBentryALTinterwordspacing
A.~Casanova, J.-C. Faug{\`e}re, G.~Macario-Rat, J.~Patarin, L.~Perret, and J.~Ryckeghem, ``Gemss: A great multivariate short signature,'' 2017. [Online]. Available: \url{https://api.semanticscholar.org/CorpusID:8432066}
\BIBentrySTDinterwordspacing

\bibitem{Rainbowbroken}
\BIBentryALTinterwordspacing
W.~Beullens, ``Breaking rainbow takes a weekend on a laptop,'' Cryptology ePrint Archive, Paper 2022/214, 2022, \url{https://eprint.iacr.org/2022/214}. [Online]. Available: \url{https://eprint.iacr.org/2022/214}
\BIBentrySTDinterwordspacing

\bibitem{MQ-sign}
\BIBentryALTinterwordspacing
K.-A. Shim, J.~Kim, and Y.~An, ``Mq-sign: A new post-quantum signature scheme based on multivariate quadratic equations: Shorter and faster.'' [Online]. Available: \url{www.kpqc.or.kr}
\BIBentrySTDinterwordspacing

\bibitem{Rainbow1}
A.~Ferozpuri and K.~Gaj, ``High-speed fpga implementation of the nist round 1 rainbow signature scheme,'' in \emph{2018 International Conference on ReConFigurable Computing and FPGAs (ReConFig)}, 2018, pp. 1--8.

\bibitem{Miller}
V.~S. Miller, ``Use of elliptic curves in cryptography,'' in \emph{Advances in Cryptology --- CRYPTO '85 Proceedings}, H.~C. Williams, Ed.\hskip 1em plus 0.5em minus 0.4em\relax Berlin, Heidelberg: Springer Berlin Heidelberg, 1986, pp. 417--426.

\bibitem{KoblitzECC}
\BIBentryALTinterwordspacing
N.~Koblitz, ``Elliptic curve cryptosystems,'' \emph{Mathematics of Computation}, vol.~48, pp. 203--209, 1987. [Online]. Available: \url{https://api.semanticscholar.org/CorpusID:14288427}
\BIBentrySTDinterwordspacing

\bibitem{Hasse1936}
\BIBentryALTinterwordspacing
H.~Hasse, ``Zur theorie der abstrakten elliptischen funktionenkörper iii. die struktur des meromorphismenrings. die riemannsche vermutung.'' \emph{Journal für die reine und angewandte Mathematik}, vol. 175, pp. 193--208, 1936. [Online]. Available: \url{http://eudml.org/doc/149968}
\BIBentrySTDinterwordspacing

\bibitem{SIDH2}
D.~Jao and L.~De~Feo, ``Towards quantum-resistant cryptosystems from supersingular elliptic curve isogenies,'' in \emph{Post-Quantum Cryptography}, B.-Y. Yang, Ed.\hskip 1em plus 0.5em minus 0.4em\relax Berlin, Heidelberg: Springer Berlin Heidelberg, 2011, pp. 19--34.

\bibitem{SIDH1}
\BIBentryALTinterwordspacing
A.~Rostovtsev and A.~Stolbunov, ``Public-key cryptosystem based on isogenies,'' Cryptology ePrint Archive, Paper 2006/145, 2006. [Online]. Available: \url{https://eprint.iacr.org/2006/145}
\BIBentrySTDinterwordspacing

\bibitem{HHS}
\BIBentryALTinterwordspacing
J.-M. Couveignes, ``Hard homogeneous spaces,'' Cryptology ePrint Archive, Paper 2006/291, 2006. [Online]. Available: \url{https://eprint.iacr.org/2006/291}
\BIBentrySTDinterwordspacing

\bibitem{Tani_2009}
\BIBentryALTinterwordspacing
S.~Tani, ``Claw finding algorithms using quantum walk,'' \emph{Theoretical Computer Science}, vol. 410, no.~50, pp. 5285--5297, nov 2009. [Online]. Available: \url{https://doi.org/10.1016%2Fj.tcs.2009.08.030}
\BIBentrySTDinterwordspacing

\bibitem{AttackSIKE}
\BIBentryALTinterwordspacing
P.~C. Oorschot and M.~J. Wiener, ``Parallel collision search with cryptanalytic applications,'' \emph{J. Cryptol.}, vol.~12, no.~1, p. 1–28, jan 1999. [Online]. Available: \url{https://doi.org/10.1007/PL00003816}
\BIBentrySTDinterwordspacing

\bibitem{SIKE}
\BIBentryALTinterwordspacing
B.~Koziel, A.-B. Ackie, R.~E. Khatib, R.~Azarderakhsh, and M.~Mozaffari-Kermani, ``Sike'd up: Fast and secure hardware architectures for supersingular isogeny key encapsulation,'' Cryptology ePrint Archive, Paper 2019/711, 2019. [Online]. Available: \url{https://eprint.iacr.org/2019/711}
\BIBentrySTDinterwordspacing

\bibitem{FIBS}
\BIBentryALTinterwordspacing
S.~Kim, Y.~Lee, and K.~Yoon. Fibs : Fast isogeny based digital signature. (Accessed on 03/2024). [Online]. Available: \url{https://www.kpqc.or.kr/images/pdf/FIBS.pdf}
\BIBentrySTDinterwordspacing

\bibitem{CompressedSIKE}
\BIBentryALTinterwordspacing
R.~Azarderakhsh, D.~Jao, K.~Kalach, B.~Koziel, and C.~Leonardi, ``Key compression for isogeny-based cryptosystems,'' Cryptology ePrint Archive, Paper 2016/229, 2016, \url{https://eprint.iacr.org/2016/229}. [Online]. Available: \url{https://eprint.iacr.org/2016/229}
\BIBentrySTDinterwordspacing

\bibitem{SIKEARM}
M.~Anastasova, R.~Azarderakhsh, and M.~M. Kermani, ``Fast strategies for the implementation of sike round 3 on arm cortex-m4,'' \emph{IEEE Transactions on Circuits and Systems I: Regular Papers}, vol.~68, no.~10, pp. 4129--4141, 2021.

\bibitem{SIKEFPGA}
\BIBentryALTinterwordspacing
R.~Elkhatib, B.~Koziel, R.~Azarderakhsh, and M.~Mozaffari~Kermani, ``Cryptographic engineering a fast and efficient sike in fpga,'' \emph{ACM Trans. Embed. Comput. Syst.}, vol.~23, no.~2, mar 2024. [Online]. Available: \url{https://doi.org/10.1145/3584919}
\BIBentrySTDinterwordspacing

\bibitem{NEON-SIDH}
\BIBentryALTinterwordspacing
B.~Koziel, A.~Jalali, R.~Azarderakhsh, M.~M. Kermani, and D.~Jao, ``Neon-sidh: Efficient implementation of supersingular isogeny diffie-hellman key-exchange protocol on arm,'' Cryptology ePrint Archive, Paper 2016/669, 2016, \url{https://eprint.iacr.org/2016/669}. [Online]. Available: \url{https://eprint.iacr.org/2016/669}
\BIBentrySTDinterwordspacing

\bibitem{MPC}
\BIBentryALTinterwordspacing
Y.~Ishai, E.~Kushilevitz, R.~Ostrovsky, and A.~Sahai, ``Zero-knowledge from secure multiparty computation,'' in \emph{Proceedings of the Thirty-Ninth Annual ACM Symposium on Theory of Computing}, ser. STOC '07.\hskip 1em plus 0.5em minus 0.4em\relax New York, NY, USA: Association for Computing Machinery, 2007, p. 21–30. [Online]. Available: \url{https://doi.org/10.1145/1250790.1250794}
\BIBentrySTDinterwordspacing

\bibitem{MPCapplication}
I.~Giacomelli, J.~Madsen, and C.~Orlandi, ``Zkboo: Faster zero-knowledge for boolean circuits,'' in \emph{Proceedings of the 25th USENIX Conference on Security Symposium}, ser. SEC'16.\hskip 1em plus 0.5em minus 0.4em\relax USA: USENIX Association, 2016, p. 1069–1083.

\bibitem{Picnic}
\BIBentryALTinterwordspacing
M.~Chase, D.~Derler, S.~Goldfeder, C.~Orlandi, S.~Ramacher, C.~Rechberger, D.~Slamanig, and G.~Zaverucha, ``Post-quantum zero-knowledge and signatures from symmetric-key primitives,'' Cryptology ePrint Archive, Paper 2017/279, 2017. [Online]. Available: \url{https://eprint.iacr.org/2017/279}
\BIBentrySTDinterwordspacing

\bibitem{AIMer}
\BIBentryALTinterwordspacing
S.~Kim, J.~Ha, M.~Son, B.~Lee, D.~Moon, J.~Lee, S.~Lee, J.~Kwon, J.~Cho, H.~Yoon, and J.~Lee, ``Aim: Symmetric primitive for shorter signatures with stronger security (full version),'' Cryptology ePrint Archive, Paper 2022/1387, 2022. [Online]. Available: \url{https://eprint.iacr.org/2022/1387}
\BIBentrySTDinterwordspacing

\bibitem{Koblitz}
M.~Fellows and N.~Koblitz, ``Kid krypto,'' in \emph{Advances in Cryptology --- CRYPTO' 92}, E.~F. Brickell, Ed.\hskip 1em plus 0.5em minus 0.4em\relax Berlin, Heidelberg: Springer Berlin Heidelberg, 1993, pp. 371--389.

\bibitem{PCC-HP}
\BIBentryALTinterwordspacing
J.~Kratochvíl, ``Perfect codes over graphs,'' \emph{Journal of Combinatorial Theory, Series B}, vol.~40, no.~2, pp. 224--228, 1986. [Online]. Available: \url{https://www.sciencedirect.com/science/article/pii/0095895686900791}
\BIBentrySTDinterwordspacing

\bibitem{IPCC}
\BIBentryALTinterwordspacing
J.~Ryu, Y.~Kim, S.~Yoon, J.-S. Kang, and Y.~Yeom. Ipcc - improved perfect code cryptosystems. (Accessed on 03/2024). [Online]. Available: \url{https://www.kpqc.or.kr/images/pdf/IPCC.pdf}
\BIBentrySTDinterwordspacing

\bibitem{NTRUECCperf}
I.~Upasana, N.~Nandanavanam, A.~Nandanavanam, and N.~Naaz, ``Performance characteristics of ntru and ecc cryptosystem in context of iot environment,'' in \emph{2020 IEEE International Conference on Distributed Computing, VLSI, Electrical Circuits and Robotics (DISCOVER)}, 2020, pp. 23--28.

\bibitem{NTRUHRSSPerf}
\BIBentryALTinterwordspacing
Z.~Liang, B.~Fang, J.~Zheng, and Y.~Zhao, ``Compact and efficient kems over ntru lattices,'' Cryptology ePrint Archive, Paper 2022/579, 2022. [Online]. Available: \url{https://eprint.iacr.org/2022/579}
\BIBentrySTDinterwordspacing

\bibitem{NTRUPrimePerf}
\BIBentryALTinterwordspacing
Ntru prime speed. (Accessed on 03/2024). [Online]. Available: \url{https://ntruprime.cr.yp.to/speed.html}
\BIBentrySTDinterwordspacing

\bibitem{KorPerf}
\BIBentryALTinterwordspacing
H.~Kwon, M.~Sim, G.~Song, M.~Lee, and H.~Seo, ``Evaluating kpqc algorithm submissions: Balanced and clean benchmarking approach,'' Cryptology ePrint Archive, Paper 2023/1163, 2023. [Online]. Available: \url{https://eprint.iacr.org/2023/1163}
\BIBentrySTDinterwordspacing

\bibitem{SPHINCSKyberPerf}
\BIBentryALTinterwordspacing
K.~Bürstinghaus-Steinbach, C.~Krauß, R.~Niederhagen, and M.~Schneider, ``Post-quantum tls on embedded systems,'' Cryptology ePrint Archive, Paper 2020/308, 2020. [Online]. Available: \url{https://eprint.iacr.org/2020/308}
\BIBentrySTDinterwordspacing

\bibitem{Aigis-Enc2}
Y.~Hu, S.~Dong, and X.~Dong, ``Analysis on aigis‐enc: Asymmetrical and symmetrical,'' \emph{IET Information Security}, vol.~15, 03 2021.

\bibitem{TALE2}
\BIBentryALTinterwordspacing
S.~Zhou, H.~Xue, D.~Zhang, K.~Wang, X.~Lu, B.~Li, and J.~He, ``Preprocess-then-ntt technique and its applications to kyber and newhope,'' Cryptology ePrint Archive, Paper 2018/995, 2018. [Online]. Available: \url{https://eprint.iacr.org/2018/995}
\BIBentrySTDinterwordspacing

\bibitem{GCKSign}
\BIBentryALTinterwordspacing
J.~Woo, K.~Lee, and J.~H. Park, ``Gcksign: Simple and efficient signatures from generalized compact knapsacks,'' Cryptology ePrint Archive, Paper 2022/1665, 2022. [Online]. Available: \url{https://eprint.iacr.org/2022/1665}
\BIBentrySTDinterwordspacing

\bibitem{Levels}
\BIBentryALTinterwordspacing
T.~N.~I. of~Standards and T.~(NIST). (2016, 12) Submission requirements and evaluation criteria for the post-quantum cryptography standardization process. (Accessed on 03/2024). [Online]. Available: \url{https://csrc.nist.gov/CSRC/media/Projects/Post-Quantum-Cryptography/documents/call-for-proposals-final-dec-2016.pdf}
\BIBentrySTDinterwordspacing

\bibitem{Piglet}
L.~Wang and J.~Hu, ``Two new module-code-based kems with rank metric,'' 05 2019, pp. 176--191.

\bibitem{Rollo}
\BIBentryALTinterwordspacing
C.~Kim, Y.-S. Kim, and J.-S. No, ``Layered rollo-i: Faster rank-metric code-based kem using ideal lrpc codes,'' Cryptology ePrint Archive, Paper 2022/1572, 2022. [Online]. Available: \url{https://eprint.iacr.org/2022/1572}
\BIBentrySTDinterwordspacing

\bibitem{pqsigRM}
\BIBentryALTinterwordspacing
J.~Cho, J.-S. No, Y.~Lee, Z.~Koo, and Y.-S. Kim, ``Enhanced pqsigrm: Code-based digital signature scheme with short signature and fast verification for post-quantum cryptography,'' Cryptology ePrint Archive, Paper 2022/1493, 2022. [Online]. Available: \url{https://eprint.iacr.org/2022/1493}
\BIBentrySTDinterwordspacing

\bibitem{Perform-lattice-multivariate}
Y.-A. Chang, M.-S. Chen, J.-S. Wu, and B.-Y. Yang, ``Postquantum ssl/tls for embedded systems,'' in \emph{2014 IEEE 7th International Conference on Service-Oriented Computing and Applications}, 2014, pp. 266--270.

\bibitem{LWEQuantum}
\BIBentryALTinterwordspacing
Y.~Chen, ``Quantum algorithms for lattice problems,'' Cryptology ePrint Archive, Paper 2024/555, 2024, \url{https://eprint.iacr.org/2024/555}. [Online]. Available: \url{https://eprint.iacr.org/2024/555}
\BIBentrySTDinterwordspacing

\bibitem{ETSI}
\BIBentryALTinterwordspacing
European telecommunications standards institute. (Accessed on 03/2024). [Online]. Available: \url{https://www.etsi.org/}
\BIBentrySTDinterwordspacing

\bibitem{CACR}
\BIBentryALTinterwordspacing
Chinese association for cryptologic research. (Accessed on 03/2024). [Online]. Available: \url{https://www.cacrnet.org.cn/}
\BIBentrySTDinterwordspacing

\bibitem{cryptrec}
\BIBentryALTinterwordspacing
Cryptography research and evaluation committees. (Accessed on 03/2024). [Online]. Available: \url{https://www.cryptrec.go.jp/en/}
\BIBentrySTDinterwordspacing

\bibitem{QURECA23}
\BIBentryALTinterwordspacing
Overview of quantum initiatives worldwide 2023. (Accessed on 03/2024). [Online]. Available: \url{https://qureca.com/es/overview-of-quantum-initiatives-worldwide-2023/}
\BIBentrySTDinterwordspacing

\bibitem{QURECA22}
\BIBentryALTinterwordspacing
(2022) Overview of quantum initiatives worldwide 2022. (Accessed on 03/2024). [Online]. Available: \url{https://qureca.com/overview-on-quantum-initiatives-worldwide-update-2022/}
\BIBentrySTDinterwordspacing

\bibitem{RSAcrackIBM}
\BIBentryALTinterwordspacing
How quantum computers can break the internet... starting now no secret is safe. (Accessed on 03/2024). [Online]. Available: \url{https://www.veritasium.com/videos/2023/4/14/how-quantum-computers-break-the-internet-starting-now/}
\BIBentrySTDinterwordspacing

\bibitem{Cloudfare}
\BIBentryALTinterwordspacing
The tls post-quantum experiment. (Accessed on 03/2024). [Online]. Available: \url{https://blog.cloudflare.com/the-tls-post-quantum-experiment}
\BIBentrySTDinterwordspacing

\bibitem{SCA}
N.~Tsalis, E.~Vasilellis, D.~Mentzelioti, and T.~Apostolopoulos, ``A taxonomy of side channel attacks on critical infrastructures and relevant systems,'' \emph{Advanced Sciences and Technologies for Security Applications}, 2019.

\bibitem{SCA-PLC}
D.~Tychalas and M.~Maniatakos, ``Special session: Potentially leaky controller: Examining cache side-channel attacks in programmable logic controllers,'' in \emph{2020 IEEE 38th International Conference on Computer Design (ICCD)}, 2020, pp. 33--36.

\bibitem{Pomaranch}
M.~Mozaffari-Kermani, R.~Azarderakhsh, and A.~Aghaie, ``Reliable and error detection architectures of pomaranch for false-alarm-sensitive cryptographic applications,'' \emph{IEEE Transactions on Very Large Scale Integration (VLSI) Systems}, vol.~23, no.~12, pp. 2804--2812, 2015.

\bibitem{NTT1}
K.~Ahmadi, S.~Aghapour, M.~M. Kermani, and R.~Azarderakhsh, ``Efficient algorithm level error detection for number-theoretic transform assessed on fpgas,'' 2024.

\bibitem{DVFS}
\BIBentryALTinterwordspacing
T.~Yu, C.~Cheng, Z.~Yang, Y.~Wang, Y.~Pan, and J.~Weng, ``Hints from hertz: Dynamic frequency scaling side-channel analysis of number theoretic transform in lattice-based kems,'' Cryptology ePrint Archive, Paper 2024/070, 2024, \url{https://eprint.iacr.org/2024/070}. [Online]. Available: \url{https://eprint.iacr.org/2024/070}
\BIBentrySTDinterwordspacing

\bibitem{Faultdet}
\BIBentryALTinterwordspacing
M.~Mozaffari-Kermani, R.~Azarderakhsh, and A.~Aghaie, ``Fault detection architectures for post-quantum cryptographic stateless hash-based secure signatures benchmarked on asic,'' \emph{ACM Trans. Embed. Comput. Syst.}, vol.~16, no.~2, dec 2016. [Online]. Available: \url{https://doi.org/10.1145/2930664}
\BIBentrySTDinterwordspacing

\bibitem{Faultdet1}
M.~Mozaffari-Kermani and R.~Azarderakhsh, ``Reliable hash trees for post-quantum stateless cryptographic hash-based signatures,'' in \emph{2015 IEEE International Symposium on Defect and Fault Tolerance in VLSI and Nanotechnology Systems (DFTS)}, 2015, pp. 103--108.

\bibitem{PQCIoT}
\BIBentryALTinterwordspacing
Z.~Ye, R.~Song, H.~Zhang, D.~Chen, R.~C.-C. Cheung, and K.~Huang, ``A highly-efficient lattice-based post-quantum cryptography processor for iot applications,'' \emph{IACR Transactions on Cryptographic Hardware and Embedded Systems}, vol. 2024, no.~2, p. 130–153, Mar. 2024. [Online]. Available: \url{https://tches.iacr.org/index.php/TCHES/article/view/11423}
\BIBentrySTDinterwordspacing

\bibitem{ETSI-Whitepaper15}
\BIBentryALTinterwordspacing
M.~Campagna, L.~Chen, {\" O}.~Dagdelen, J.~Ding, J.~K. Fernick, N.~Gisin, D.~Hayford, T.~Jennewein, N.~L{\" u}tkenhaus, M.~Mosca, B.~Neill, M.~Pecen, R.~Perlner, G.~Ribordy, J.~M. Schanck, D.~Stebila, N.~Walenta, W.~Whyte, and Z.~Zhang, ``Quantum safe cryptography and security: An introduction, benefits, enablers and challengers,'' ETSI (European Telecommunications Standards Institute), Tech. Rep., June 2015. [Online]. Available: \url{http://www.etsi.org/images/files/ETSIWhitePapers/QuantumSafeWhitepaper.pdf}
\BIBentrySTDinterwordspacing

\bibitem{KyberControversy}
\BIBentryALTinterwordspacing
D.~Bernstein. Nsa, nist, and post-quantum cryptography. (Accessed on 03/2024). [Online]. Available: \url{https://www.muckrock.com/foi/united-states-of-america-10/nsa-nist-and-post-quantum-cryptography-126349/}
\BIBentrySTDinterwordspacing

\bibitem{lowpowermidori}
M.~Yoshikawa and Y.~Nozaki, ``Electromagnetic analysis method for ultra low power cipher midori,'' in \emph{2017 IEEE 8th Annual Ubiquitous Computing, Electronics and Mobile Communication Conference (UEMCON)}, 2017, pp. 70--75.

\bibitem{iotSensEnergy}
\BIBentryALTinterwordspacing
H.~N.~S. Aldin, M.~R. Ghods, F.~Nayebipour, and M.~N. Torshiz, ``A comprehensive review of energy harvesting and routing strategies for iot sensors sustainability and communication technology,'' \emph{Sensors International}, vol.~5, p. 100258, 2024. [Online]. Available: \url{https://www.sciencedirect.com/science/article/pii/S2666351123000323}
\BIBentrySTDinterwordspacing

\end{thebibliography}

\end{document}